\documentclass[conference]{IEEEtran}
\IEEEoverridecommandlockouts
\usepackage[hyphens]{url}
\usepackage{hyperref}
\usepackage[hyphenbreaks]{breakurl}
\usepackage{cite}
\usepackage[numbers,sort&compress]{natbib}
\usepackage{amsthm}
\usepackage{amsmath,amssymb,amsfonts}
\usepackage{algorithm}
\usepackage{float}
\usepackage[caption=false,font=footnotesize]{subfig}
\usepackage[noend]{algorithmic}
\usepackage{graphicx}
\usepackage{wrapfig}
\usepackage{textcomp}
\usepackage{color}
\usepackage{tcolorbox}
\usepackage{xhfill}
\usepackage{marvosym}
\usepackage{paralist}
\usepackage{makecell}
\usepackage{multirow}
\usepackage{multicol}
\usepackage{diagbox}
\usepackage{lipsum}
\usepackage{textcomp}
\usepackage{xcolor}

\def\BibTeX{{\rm B\kern-.05em{\sc i\kern-.025em b}\kern-.08em
    T\kern-.1667em\lower.7ex\hbox{E}\kern-.125emX}}
\renewcommand{\algorithmiccomment}[1]{\bgroup\hfill\scriptsize$\triangleright$~#1\egroup} 
\begin{document}
\newcommand{\ignore}[1]{}
\newcommand{\eat}[1]{}

\def\bitcoinA{%
  \leavevmode
  \vtop{\offinterlineskip 
    \setbox0=\hbox{B}%
    \setbox2=\hbox to\wd0{\hfil\hskip-.03em
    \vrule height .3ex width .15ex\hskip .08em
    \vrule height .3ex width .15ex\hfil}
    \vbox{\copy2\box0}\box2}}

\newcommand{\system}{$\mathsf{FairThunder}$}
\newcommand{\systemd}{$\mathsf{FairDownload}$}
\newcommand{\systems}{$\mathsf{FairStream}$}
\newcommand{\MT}{\mathsf{MT}}
\newcommand{\cR}{\mathsf{root}}
\newcommand{\adv}{\mathcal{A}}
\newcommand{\VFD}{\mathcal{F}_{\mathsf{VFD}}}
\newcommand{\PoD}{\mathcal{F}_\mathsf{PoD}}
\newcommand{\PoDD}{\mathcal{F}_\mathsf{dPoD}}
\newcommand{\PoDS}{\mathcal{F}_{\mathsf{sPoD}}}
\newcommand{\PoM}{\mathsf{ValidatePoM}}

\newtheorem{definition}{Definition}
\newtheorem{theorem}{Theorem}
\newtheorem{lemma}{{\bf Lemma}}
\newtheorem{corollary}{Corollary}

\newcommand{\yuan}[1]{\textcolor{red}{#1}}
\newcommand{\qiang}[1]{\textcolor{blue}{#1}}
\newcommand{\songlin}[1]{\textcolor{blue}{#1}}

\newcommand{\tabincell}[2]{\begin{tabular}{@{}#1@{}}#2\end{tabular}}  

\title{Fair Peer-to-Peer Content Delivery via Blockchain}
\author{\IEEEauthorblockN{Songlin He$^1$, Yuan Lu$^{2, *}$, Qiang Tang$^3$, Guiling Wang$^1$, Chase Qishi Wu$^1$}
	\IEEEauthorblockA{$^1$New Jersey Institute of Technology, $^2$Institute of Software  Chinese Academy of Sciences, $^3$The University of Sydney\\
	sh553@njit.edu, luyuan@iscas.ac.cn, qiang.tang@sydney.edu.au, \{gwang, chase.wu\}@njit.edu}
}

\maketitle

\renewcommand*{\thefootnote}{\fnsymbol{footnote}}
\footnotetext[0]{* Correspondence: Yuan Lu, Ph.D., Institute of Software, CAS}
\renewcommand*{\thefootnote}{\arabic{footnote}}

\begin{abstract}
	Peer-to-peer (p2p) content delivery is promising to provide benefits including cost-saving and scalable peak-demand handling in comparison with conventional content delivery networks (CDNs), and it can complement the decentralized storage networks such as Filecoin. However, reliable p2p delivery requires proper enforcement of delivery fairness, i.e., the deliverers should be rewarded according to their in-time delivery. Unfortunately, most existing studies on delivery fairness are based on non-cooperative game-theoretic assumptions that are arguably unrealistic in the ad-hoc p2p setting.
	
	We for the first time put forth an expressive yet still minimalist security requirement for desired fair p2p content delivery,  and  give two efficient solutions \systemd{} and \systems{} via the blockchain for p2p downloading and p2p streaming scenarios, respectively. Our designs not only guarantee delivery fairness to ensure deliverers be paid (nearly) proportional to their in-time delivery but also ensure the content consumers and content providers are fairly treated. The fairness of each party can be guaranteed when the other two parties collude to arbitrarily misbehave. Moreover, the systems are efficient in the sense of attaining asymptotically optimal on-chain costs and optimal deliverer communication.

We implement the protocols to build the prototype systems atop the Ethereum Ropsten network. Extensive experiments done in LAN and WAN settings showcase their high practicality.
\end{abstract}



\section{Introduction}
\label{sec:Introduction}

The peer-to-peer (p2p) content delivery systems are permissionless decentralized services  to seamlessly replicate contents to the end consumers. Typically these systems~\cite{Cohen-2003-P2P,kulbak2005emule} encompass a large ad-hoc network of deliverers such as normal Internet users or small organizations, 
thus overcoming the bandwidth bottleneck of the original content providers. In contrast to  giant pre-planned content delivery networks (i.e., CDNs such as Akamai~\cite{Akamai-2021-COM} and CloudFlare~\cite{CloudFlare-2021-COM}), 
p2p content delivery can crowdsource unused bandwidth resources of tremendous Internet peers, thus having a wide array of benefits including  robust service availability, bandwidth cost savings, and scalable peak-demand handling~\cite{Aalmashaqbeh-2019-Columbia,Anjum-et-al-2017-CN}.

Recently, renewed attentions to p2p content delivery are gathered  \cite{Wang-et-al-2018-ICC,Goyal-et-al-2019-Usenix,Aalmashaqbeh-2019-Columbia} due to the fast popularization of  decentralized storage networks (DSNs)~\cite{Swarm-ethereum-2020,IPFS-Benet-2014,Filecoin-2017-Online, StorJ-2018-WhiteBook, Miller-et-al-2014-SP}. Indeed, DSNs feature decentralized and robust {\em content storage}, but lack well-designed {\em content delivery} mechanisms catering for a prosperous content consumption market in the p2p setting, where the content shall not only be reliably stored but also must be always quickly {\em retrievable} despite potentially malicious participants ~\cite{Content_Consum_WARC-2020-WEB, Filecoin-Retrieval-2021-Spec}. 


The primary challenge of designing a proper delivery mechanism for complementing DSNs is to realize the strict guarantee of ``fairness'' against adversarial peers. In particular, a fair p2p content delivery system has to promise well-deserved items (e.g., retrieval of valid contents, well-paid rewards to spent bandwidth) to all participants~\cite{Fan-et-al-2008-TON}.
Otherwise, free-riding parties can abuse the system~\cite{Feldman-2004-CEC,Locher-et-al-2006-HotNets,Piatek-et-al-2007-NSDI} and cause rational ones to escape, eventually resulting in possible system collapse~\cite{Hardin-et-al-2009-Science}. We reason as follows to distinguish two types of quintessential fairness, namely {\em delivery fairness} and {\em exchange fairness}, in the p2p content delivery setting where three parties, i.e., content {\em provider}, content {\em deliverer} and content {\em consumer}, are involved.

\smallskip
\noindent
{\bf Exchange fairness is not delivery fairness}.
Exchange fairness~\cite{Blum-1983-STOC,Damgaard-1995-Cryptology,Asokan-et-al-2000-JSAC,Kupccu--et-al-2010-RSAC,Dziembowski-et-al-2018-CCS,Maxwell-2016-BitcoinCore}, specifically for digital goods (such as signatures and videos), refers to ensuring one party's input keep {\em confidential} until it does learn the other party's input. Unfortunately, in the p2p content delivery setting, it is insufficient to merely consider exchange fairness because a content deliverer would expect to receive rewards proportional to the bandwidth resources it spends. Noticeably, exchange fairness fails to capture such new desiderata related to bandwidth cost, as it does not rule out that a deliverer may receive no reward after transferring a huge amount of {\em encrypted} data to the other party, which clearly breaks the deliverer's expectation on being well-paid but does not violate  exchange fairness at all.

Consider FairSwap \cite{Dziembowski-et-al-2018-CCS} as a concrete example: the deliverer first  sends  the encrypted content and semantically secure digest to the consumer, then waits for a confirmation message from the consumer (through the blockchain) to confirm her receiving of these ciphertext, so the deliverer can reveal his encryption key to the content consumer via the blockchain; but, in case the consumer aborts, all bandwidth used  to send ciphertext is wasted, causing no reward for the deliverer. A seemingly enticing way to mitigate the above attack on delivery fairness in FairSwap could be splitting the content into $n$ smaller chunks and run FairSwap protocols for each chunk, but the on-chain cost would grow linear in $n$, resulting in prohibitive on-chain cost for large contents such as movies. Adapting other fair exchange protocols for delivery fairness would encounter similar issues like FairSwap. Hence, the efficient construction satisfying delivery fairness remains unclear.

To capture the ``special" exchanged item for deliverers, we formulate delivery fairness (in Sec.~\ref{sec:ProblemFormulation}), stating that deliverers can receive rewards (nearly) proportional to the contributed bandwidth for delivering data to the consumers.

\smallskip
\noindent
{\bf Insufficiencies of existing ``delivery fairness''}. A range of existing literature~\cite{Sherman-et-al-2012-TON,Kamvar-et-al-2003-WWW,Sirivianos-2007-Usenix,Shin-et-al-2017-TON,Levin-et-al-2008-SIGCOMM} involve delivery fairness for p2p delivery. However, to our knowledge, no one assures delivery fairness in the {\em cryptographic} sense, as we seek to do. Specifically, they~\cite{Sherman-et-al-2012-TON,Kamvar-et-al-2003-WWW,Sirivianos-2007-Usenix,Shin-et-al-2017-TON,Levin-et-al-2008-SIGCOMM} are presented in {\em non-cooperative game-theoretic} settings where independent attackers free ride spontaneously without communication of their strategies, and the attackers are rational with the intentions to maximize their own benefits. Therefore, these works boldly ignore that the adversary intends to break the system. Unfortunately, such rational assumptions are particularly elusive  to stand in ad-hoc open systems accessible by all malicious evils. The occurrences of  tremendous  real-world  attacks in ad-hoc open systems~\cite{Mehar-et-al-IGI-2019, Botnet-2020-ENISA}   hint us how vulnerable  the prior  studies' heavy  assumptions can be and further weaken the confidence of using  them in real-world p2p content delivery.

\smallskip
\noindent
{\bf Lifting for ``exchange fairness" between provider and consumer}. Besides  the natural delivery fairness, it is equally vital to ensure exchange fairness for providers and consumers in a basic context of p2p content delivery, especially with the end goal to complement DSNs and enable some content providers to sell contents to consumers with delegating costly  delivery/storage to a p2p network. In particular, the content provider should be guaranteed to receive payments proportional to the amount of correct data learned by the consumer; vice versa, the consumer only has to pay if indeed receiving qualified content.

Na\"\i ve attempts of tuning a fair exchange protocol~\cite{Asokan-et-al-2000-JSAC,Belenkiy-et-al-2007-WPES,Kupccu--et-al-2010-RSAC,Dziembowski-et-al-2018-CCS,Maxwell-2016-BitcoinCore,Eckey-et-al-2019-Arxiv} into p2p content delivery can guarantee neither delivery fairness (as analyzed earlier) nor exchange fairness: simply running fair exchange protocols twice between the deliverers and the content providers and between the deliverers and the consumers, respectively, would leak valuable contents, raising the threat of massive content leakage. Even worse, this idea disincentivizes the deliverers as they have to pay for the whole content before making a life by delivering the content to consumers.


\smallskip
\noindent
{\bf Our contributions.} Overall, it remains an open problem to realize such strong fairness guarantees in p2p content delivery  to protect {\em all} providers, deliverers, and consumers. We for the first time formalize such security intuitions into a well-defined cryptographic problem on fairness, and present a couple of efficient blockchain-based protocols to solve it. In sum, our contributions are:

\begin{enumerate}
	\item  \underline{\smash{\em Formalizing p2p content delivery with delivery fairness.}} We formulate the problem of p2p content delivery with desired security goals, where fairness ensures that every party is fairly treated even others arbitrarily collude or are corrupted.
	
	\item \underline{\smash{\em  Verifiable fair delivery.}} We put forth a novel delivery fairness notion between a sender and a receiver dubbed verifiable fair delivery ($\mathsf{VFD}$): a non-interactive honest verifier  can check whether a sender indeed sends a sequence of qualified data chunks to a receiver as long as not both the sender and the receiver are corrupted.
	
	This primitive is powerful in the sense that: (i) the verifier only has to be non-interactive and honest, so it can be easily instantiated via the blockchain; 
	(ii) qualified data can be flexibly specified through a global predicate known by the sender, the receiver and the verifier,
	so the predicate validation can be tuned to augment $\mathsf{VFD}$ in a certain way  for the full-fledged p2p delivery scenario.
	
	\item  \underline{\smash{\em Lifting $\mathsf{VFD}$ for full-fledged p2p content delivery.}} We specify $\mathsf{VFD}$ to validate that each data chunk is signed by the original content provider, and wrap up the concrete instantiation to design an efficient blockchain-enabled fair p2p content delivery protocol \systemd{}, which allows: (i) the consumer can retrieve content via downloading, i.e., {\em view-after-delivery}; (ii) minimal involvement of the content provider in the sense that only two messages are needed from the provider during the whole course of the protocol execution; (iii) one-time contract deployment and preparation while repeatable delivery of the same content to different consumers.
	
	Thanks to the carefully instantiated $\mathsf{VFD}$, 
	the provider's content cannot be modified by the deliverer, so we essentially can view the fairness of consumer and provider as a fair exchange problem for digital goods between two parties.
	To facilitate the ``two-party'' exchange fairness, we leverage the  proof-of-misbehavior method (instead of using heavy cryptographic proofs for honesty \cite{Maxwell-2016-BitcoinCore}),
	thus launching a simple mechanism to allow the consumer to dispute and prove that the provider indeed sells wrong content inconsistent to a certain digest;
	along the way, we dedicatedly tune this component for better efficiency: 
	(i)~universal composability security  \cite{Dziembowski-et-al-2018-CCS} is explicitly given up to employ {\em one-way security} in the stand-alone setting; 
	(ii) the generality of supporting any form of dispute on illegitimate contents  \cite{Dziembowski-et-al-2018-CCS} is weaken to those inconsistent to digest in form of Merkle tree root. 
	
	\item \underline{\smash{\em Less latency for streaming delivery.}}
	Though the protocol \systemd{} is efficient as well as minimize the provider's activities, it also incurs considerable latency since the consumer can obtain the content only after all data chunks are delivered.
	To accommodate the streaming scenario where the consumer can {\em view-while-delivery}, we propose another simple while efficient protocol \systems{}, where each data chunk can be retrieved in $O(1)$ communication rounds. Though the design requires more involvement of the content provider, whose overall communication workload, however, remains much smaller than the content itself. \systems{} realizes the same security goals as the \systemd{} protocol.
	
	\item \underline{\smash{\em Optimal on-chain and deliverer complexities.}} Both the downloading and streaming protocols achieve asymptotically optimal  $\Tilde{O}(\eta + \lambda)$ on-chain computational costs even when dispute occurs. The on-chain costs only relates to the small chunk size parameter  $\eta$  and the even smaller security parameter $\lambda$. This becomes critical to preserve low-cost of blockchain-based p2p content delivery.
	Moreover, in both protocols, the deliverer only sends ${O}(\eta + \lambda)$ bits amortized for each chunk. Considering the fact that $\lambda$ is much less than $\eta$, this corresponds to asymptotically optimal deliverer communication, and is the key to keep p2p downloading and p2p streaming highly efficient.
	
	\item \underline{\smash{\em Optimized implementations.}} We implement\footnote{Code availability: https://github.com/Blockchain-World/FairThunder.git} and optimize  \systemd{} and \systems{}.
	Various non-trivial optimizations are performed to improve  the critical on-chain performance including efficient on-chain implementation of ElGamal verifiable decryption  over bn-128 curve. 
	Extensive experiments are also conducted atop Ethereum Ropsten network, showing real-world applicability.
\end{enumerate} 

\smallskip
\noindent
{\bf Structure.} Section~\ref{sec:Preliminaries} presents the notations and involved cryptographic primitives. In Section~\ref{sec:vfd}, we introduce a building block, viz. verifiable fair delivery ($\mathsf{VFD}$). Section~\ref{sec:ProblemFormulation} gives the formulation of fair p2p content delivery and the desired security goals. In Section~\ref{sec:ProtocolDesign_Downloading}, we present the fair p2p content delivery in the downloading setting with the instantiation of the $\mathsf{VFD}$ module, and in Section~\ref{sec:ProtocolDesign_Streaming}, we further present the fair p2p content delivery for the streaming setting. Section~\ref{sec:ImplementationandEvaluation} provides the details of the protocols' implementation and evaluation. Section~\ref{sec:RelatedWork} elaborates the comparison with related works, and we conclude in Section~\ref{sec:Conclusion}.


\section{Preliminaries}
\label{sec:Preliminaries}

In this section, we briefly describe the notations and relevant cryptographic primitives.

\smallskip
\noindent {\bf Notations.}  We use $[n]$ to denote the set of integers $\{1,\dots,n\}$, $[a,b]$ to denote the set $\{a,\dots,b\}$, $x||y$ to denote a string concatenating $x$ and $y$,  $\leftarrow_{\$}$ to denote uniformly sampling, and $\preceq$ to denote the prefix relationship.

\smallskip
\noindent \textbf{Global $\mathsf{ledger}$.}
It  provides the   primitive of  cryptocurrency that can deal with ``coin'' transfers transparently. 
Detailedly, each entry of the dictionary $\mathsf{ledger}[\mathcal{P}_i]$ records the balance of the party $\mathcal{P}_i$, and is global (which means it is accessible by all system participants including the adversary).
Moreover, the global dictionary $\mathsf{ledger}$ can be a subroutine of the so-called \textit{smart contract} -- a pre-defined piece of automatically executing code -- that can transact ``coins'' to a designated party by invoking the $\mathsf{ledger}$ dictionary when some conditions are met. For example, if a smart contract (which can be seen as a certain ideal functionality) executes $\mathsf{ledger}[\mathcal{P}_i]=\mathsf{ledger}[\mathcal{P}_i]+\bitcoinA$, the balance of $\mathcal{P}_i$ would increase by $\bitcoinA$.

\smallskip
\noindent \textbf{Merkle tree.}
This  consists of a tuple of algorithms $(\mathsf{BuildMT}, \mathsf{GenMTP},
\mathsf{VerifyMTP})$.
$\mathsf{BuildMT}$ accepts as input a sequence of elements  $m= (m_1, m_2, \cdots, m_n)$  and outputs the Merkle tree $\mathsf{MT}$ with  $\mathsf{root}$ that commits $m$. Note   we let $\mathsf{root}(\mathsf{MT})$ to denote the Merkle tree $\mathsf{MT}$'s  $\mathsf{root}$. $\mathsf{GenMTP}$ takes as input the Merkle tree $\mathsf{MT}$ (built for $m$) and  the $i$-th element $m_i$ in $m$,   and outputs a proof $\pi_i$ to attest the inclusion of $m_i$ at the position $i$ of $m$. $\mathsf{VerifyMTP}$ takes as input the $\mathsf{root}$ of Merkle tree $\mathsf{MT}$, the index $i$, the Merkle proof $\pi_i$, and $m_i$, and outputs either 1 (accept) or 0 (reject). The security of Merkle tree scheme ensures that: for any probabilistic polynomial-time (P.P.T.) adversary $\mathcal{A}$, any sequence $m$ and any index $i$, conditioned on $\mathsf{MT}$ is a Merkle tree built for $m$, $\mathcal{A}$  cannot produce a fake Merkle tree proof fooling  $\mathsf{VerifyMTP}$ to accept $m'_i \ne m_i\in m$ except with negligible probability given $m$, $\mathsf{MT}$ and security parameters.

\smallskip
\noindent \textbf{Verifiable decryption.} We consider a specific verifiable public key encryption ($\mathsf{VPKE}$) scheme consisting of a tuple of algorithms $(\mathsf{VPKE.KGen},\mathsf{VEnc}, \mathsf{VDec}, \mathsf{ProvePKE}, \mathsf{VerifyPKE})$ and allowing the decryptor to produce the plaintext along with a proof attesting the correct decryption~\cite{Camenisch-et-al-2003-Crypto}. Specifically, $\mathsf{KGen}$   outputs a public-private key pair, i.e., $(h,k) \leftarrow \mathsf{VPKE.KGen}(1^{\lambda})$ where $\lambda$ is a security parameter. The public key encryption   satisfies semantic security. Furthermore, for any $(h,k)\leftarrow \mathsf{VPKE.KGen}(1^{\lambda})$, the $\mathsf{ProvePKE}_{k}$ algorithm takes as input the private key $k$ and the cipher $c$, and outputs a message $m$ with a proof $\pi$; while the $\mathsf{VerifyPKE}_{h}$ algorithm takes as input the public key $h$ and $(m,c,\pi)$, and outputs $1/0$ to accept/reject the statement that $m=\mathsf{VDec}_{k}(c)$. Besides the semantic security, the verifiable decryption scheme need satisfy the following extra properties:
\begin{compactenum}
	\item[$\bullet$] \textit{Completeness}. $Pr[\mathsf{VerifyPKE}_{h}(m,c,\pi)=1 | (m,\pi)\leftarrow \mathsf{ProvePKE}_{k}(c)]=1$, for $\forall$ $c$ and $(h,k)\leftarrow\mathsf{KGen}(1^{\lambda})$;
	\item[$\bullet$] \textit{Soundness}. For any $(h,k)\leftarrow\mathsf{KGen}(1^{\lambda})$ and $c$, no probabilistic poly-time (P.P.T.) adversary $\mathcal{A}$ can produce a proof $\pi$ fooling $\mathsf{VerifyPKE}_{h}$ to accept that $c$ is decrypted to $m'$ if $m' \neq \mathsf{VDec}_{k}(c)$ except with negligible probability;
	\item[$\bullet$] \textit{Zero-Knowledge}. The proof $\pi$ can be simulated by a P.P.T. simulator $\mathcal{S}_{\mathsf{VPKE}}$ taking as input only public knowledge $m,h,c$, hence  nothing more than the truthness of the statement $(m,c) \in \{(m,c)|m = \mathsf{VDec}_{k}(c)\}$ is leaked.
\end{compactenum}

\smallskip
\noindent \textbf{Cryptographic primitives.} We also  consider: (i) a cryptographic hash function $\mathcal{H}: \{0,1\}^*\rightarrow\{0,1\}^{\lambda}$ in the random oracle model \cite{Bellare-and-Rogaway-1993-CCS}; (ii) a  {\em semantically  secure} (fixed-length) symmetric encryption made of $(\mathsf{SE.KGen}, \mathsf{SEnc}, \mathsf{SDec})$;  (iii) an {\em existential unforgeability under chosen message attack} (EU-CMA) secure digital signature scheme consisting of the polynomial-time algorithms $(\mathsf{SIG.KGen}, \mathsf{Sign}, \mathsf{Verify})$.

\section{Warm-up: Verifiable Fair Delivery}
\label{sec:vfd}

We first warm up and set forth a building block termed {\em verifiable fair delivery} ($\mathsf{VFD}$), which enables an honest verifier to verify that a sender indeed transfers some amount of data to a receiver. It later acts as a key module in the fair p2p content delivery protocol. The high level idea of $\mathsf{VFD}$ lies in: the receiver needs to send back a signed ``receipt" in order to acknowledge the sender's bandwidth contribution and continuously receives the next data chunk. Consider that the data chunks of same size $\eta$ are transferred {\em sequentially} starting from the first chunk, the sender can always use the latest receipt containing the chunk index to prove to the verifier about the total contribution. Intuitively the sender {\em at most} wastes bandwidth of transferring one chunk.

\smallskip
\noindent
\textbf{Syntax of $\mathsf{VFD}$}.
The $\mathsf{VFD}$ protocol is among an interactive poly-time Turing-machine (ITM) sender denoted by $\mathcal{S}$, an ITM receiver denoted by $\mathcal{R}$, and a non-interactive Turing-machine verifier denoted by $\mathcal{V}$, and follows the syntax: 

\begin{itemize}
	\item {\bf  Sender.} The sender $\mathcal{S}$ can be activated by calling an interface  $\mathcal{S}.\mathsf{send}()$
	with inputting a sequence of $n$ data chunks and their corresponding validation strings denoted by $((c_1,\sigma_{c_1}), \dots, (c_n,\sigma_{c_n}))$, and there exists an efficient and global predicate $\Psi(i, c_i, \sigma_{c_i})\rightarrow\{0,1\}$ to check whether   $c_i$ is the $i$-th valid chunk due to $\sigma_{c_i}$; once activated, the sender $\mathcal{S}$ interacts with the receiver $\mathcal{R}$, and opens an interface $\mathcal{S}.\mathsf{prove()}$ that can be invoked to return a proof string $\pi$ indicating the number of sent chunks;
	
	\item {\bf Receiver.} The receiver $\mathcal{R}$ can be activated by calling an interface $\mathcal{R}.\mathsf{recv}()$ with taking as input the description of the global predicate $\Psi(\cdot)$ to  interact with $\mathcal{S}$, and outputs a sequence of $((c_1,\sigma_{c_1}), \dots, (c_{n'},\sigma_{c_{n'}}))$, where $n' \in [n]$ and every $(c_i,\sigma_{c_i})$ is valid due to $\Psi(\cdot)$;
	
	\item {\bf Verifier.} The verifier $\mathcal{V}$ inputs the proof $\pi$ generated by $\mathcal{S}.\mathsf{prove()}$, and outputs  an integer  $\mathsf{ctr} \in \{0,\cdots,n\}$.
\end{itemize} 

\smallskip
\noindent
{\bf Security of $\mathsf{VFD}$.} The $\mathsf{VFD}$ protocol must satisfy the following security properties:

\begin{itemize}
	\item {\bf Termination}. If at least one of $\mathcal{S}$ and $\mathcal{R}$ is honest, the $\mathsf{VFD}$ protocol terminates within {\em at most} $2n$ rounds, where $n$ is the number of content chunks;
	
	\item {\bf Completeness}. If $\mathcal{S}$ and $\mathcal{R}$ are both honest and activated, after $2n$ rounds, $\mathcal{S}$ is able to generate a proof $\pi$ which $\mathcal{V}$ can take as input and output $\mathsf{ctr}\equiv n$, while $\mathcal{R}$ can output $((c_1,\sigma_{c_1}), \dots, (c_n,\sigma_{c_n}))$, which is same to $\mathcal{S}$'s input;
	
	\item {\bf Verifiable $\eta$ delivery fairness}. When one of $\mathcal{S}$ and $\mathcal{R}$ maliciously aborts, $\mathsf{VFD}$ shall satisfy the following delivery fairness requirements:
	
	\begin{itemize}
		\item \underline{\smash{\em Verifiable delivery fairness against $\mathcal{S}^*$.}} For any corrupted probabilistic poly-time (P.P.T.) sender $\mathcal{S}^*$ controlled by $\adv$, it is guarantee that: the honest receiver $\mathcal{R}$ will always receive the valid sequence $(c_1,\sigma_{c_1}),\dots,(c_\mathsf{ctr},\sigma_{c_\mathsf{ctr}})$ if $\mathcal{A}$ can produce the proof $\pi$ that enables $\mathcal{V}$ to output $\mathsf{ctr}$.
		
		\item \underline{\smash{\em Verifiable delivery fairness against $\mathcal{R}^*$.}} For any corrupted P.P.T. receiver $\mathcal{R}^*$ controlled by $\adv$, it is ensured that: the honest sender $\mathcal{S}$ can always generate a proof $\pi$, which enables $\mathcal{V}$ to output {\em at least} $(\mathsf{ctr}-1)$ if $\mathcal{A}$ receives the valid sequence $(c_1,\sigma_{c_1}),\dots,(c_\mathsf{ctr},\sigma_{c_\mathsf{ctr}})$. At most $\mathcal{S}$ wastes bandwidth for delivering one content chunk of size $\eta$.
	\end{itemize}
\end{itemize}

\smallskip
\noindent
\textbf{$\mathsf{VFD}$ protocol $\Pi_\mathsf{VFD}$}. We consider the authenticated setting that the sender $\mathcal{S}$ and the receiver $\mathcal{R}$ have generated public-private key pairs $(pk_{\mathcal{S}}, sk_{\mathcal{S}})$ and $(pk_{\mathcal{R}}, sk_{\mathcal{R}})$ for digital signature, respectively;  and they have announced the public keys to bind to themselves. Then, $\mathsf{VFD}$ with the global predicate $\Psi(\cdot)$ can be realized by the protocol $\Pi_\mathsf{VFD}$ hereunder among $\mathcal{S}$, $\mathcal{R}$ and $\mathcal{V}$ against P.P.T. and static adversary in the \textit{stand-alone} setting\footnote{We omit the \textit{session id} (denoted as $sid$) in the stand-alone context for brevity. To defend against \textit{replay} attack in concurrent sessions, it is trivial to let the authenticated messages include an $sid$ field, which, for example, can be instantiated by the hash of the transferred data identifier $\mathsf{root}_m$, the involved parties' addresses and an increasing-only nonce, namely $sid := \mathcal{H}(\mathsf{root}_m||\mathcal{V}\_address||pk_{\mathcal{S}}||pk_{\mathcal{R}}||nonce)$.} with the synchronous network assumption:

\begin{itemize}
    \item {\bf Construction of sender.} The sender $\mathcal{S}$, after activated via $\mathcal{S}.\mathsf{send}()$ with the input $((c_1,\sigma_{c_1}), \dots, (c_n,\sigma_{c_n}))$, $pk_{\mathcal{S}}$ and $pk_{\mathcal{R}}$, starts a timer $\mathcal{T}_{\mathcal{S}}$ lasting two synchronous rounds, initializes a variable $\pi_\mathcal{S}:=null$, and executes as follows:
    \begin{itemize}
		\item For each $i \in [n]$: the sender sends $(\mathsf{deliver}, i, c_i, \sigma_{c_i})$ to $\mathcal{R}$, and waits for the response message $(\mathsf{receipt}, i, \sigma_{\mathcal{R}}^{i})$ from $\mathcal{R}$. If $\mathcal{T}_{\mathcal{S}}$ expires before receiving the response, breaks the iteration; otherwise $\mathcal{S}$ verifies whether $\mathsf{Verify}(\mathsf{receipt}||i||pk_{\mathcal{R}}||pk_{\mathcal{S}}, \sigma_{\mathcal{R}}^{i}, pk_{\mathcal{R}})\equiv 1$ or not, if {\em true}, resets $\mathcal{T}_{\mathcal{S}}$, outputs $\pi_\mathcal{S}:=(i, \sigma_{\mathcal{R}}^{i})$, and continues to run the next iteration (i.e., increasing $i$ by one); if {\em false}, breaks the iteration;
		\item Upon $\mathcal{S}.\mathsf{prove}()$ is invoked, it returns $\pi_\mathcal{S}$ as the $\mathsf{VFD}$ proof and halts.
	\end{itemize}
	
	\item {\bf Construction of receiver.} The receiver $\mathcal{R}$, after activated via $\mathcal{R}.\mathsf{recv}()$ with the input $pk_{\mathcal{S}}$ and $(pk_{\mathcal{R}}, sk_{\mathcal{R}})$, starts a timer $\mathcal{T}_{\mathcal{R}}$ lasting two synchronous rounds and operates as: for each $j  \in [n]$: $\mathcal{R}$ waits for $(\mathsf{deliver}, j, c_j, \sigma_{c_j})$ from  $\mathcal{S}$ and halts if $\mathcal{T}_{\mathcal{R}}$ expires before receiving the $\mathsf{deliver}$ message; otherwise $\mathcal{R}$ verifies whether $\Psi(j, c_j, \sigma_{c_j})\equiv1$ or not; if {\em true},  resets $\mathcal{T}_{\mathcal{R}}$,  outputs $(c_j, \sigma_{c_j})$, and sends  $(\mathsf{receipt}, i, \sigma^{i}_{\mathcal{R}})$ to $\mathcal{S}$ where $\sigma^{i}_{\mathcal{R}}\leftarrow \mathsf{Sign}(\mathsf{receipt}||i||pk_{\mathcal{R}}||pk_{\mathcal{S}}, sk_{\mathcal{R}})$, halts if {\em false}. Note that the global predicate $\Psi(\cdot)$ is efficient as essentially it just performs a signature verification.
	
	\item {\bf Construction of verifier}. Upon the input $\pi_{\mathcal{S}}$, the verifier $\mathcal{V}$ parses it into $(\mathsf{ctr}, \sigma_{\mathcal{R}}^{\mathsf{ctr}})$, and checks whether $\mathsf{Verify}(\mathsf{receipt}||\mathsf{ctr}||pk_{\mathcal{R}}||pk_{\mathcal{S}},  \sigma_{\mathcal{R}}^{\mathsf{ctr}}, pk_{\mathcal{R}})\equiv1$ or not; if {\em true}, it outputs $\mathsf{ctr}$, or else outputs 0. Recall that $\mathsf{Verify}$ is to verify signatures.
	
\end{itemize}

\begin{lemma}
	\label{lemma:VFD_completeness_fairness}
	{In the synchronous authenticated network and stand-alone setting, the protocol $\Pi_\mathsf{VFD}$ satisfies termination, completeness and the verifiable $\eta$ delivery fairness against non-adaptive  P.P.T. adversary corrupting one of the sender and the receiver.}
\end{lemma}

\noindent{\em Proof.} If both the sender and the receiver are honest, there would be $2n$ communication rounds since for every delivered chunk, the sender obtains a ``receipt" from the receiver for acknowledging bandwidth contribution. If one malicious party aborts, the other honest one would also terminate after its maintained timer expires, resulting in less than $2n$ communication rounds. Therefore, the termination property is guaranteed. 

In addition, when both the sender $\mathcal{S}$ and the receiver $\mathcal{R}$ are honest, the {\em completeness} of $\mathsf{VFD}$ is immediate to see: in each round, $\mathcal{S}$ would start to deliver the next chunk after receiving the receipt from $\mathcal{R}$ within 2 rounds, i.e., a round-trip time. After 2$n$ synchronous rounds, $\mathcal{R}$ receives the chunk-validation pairs $((c_1, \sigma_{c_1}),\cdots,(c_n,\sigma_{c_n}))$ and $\mathcal{S}$ outputs the last receipt as a proof $\pi$, which is taken as input by the verifier $\mathcal{V}$ to output $n$ demonstrating $\mathcal{S}$'s delivery contribution. 

For the $\eta$ delivery fairness of $\mathsf{VFD}$, on one hand, the malicious $\mathcal{S}^*$ corrupted by $\mathcal{A}$ may abort after receiving the $\mathsf{ctr}$-th ($1\leq \mathsf{ctr}\leq n$) receipt. In that case, $\mathcal{R}$ is also guaranteed to receive a valid sequence of $((c_1,\cdots,\sigma_{c_1}),\cdots, (c_{\mathsf{ctr}},\sigma_{c_{\mathsf{ctr}}}))$ with overwhelming probability, unless $\mathcal{A}$ can forge $\mathcal{R}$'s signature. However, it requires $\mathcal{A}$ to break the underlying EU-CMA signature scheme, which is of negligible probability. 
On the other hand, 
for the malicious $\mathcal{R}^*$ corrupted by $\mathcal{A}$,
if $\mathcal{V}$ takes the honest $\mathcal{S}$'s proof and can output $\mathsf{ctr}$, then $\mathcal{S}$ {\em at most} has sent $(\mathsf{ctr}+1)$ chunk-validation pairs, i.e., $(c_i, \sigma_{c_i})$, to $\mathcal{A}$. Overall, $\mathcal{S}$ {\em at most} wastes bandwidth of delivering one chunk of size $\eta$. Hence, the $\eta$ delivery fairness of $\mathsf{VFD}$ is rigorously guaranteed.


\section{Formalizing p2p Content Delivery}
\label{sec:ProblemFormulation}

Here we extend delivery fairness between a sender and a receiver and define the needed properties of p2p content delivery required by the provider, the deliverer and the consumer.

\subsection{System Model}

\noindent
{{\bf Participating Parties}}. We consider the following explicit entities (i.e., interactive Turing machines by cryptographic convention) in the context of p2p content delivery: 

\begin{itemize}
	\item \underline{\smash{\em Content Provider}} is an entity (denoted by $\mathcal{P}$) that owns the original content $m$ composed of $n$ data chunks,\footnote{Remark that the content $m$ is {\em dividable} in the sense that each chunk is independent to other chunks, e.g., each chunk is a small 10-second video.} satisfying a public known predicate $\phi(\cdot)$,\footnote{Throughout the paper, we consider that the predicate $\phi$ is in the form of $\phi(m) = [\mathsf{root}(\mathsf{BuildMT}(m)) \equiv \mathsf{root}_m]$, where  $\mathsf{root}$ is the Merkle tree root of the content $m$. In practice, it can be aquired from a semi-trusted third party, such as BitTorrent forum sites~\cite{Kupccu--et-al-2010-RSAC} or VirusTotal~\cite{Janin-et-al-2020-EuroSPW}.} and $\mathcal{P}$ is willing to sell to the users of interest. Meanwhile, the provider would like to delegate the delivery of its content to a third-party (viz. a deliverer) with promise to pay $\bitcoinA_{\mathcal{P}}$ for each successfully delivered chunk.
	
	\item \underline{\smash{\em Content Deliverer}} (denoted by $\mathcal{D}$) contributes its idle bandwidth resources to deliver  the content on behalf of the content provider $\mathcal{P}$ and would receive the payment proportional to the amount of delivered data. In the p2p delivery scenario, deliverers can be some small organizations or individuals, e.g., the {\em RetrievalProvider} in Filecoin~\cite{Filecoin-Retrieval-2021-Spec}.
	
	\item \underline{\smash{\em Content Consumer}} is an entity (denoted by $\mathcal{C}$) that would  pay  $\bitcoinA_{\mathcal{C}}$ for each chunk in the   content $m$ by interacting with $\mathcal{P}$ and $\mathcal{D}$.
\end{itemize}

\noindent
{\bf Adversary}. Following   modern cryptographic practices~\cite{Katz-et-al-2014-CRC}, we consider the   adversary $\adv$ with following standard abilities:
\begin{itemize}
	\item \underline{\smash{\em Static corruptions.}} The adversary $\adv$  can   corrupt some parties only before  the course of protocol executions;
	\item \underline{\smash{\em Computationally bounded.}} The adversary $\adv$ is restricted to P.P.T. algorithms;
	\item \underline{\smash{\em Synchronous authenticated  channel.}} We adopt the  {synchronous network model of authenticated point-to-point channels}  to describe the ability of $\adv$ on controlling   communications, namely, for any messages sent between honest parties, $\adv$ is   consulted to delay them up to a-priori known $\Delta$ but cannot drop, reroute or modify them. W.l.o.g., we consider a global clock in the system, and $\adv$ can delay the messages up to a clock round \cite{Kosba-et-al-2016-SP,Kiayias-et-al-2016-Crypto}.
\end{itemize}

\noindent
{{\bf Arbiter smart contract $\mathcal{G}$}}. The system is in a hybrid model with oracle access to an arbiter smart contract $\mathcal{G}$. The contract $\mathcal{G}$ is a stateful ideal functionality that leaks all its internal states to the adversary $\adv$ and all parties, while allowing to pre-specify some immutable conditions (that can be triggered through interacting with $\mathcal{P}$, $\mathcal{D}$, and $\mathcal{C}$) to transact ``coins'' over the cryptocurrency $\mathsf{ledger}$, thus ``mimicking''  the contracts  in  real  life transparently. In practice, the contract can be instantiated through many real-world blockchains such as Ethereum~\cite{Wood-et-al-2014-Ethereum-Yellow-Paper}. Throughout this paper, the details of the arbiter contracts $\mathcal{G}$ follow the conventional pseudo-code notations in the seminal work due to Kosba {\em et al}.~\cite{Kosba-et-al-2016-SP}.

\subsection{Design Goals}
Now we formulate the problem of fair content delivery with an emphasis on the delivery fairness, which to our knowledge is the first formal definition to abstract the necessary security/utility requirements of delegated p2p content delivery.

\smallskip
\noindent
{{\bf Syntax}}. 
A fair p2p content delivery protocol $\Pi=(\mathcal{P}, \mathcal{D}, \mathcal{C})$ is a tuple of three  P.P.T. interactive Turing machines (ITMs) consisting of two explicit phases:
\begin{itemize}
	\item \underline{\smash{\em Preparation phase.}}
	The provider $\mathcal{P}$ takes as input public parameters and the content $m=(m_1,\dots, m_n) \in \{0,1\}^{\eta\times n}$ that satisfies $\phi(m)\equiv1$, where $\eta$ is chunk size in bit and $n$ is the number of chunks and it outputs some auxiliary data, e.g., encryption keys;
	the deliverer $\mathcal{D}$ takes as input public parameters and outputs some auxiliary data, e.g., encrypted content; the consumer $\mathcal{C}$ does not involve in this phase. Note $\mathcal{P}$ deposits a budget of $n \cdot \bitcoinA_{\mathcal{P}}$ in $\mathsf{ledger}$ to incentivize $\mathcal{D}$ so it can {\em minimize} bandwidth usage in the next phase.
	
	\item \underline{\smash{\em Delivery phase.}}
	The provider $\mathcal{P}$ and the   deliverer $\mathcal{D}$ take as input their auxiliary data obtained in the preparation phase, respectively, and they would receive the deserved payment; the consumer $\mathcal{C}$ takes as input public parameters and outputs the content $m$ with $\phi(m)\equiv 1$. Note $\mathcal{C}$ has a budget of $n \cdot \bitcoinA_{\mathcal{C}}$ in $\mathsf{ledger}$ to ``buy'' the content $m$ satisfying $\phi(m)\equiv1$, where $\bitcoinA_{\mathcal{C}} > \bitcoinA_{\mathcal{P}}$. 
\end{itemize}

Furthermore, the fair p2p content delivery  protocol $\Pi$ shall meet the following security requirements.

\smallskip
\noindent
{{\bf Completeness}}. For any content predicate $\phi(\cdot)$ in the form of $\phi(m) = [\mathsf{root}(\mathsf{BuildMT}(m)) \equiv \mathsf{root}_m]$, 
conditioned on $\mathcal{P}, \mathcal{D}$ and $\mathcal{C}$ are all honest, the protocol $\Pi$ attains:
\begin{itemize}
	\item The consumer $\mathcal{C}$ would obtain the qualified content $m$ satisfying $\phi(m)\equiv1$, and its balance in the global   $\mathsf{ledger}[\mathcal{C}]$ would decrease by $n\cdot\bitcoinA_{\mathcal{C}}$, where $\bitcoinA_{\mathcal{C}}$ represents the amount paid by $\mathcal{C}$ for each content chunk.
	\item The deliverer $\mathcal{D}$ would receive the payment $n\cdot\bitcoinA_{\mathcal{P}}$ over the global $\mathsf{ledger}$, where $\bitcoinA_{\mathcal{P}}$ represents the amount paid by $\mathcal{P}$ to $\mathcal{D}$ for delivering a content chunk to the consumer.
	\item The provider $\mathcal{P}$ would receive its well-deserved payments over the ledger, namely,  $\mathsf{ledger}[\mathcal{P}]$ would increase by $n\cdot(\bitcoinA_{\mathcal{C}}-\bitcoinA_{\mathcal{P}})$ as it receives $n\cdot\bitcoinA_{\mathcal{C}}$ from the consumer while it pays out $n\cdot\bitcoinA_{\mathcal{P}}$ to the deliverer.
\end{itemize}

\smallskip
\noindent
{\bf Fairness}. The protocol $\Pi$ shall satisfy the following fairness requirements:

\begin{itemize}
	\item \underline{\smash{\em Consumer Fairness.}} For $\forall$ corrupted P.P.T. $\mathcal{D}^*$ and $\mathcal{P}^*$ (fully controlled by $\adv$), it is guaranteed to the honest consumer $\mathcal{C}$ with overwhelming probability that: the $\mathsf{ledger}[\mathcal{C}]$ decreases by $\ell \cdot \bitcoinA_{\mathcal{C}}$  only if $\mathcal{C}$ receives a sequence of chunks $(m_1,\dots,m_\ell) \preceq m$ where $\phi(m)\equiv1$, Intuitively, this property states that $\mathcal{C}$ pays proportional to valid chunks it {\em de facto} receives.
	
	\item \underline{\smash{\em Delivery $\eta$-Fairness.}} For $\forall$  malicious P.P.T. $\mathcal{C}^*$ and $\mathcal{P}^*$ corrupted by $\adv$, it is assured to the honest deliverer $\mathcal{D}$ that: if $\mathcal{D}$ sent overall $O(\ell\cdot\eta + 1)$ bits during the protocol, $\mathcal{D}$ should {\em at least} obtain the payment of $(\ell-1)\cdot\bitcoinA_{\mathcal{P}}$. In other words, the unpaid delivery is bounded by $O(\eta)$ bits.
	
	\item \underline{\smash{\em Provider $\eta$-Fairness.}} For $\forall$  corrupted P.P.T. $\mathcal{C}^*$ and $\mathcal{D}^*$ controlled by $\adv$, it is ensured to the honest   provider $\mathcal{P}$ that: if   $\adv$ can output  ${\eta\cdot\ell}$ bits consisted in the content $m$, the provider $\mathcal{P}$ shall  receive at least $(\ell-1)\cdot(\bitcoinA_{\mathcal{C}}-\bitcoinA_{\mathcal{P}})$ net income,
	namely, $\mathsf{ledger}[\mathcal{P}]$  increases by $(\ell-1)\cdot(\bitcoinA_{\mathcal{C}}-\bitcoinA_{\mathcal{P}})$,
	with all except negligible probability.
	i.e., $\mathcal{P}$ is ensured that {\em at most} $O(\eta)$-bit content are revealed without being well paid.
	
\end{itemize}


\smallskip
\noindent
{{\bf Confidentiality against deliverer}}. This  is needed to protect copyrighted data against probably corrupted deliverers, otherwise a malicious consumer can pretend to be or collude with a deliverer to obtain the plaintext content without paying for the provider, which violates the exchange fairness for $\mathcal{P}$. Informally, we require that the corrupted $\mathcal{D}^*$ on receiving protocol scripts (e.g., the delegated content chunks from the provider) cannot produce the provider's input content with all but negligible probability in a delivery session.\footnote{To preserve the content digital rights across multiple delivery sessions in the p2p content delivery setting, it is feasible to integrate digital rights management (DRM) schemes~\cite{Ma-et-al-2018-FGCS}, which can be an interesting future work.}

We remark that confidentiality is not captured by fairness, as it is trivial to see   a protocol satisfying fairness might not have confidentiality: upon all payments are cleared and the consumers receives the whole content, the protocol lets the consumer send the content to the deliverer.

\smallskip
\noindent
{{\bf Timeliness}}.  When at least one of the parties $\mathcal{P}$, $\mathcal{D}$ and $\mathcal{C}$ is honest (i.e., others can be corrupted by $\adv$), the   honest ones are guaranteed to halt in $O(n)$ synchronous   rounds where $n$ is the number of content chunks. At the completion or abortion of the protocol, the aforementioned fairness and confidentiality are always guaranteed.

\smallskip
\noindent
{{\bf Non-trivial efficiency}}. We require the necessary non-trivial efficiency to rule out possible trivial solutions:
\begin{itemize}
	\item The messages sent to $\mathcal{G}$ from honest parties are uniformly bounded by $\Tilde{O}(1)$ bits, which excludes a trivial way of using the smart contract to directly deliver the content.
	
	\item In the delivery phase, the messages sent by  honest  $\mathcal{P}$ are uniformly bounded by $n\cdot\lambda$ bits, where $\lambda$ is a small cryptographic parameter, thus ensuring $n\cdot\lambda$ much smaller than the size of content $|m|$.
	This indicates that $\mathcal{P}$ can save its bandwidth upon the completion of preparation phase and excludes the idea of delivering by itself.
\end{itemize}

\smallskip
\noindent
{{\bf Remarks}}.
We make the following discussions about the above definitions: 
(i) $\phi(\cdot)$ is a public parameter known to all parties before the protocol execution; 
(ii) our fairness requirements have already implied the case where the adversary corrupts one party of $\mathcal{P}$, $\mathcal{D}$ and $\mathcal{C}$ instead of two, since whenever the adversary corrupts two parties, it can let one of these corrupted two  follow the original  protocol; (iii) like all cryptographic protocols, it does not make sense to consider all parties are corrupted, so do we not;
(iv) the deliverer and the provider might lose well-deserved payment, but {\em at most} lose that for one chunk, i.e., the level of unfairness is strictly bounded; (v) though we focus on the case of one \textit{single} content deliverer, our formalism and design can be extended to capture \textit{multiple} deliverers, for example, when the whole content is cut to multiple pieces and each piece is delegated to a distinct deliverer. The extension with strong fairness guarantee forms an interesting future work.

In addition, one might wonder that a probably corrupted content provider fails in the middle of a transmission, causing that the consumer does not get the entire content but has to pay a lot. Nevertheless, this actually is not a serious worry in the peer-to-peer content delivery setting that aims to complement decentralized content storage networks because there essentially would be a large number of deliverers, and at least some of them can be honest. As such, if a consumer encounters failures in the middle of retrieving the content, it can iteratively ask another deliverer to start a new delivery session to fetch the remaining undelivered chunks. Moreover, our actual constructions in Section \ref{sec:ProtocolDesign_Downloading} and \ref{sec:ProtocolDesign_Streaming}   essentially allow the consumers to fetch the content from any specific chunk instead of beginning with the first one.


\section{\systemd{}: Fair p2p Downloading}
\label{sec:ProtocolDesign_Downloading}
This section presents the p2p fair delivery protocol $\Pi_{\mathsf{FD}}$, allowing the consumers to view the content after downloading (partial or all) the chunks, termed as {\em view-after-delivery}.

\subsection{\systemd{} Overview}
\begin{figure}[!t]
	\centering
	\includegraphics[width=.49\textwidth]{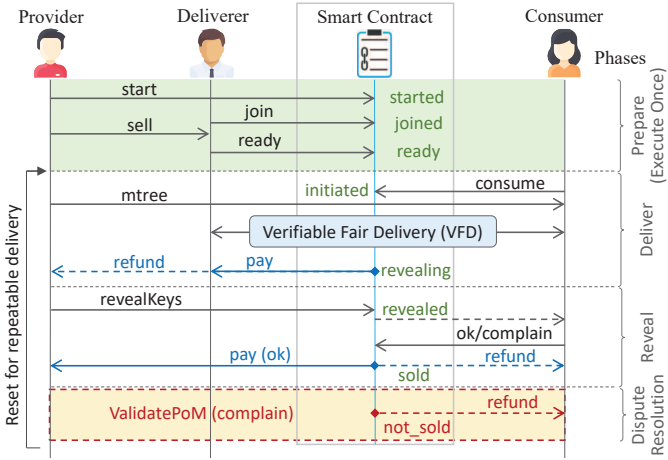}
	\vspace{-6mm}
	\caption{The overview of $\mathsf{FairDownload}$ protocol $\Pi_{\mathsf{FD}}$.}
	\label{fig:protocol_overview}
	\vspace{-4mm}
\end{figure}

At a high level, our protocol $\Pi_{\mathsf{FD}}$ can be constructed around the module of verifiable fair delivery ($\mathsf{VFD}$) and proceeds in {\em Prepare}, {\em Deliver} and {\em Reveal} phases as illustrated in Figure \ref{fig:protocol_overview}. 
The core ideas of $\Pi_{\mathsf{FD}}$ can be over-simplified as follows:
\begin{itemize}
	\item The content provider $\mathcal{P}$ encrypts each chunk, signs the encrypted chunks, and delegates to the deliverer $\mathcal{D}$; as such, the deliverer (as the sender $\mathcal{S}$) and the consumer $\mathcal{C}$ (as the receiver $\mathcal{R}$) can run a specific instance of $\mathsf{VFD}$, in which the global predicate $\Psi(\cdot)$ is instantiated to verify that each chunk must be correctly signed by $\mathcal{P}$; additionally, the non-interactive honest verifier $\mathcal{V}$ in $\mathsf{VFD}$ is instantiated via a smart contract, hence upon the contract receives a $\mathsf{VFD}$ proof from $\mathcal{D}$ claiming the in-time delivery of $\mathsf{ctr}$ chunks, it can assert that $\mathcal{C}$ indeed received $\mathsf{ctr}$ encrypted chunks signed by the provider, who can then present to reveal the decryption keys of these $\mathsf{ctr}$ chunks (via the smart contract).
	\item Nevertheless, trivial disclosure of decryption keys via the contract would cause significant on-chain cost up to linear in the number of chunks; we therefore propose a {\em structured key generation scheme} composed of Alg.~\ref{alg:KeyGroupGen},~\ref{alg:RevealKey},~\ref{alg:ExtractKey} that allows the honest provider to reveal all $\mathsf{ctr}$ decryption keys via a short $\Tilde{O}(\lambda)$-bit message; furthermore, to ensure confidentiality against the deliverer, the script to reveal decryption keys is encrypted by the consumer's public key; in case the revealed keys cannot decrypt the cipher chunk signed by $\mathcal{P}$ itself to obtain the correct data chunk, the consumer can complain to the contract via a short $\Tilde{O}(\eta+\lambda)$-bit message to prove the error of decrypted chunk and get refund.
\end{itemize}

The protocol design of $\Pi_{\mathsf{FD}}$ can ensure the fairness for each participating party even others are all corrupted by non-adaptive P.P.T. adversary. The on-chain cost keeps {\em constant} regardless of the content size $|m|$ in the optimistic mode where no dispute occurs. While in the pessimistic case, the protocol also realizes asymptotically optimal on-chain cost, which is related to the chunk size $\eta$. Moreover, the deliverer $\mathcal{D}$ can achieve asymptotically optimal communication in the sense that $\mathcal{D}$ only sends $O(\eta+\lambda)$ bits amortized for each chunk, where $\eta$ is the chunk size and $\lambda$ is a small security parameter with $\lambda \ll \eta$. These properties contribute significantly to the efficiency and practicability of applying $\Pi_{\mathsf{FD}}$ to the p2p content delivery setting.

\subsection{Arbiter Contract $\mathcal{G}_{d}^{\mathsf{ledger}}$ for Downloading}

The arbiter contract $\mathcal{G}_{d}^{\mathsf{ledger}}$ (abbr. $\mathcal{G}_{d}$) shown in Fig.~\ref{fig:downloading_contract_ideal_functionality} is a stateful ideal functionality having accesses to $\mathsf{ledger}$ to assist the fair content delivery via downloading. We make the following remarks about the contract functionality:
\begin{itemize}
	\item \underline{\smash{\em Feasibility.}} To demonstrate the feasibility of $\mathcal{G}_{d}$, we describe it by following the conventional pseudocode notation of smart contracts~\cite{Kosba-et-al-2016-SP}. The description captures the essence of real-world smart contracts, since it: (i)~reflects that the Turing-complete smart contract can be seen as a stateful program to transparently handle pre-specified functionalities; (ii)~captures that a smart contract can access the cryptocurrency ledger to faithfully deal with conditional payments upon its own internal states.
	\item \underline{\smash{\em $\mathsf{VFD}.\mathcal{V}$ subroutine.}} $\mathcal{G}_{d}$ can invoke  the $\mathsf{VFD}$ verifier $\mathcal{V}$ as a subroutine.   $\mathsf{VFD}$'s predicate   $\Psi(\cdot)$  is instantiated to verify that each chunk is signed by the provider $\mathcal{P}$.
	\item \underline{\smash{\em $\mathsf{ValidateRKeys}$ and $\PoM$ subroutines.}} The subroutines allow the consumer to prove to the contract if the content provider $\mathcal{P}$ behaves maliciously. We defer the details to the next subsection.
\end{itemize}

\begin{figure*}[!t]
	\centering
	\footnotesize
	\fbox{%
		\parbox{.96\linewidth}{%
		\vspace{-2mm}
			\begin{center}
				{\bf The Arbiter Contract Functionality $\mathcal{G}_{d}^{\mathsf{ledger}}$ for P2P Downloading}
				
				The arbiter contract  $\mathcal{G}_{d}$ has access to the $\mathsf{ledger}$, and it interacts with the provider $\mathcal{P}$, the deliverer $\mathcal{D}$, the consumer $\mathcal{C}$ and the adversary $\adv$. It locally stores the times of repeatable delivery $\theta$, the number of content chunks $n$, the content digest $\mathsf{root}_m$, the price $\bitcoinA_{\mathcal{P}}$, $\bitcoinA_{\mathcal{C}}$ and ${\bitcoinA}_{\mathsf{pf}}$, the number of delivered chunks $\mathsf{ctr}$ (initialized as 0), addresses $pk_{\mathcal{P}}, pk_{\mathcal{D}}, pk_{\mathcal{C}}, vpk_{\mathcal{C}}$, revealed keys' hash $erk_{\mathsf{hash}}$, state $\Sigma$ and three timers $\mathcal{T}_{\mathsf{round}}$ (implicitly), $\mathcal{T}_{\mathsf{deliver}}$, and $\mathcal{T}_{\mathsf{dispute}}$.
			\end{center}
			\vspace{-5mm}

			\begin{multicols}{2}
				\begin{flushleft}
					
					\noindent
					\xrfill[0.5ex]{0.5pt} {} {\bf Phase 1: Prepare} \xrfill[0.5ex]{0.5pt}
					\begin{itemize}
						\item[$\bullet$] {\color{blue}On receive} $(\mathsf{start}, pk_\mathcal{P}, \mathsf{root}_m, \theta, n, \bitcoinA_\mathcal{P}, \bitcoinA_\mathcal{C}, {\bitcoinA}_{\mathsf{pf}})$ from $\mathcal{P}$:
						\begin{itemize}[-]
							\item[-] assert $\mathsf{ledger}[\mathcal{P}]\ge (\theta\cdot( n\cdot\bitcoinA_{\mathcal{P}}+{\bitcoinA}_{\mathsf{pf}}))$ $\wedge$ $\Sigma\equiv\emptyset$
							\item[-] store $pk_\mathcal{P}, \mathsf{root}_m, \theta, n, \bitcoinA_\mathcal{P}, \bitcoinA_\mathcal{C}, {\bitcoinA}_{\mathsf{pf}}$ 
							\item[-] let $\mathsf{ledger}[\mathcal{P}] := \mathsf{ledger}[\mathcal{P}]-$ $\theta\cdot(n\cdot\bitcoinA_{\mathcal{P}}+\bitcoinA_{\mathsf{pf}})$ and $\Sigma := \mathsf{started}$
							\item[-] send $(\mathsf{started}, pk_\mathcal{P}, \mathsf{root}_m, \theta, n, \bitcoinA_\mathcal{P},$ $\bitcoinA_\mathcal{C}, {\bitcoinA}_{\mathsf{pf}})$ to all entities
						\end{itemize}
						
						\item[$\bullet$] {\color{blue}On receive} $(\mathsf{join},pk_{\mathcal{D}})$ from $\mathcal{D}$:
						\begin{itemize}
							\item[-] assert $\Sigma \equiv \mathsf{started}$
							\item[-] store $pk_{\mathcal{D}}$ and let $\Sigma := \mathsf{joined}$
							\item[-] send $(\mathsf{joined},pk_{\mathcal{D}})$ to all entities
						\end{itemize}
						
						\item[$\bullet$] {\color{blue}On receive} $(\mathsf{prepared})$ from $\mathcal{D}$:
						\begin{itemize}[-]
							\item[-] assert $\Sigma \equiv \mathsf{joined}$, and let $\Sigma := \mathsf{ready}$ 
							\item[-] send $(\mathsf{ready})$ to all entities
						\end{itemize} 
						
					\end{itemize}
					
					\noindent
					\xrfill[0.5ex]{0.5pt} {} {\bf Phase 2: Deliver}  \xrfill[0.5ex]{0.5pt}

					\begin{itemize}
						\item[$\bullet$] {\color{blue}On receive} $(\mathsf{consume}, pk_{\mathcal{C}}, vpk_{\mathcal{C}})$ from $\mathcal{C}$:
						\begin{itemize}[-]
							\item[-] assert $\theta > 0$
							\item[-] assert $\mathsf{ledger}[\mathcal{C}]\ge n\cdot\bitcoinA_{\mathcal{C}}$ $\wedge$ $\Sigma\equiv\mathsf{ready}$
							\item[-] store $pk_{\mathcal{C}}$, $vpk_{\mathcal{C}}$ and let $\mathsf{ledger}[\mathcal{C}]:=\mathsf{ledger}[\mathcal{C}]-n\cdot\bitcoinA_{\mathcal{C}}$
							\item[-] start a timer $\mathcal{T}_{\mathsf{deliver}}$ and let $\Sigma := \mathsf{initiated}$
							\item[-] send $(\mathsf{initiated}, pk_{\mathcal{C}}, vpk_{\mathcal{C}})$ to all entities
						\end{itemize}
						
						\item[$\bullet$] {\color{blue}On receive} $(\mathsf{delivered})$ from $\mathcal{C}$  or $\mathcal{T}_{\mathsf{deliver}}$ times out:
						\begin{itemize}[-]
							\item[-] assert  $\Sigma\equiv\mathsf{initiated}$
							\item[-] send $(\mathsf{getVFDProof})$ to $\mathcal{D}$, and wait for two rounds to receive $(\mathsf{receipt}, i, \sigma^{i}_{\mathcal{C}})$, then execute $\mathsf{verifyVFDProof}()$ to let $\mathsf{ctr} := i$, and then assert $0 \le \mathsf{ctr}\le n$
							\item[-] let $\mathsf{ledger}[\mathcal{D}]:=\mathsf{ledger}[\mathcal{D}]+\mathsf{ctr}\cdot\bitcoinA_{\mathcal{P}}$
							\item[-] let $\mathsf{ledger}[\mathcal{P}]:=\mathsf{ledger}[\mathcal{P}]+(n-\mathsf{ctr})\cdot\bitcoinA_{\mathcal{P}}$
							\item[-] store $\mathsf{ctr}$, let $\Sigma := \mathsf{revealing}$, and send $(\mathsf{revealing}, \mathsf{ctr})$ to all entities
						\end{itemize}
						
					\end{itemize}
					
					\noindent
					\xrfill[0.5ex]{0.5pt} {} {\bf Phase 3: Reveal} \xrfill[0.5ex]{0.5pt}
					\begin{itemize} 
						\item[$\bullet$] {\color{blue}On receive} $(\mathsf{revealKeys}, erk)$ from $\mathcal{P}$:
						\begin{itemize}[-]
							\item[-] assert  $\Sigma\equiv\mathsf{revealing}$
							\item[-] store $erk$ (essentially $erk$'s hash) and start a timer $\mathcal{T}_{\mathsf{dispute}}$
							\item[-] let $\Sigma := \mathsf{revealed}$
							\item[-] send $(\mathsf{revealed}, erk)$ to all entities					
						\end{itemize}
						
						\item[$\bullet$] {\color{blue}Upon} $\mathcal{T}_{\mathsf{dispute}}$ times out:
						\begin{itemize}[-]
							\item assert $\Sigma\equiv\mathsf{revealed}$ and current time $\mathcal{T} \geq \mathcal{T}_{\mathsf{dispute}}$
							\item $\mathsf{ledger}[\mathcal{P}]:=\mathsf{ledger}[\mathcal{P}]+\mathsf{ctr}\cdot\bitcoinA_{\mathcal{C}} + {\bitcoinA}_{\mathsf{pf}}$
							\item $\mathsf{ledger}[\mathcal{C}]:=\mathsf{ledger}[\mathcal{C}]+(n-\mathsf{ctr})\cdot\bitcoinA_{\mathcal{C}}$
							\item let $\Sigma := \mathsf{sold}$ and send $(\mathsf{sold})$ to all entities
						\end{itemize}
						{\ } \\
						{\color{purple}$\triangleright$ Below is the dispute resolution}
						\\
						
						\item[$\bullet$] {\color{blue}On receive} $(\mathsf{wrongRK})$ from $\mathcal{C}$ before $\mathcal{T}_{\mathsf{dispute}}$ times out:
						\begin{itemize}[-]
							\item assert $\Sigma\equiv\mathsf{revealed}$ and current time $\mathcal{T} < \mathcal{T}_{\mathsf{dispute}}$
							\item if $(\mathsf{ValidateRKeys}(n, \mathsf{ctr},erk)\equiv false)$:
							\begin{itemize}[*]
								\item let $\mathsf{ledger}[\mathcal{C}]:=\mathsf{ledger}[\mathcal{C}]+n\cdot\bitcoinA_{\mathcal{C}} + {\bitcoinA}_{\mathsf{pf}}$
								\item let $\Sigma := \mathsf{not\_sold}$ and send $(\mathsf{not\_sold})$ to all entities
							\end{itemize}
						\end{itemize}
						
						\item[$\bullet$] {\color{blue}On receive} $(\mathsf{PoM}, i, j, c_i, \sigma_{c_i}, \mathcal{H}(m_i),$ $\pi^{i}_{\mathsf{MT}},rk, erk,\pi_{\mathsf{VD}})$ from $\mathcal{C}$ before $\mathcal{T}_{\mathsf{dispute}}$ times out:
						\begin{itemize}[-]
							\item assert $\Sigma\equiv\mathsf{revealed}$ and current time $\mathcal{T} < \mathcal{T}_{\mathsf{dispute}}$
							\item invoke the $\mathsf{ValidatePoM}(i, j, c_i, \sigma_{c_i},\newline \mathcal{H}(m_i), \pi^{i}_{\mathsf{MT}}, rk, erk,\pi_{\mathsf{VD}})$ subroutine, if $true$ is returned:
							\begin{itemize}[*]
								\item let $\mathsf{ledger}[\mathcal{C}]:=\mathsf{ledger}[\mathcal{C}]+n\cdot\bitcoinA_{\mathcal{C}} + {\bitcoinA}_{\mathsf{pf}}$
								\item let $\Sigma := \mathsf{not\_sold}$ and send $(\mathsf{not\_sold})$ to all entities
							\end{itemize}
						\end{itemize}
						
						{\ } \\
						{\color{violet}$\triangleright$ Reset to the ready state for repeatable delivery}
						\item[$\bullet$] {\color{blue}On receive} $(\mathsf{reset})$ from $\mathcal{P}$:
						\begin{itemize}[-]
							\item assert $\Sigma\equiv \mathsf{sold}$ or $\Sigma\equiv \mathsf{not\_sold}$
							\item set $\mathsf{ctr}$, $\mathcal{T}_{\mathsf{deliver}}$, $\mathcal{T}_{\mathsf{dispute}}$ as 0
							\item nullify $pk_{\mathcal{C}}$ and $vpk_{\mathcal{C}}$
							\item let $\theta := \theta - 1$, and $\Sigma := \mathsf{ready}$
							\item send $(\mathsf{ready})$ to all entities
						\end{itemize}
						
					\end{itemize}
				\end{flushleft}
			\end{multicols}
        \vspace{-3mm}
		}
	}
	\caption{The arbiter contract functionality $\mathcal{G}_{d}^{\mathsf{ledger}}$ for downloading. ``Sending to all entities" captures that the smart contract is transparent to the public.}\label{fig:downloading_contract_ideal_functionality}
	\vspace{-2mm}
\end{figure*}

\subsection{$\Pi_\mathsf{FD}$: \systemd{} Protocol}
\label{sec:protocol_details}
Now we present the  details of fair p2p downloading protocol $\Pi_{\mathsf{FD}}$. In particular, the protocol aims to deliver a content $m$ made of $n$ chunks\footnote{W.l.o.g., we assume $n = 2 ^ k$ for $k \in \mathbb{Z}$  for   presentation simplicity.} with a-priori known digest in the form of Merkle tree root, i.e., $\mathsf{root}_m$. We omit the session id \textit{sid} and the content digest $\mathsf{root}_m$ during the protocol description since they remain the same within a delivery session.

\begin{figure*}[!t]
	\centering
	\includegraphics[width=.96\textwidth]{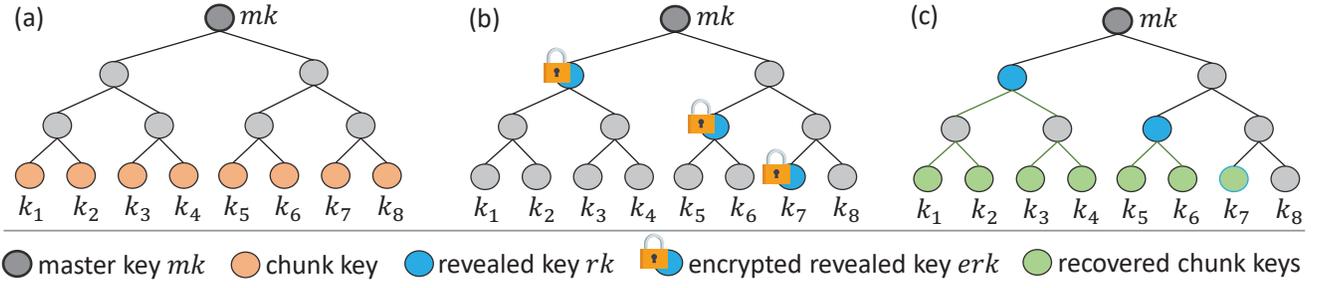}
	\vspace{-2mm}
	\caption{An example of the structured key derivation scheme in $\Pi_{\mathsf{FD}}$, where $n=8$ is the number of chunks. In (a), the encryption chunk keys $k_1,\cdots,k_8$ are  derived from a randomly sampled master key $mk$; in (b), consider the number of the delivered chunks and therefore the number of chunk keys to reveal is $\mathsf{ctr}=7$, so the three blue elements $rk$ need to be revealed; besides, $rk$ is encrypted by $\mathcal{C}$'s public key, yielding $erk$; in (c), all 7 encryption chunk keys $k_1,\cdots,k_7$ can be recovered from the revealed three blue elements $rk$, which can be used to decrypt the chunks received from $\mathcal{D}$. Note that this example shows the {\em worst} case to reveal $|erk| = \log n$ elements; in a best case, only one element, i.e., the root node $mk$, needs to be revealed.} 
	\label{fig:key_derivation}
	\vspace{-2mm}
\end{figure*}

\smallskip
\noindent
{\bf Phase I for Prepare.}
The provider $\mathcal{P}$ and the deliverer $\mathcal{D}$ interact with the contract functionality $\mathcal{G}_{d}$ in this phase as:

\begin{itemize}
	
	\item The provider $\mathcal{P}$ deploys contracts and starts~\footnote{$\mathcal{P}$ can retrieve the deposits of $\bitcoinA_{\mathcal{P}}$ and $\bitcoinA_{\mathsf{pf}}$ back if there is no deliverer responds timely.} $\Pi_{\mathsf{FD}}$ by taking as input the security parameter $\lambda$, the incentive parameters $\bitcoinA_\mathcal{P}, \bitcoinA_\mathcal{C}, \bitcoinA_{\mathsf{pf}}\in \mathbb{N}$, where $\bitcoinA_{\mathsf{pf}}$ is the {\em penalty fee}\footnote{${\bitcoinA}_{\mathsf{pf}}$ can be set proportional to $(n\times\bitcoinA_{\mathcal{C}})$ in case $\mathcal{P}$ deliberately lowers it.} in a delivery session to discrouage the misbehavior from the provider $\mathcal{P}$, the number of times $\theta$ of repeatable delivery allowed for the contract, the $n$-chunk content $m = (m_1,\dots,m_n) \in \{0,1\}^{\eta\times n}$ satisfying $\mathsf{root}(\mathsf{BuildMT}(m)) \equiv \mathsf{root}_m$ where $\mathsf{root}_m$ is the content digest in the form of Merkle tree root, and executes $(pk_{\mathcal{P}},sk_{\mathcal{P}}) \leftarrow \mathsf{SIG.KGen}(1^\lambda)$, and sends $(\mathsf{start},pk_{\mathcal{P}},\mathsf{root}_m,\theta,n,\bitcoinA_{\mathcal{P}},\bitcoinA_{\mathcal{C}}, \bitcoinA_{\mathsf{pf}})$ to $\mathcal{G}_d$.
	
	\smallskip
	\item Upon $\Sigma \equiv \mathsf{joined}$, the provider $\mathcal{P}$ would execute:
	\begin{itemize}
		\item Randomly samples a master key $mk \leftarrow_{\$} \{0,1\}^{\lambda}$, and runs Alg. \ref{alg:KeyGroupGen}, namely $\mathsf{KT} \leftarrow \mathsf{GenSubKeys}(n, mk)$; stores $mk$ and $\mathsf{KT}$ locally;
		\item Uses the leaf nodes, namely $\mathsf{KT}[n-1: 2n-2]$ (i.e., exemplified by Fig.~\ref{fig:key_derivation}a) as the encryption keys to encrypt $(m_1,\dots,m_n)$, namely $c = (c_1,\dots,c_n) \leftarrow (\mathsf{SEnc}_{\mathsf{KT}[n-1]}(m_1),\dots,\mathsf{SEnc}_{\mathsf{KT}[2n-2]}(m_n))$;
		\item Signs the encrypted chunks to obtain the sequence $((c_1, \sigma_{c_1}),\cdots, (c_n, \sigma_{c_n}))$ where the signature $\sigma_{c_i} \leftarrow \mathsf{Sign}(i||c_i, sk_{\mathcal{P}}), i\in[n]$; meanwhile, computes $\mathsf{MT}  \leftarrow \mathsf{BuildMT}(m)$ and signs the Merkle tree $\mathsf{MT}$ to obtain $\sigma^{\mathsf{MT}}_{\mathcal{P}} \leftarrow \mathsf{Sign}(\mathsf{MT}, sk_{\mathcal{P}})$, then locally stores $(\mathsf{MT}, \sigma^{\mathsf{MT}}_{\mathcal{P}})$ and sends $(\mathsf{sell}, ((c_1, \sigma_{c_1}),\cdots, (c_n, \sigma_{c_n})))$ to $\mathcal{D}$;
		\item  Waits for  $(\mathsf{ready})$ from $\mathcal{G}_{d}$ to enter the next phase.
	\end{itemize}
	
	\item The deliverer $\mathcal{D}$ executes as follows during this phase:
	\begin{itemize}
		\item Upon receiving $(\mathsf{started}, pk_\mathcal{P}, \mathsf{root}_m, \theta, n, \bitcoinA_\mathcal{P}, \bitcoinA_\mathcal{C}, \bitcoinA_{\mathsf{pf}})$ from $\mathcal{G}_{d}$, executes $(pk_{\mathcal{D}},sk_{\mathcal{D}})\leftarrow \mathsf{SIG.KGen}(1^{\lambda})$, and sends $(\mathsf{join},pk_{\mathcal{D}})$ to $\mathcal{G}_d$;
		\item Waits for $(\mathsf{sell}, ((c_1, \sigma_{c_1}),\cdots, (c_n, \sigma_{c_n})))$ from $\mathcal{P}$ and then: for every $(c_i, \sigma_{c_i})$ in the $\mathsf{sell}$ message, asserts that  $\mathsf{Verify}(i||c_i, \sigma_{c_i}, pk_{\mathcal{P}})\equiv1$; if hold, sends $(\mathsf{prepared})$ to  $\mathcal{G}_{d}$, and stores $((c_1, \sigma_{c_1}),\cdots, (c_n, \sigma_{c_n}))$ locally;
		\item Waits for  $(\mathsf{ready})$ from $\mathcal{G}_{d}$ to enter the next phase.
	\end{itemize}
	
\end{itemize}

\begin{algorithm}[!t]
	\caption{$\mathsf{GenSubKeys}$ algorithm}
	\label{alg:KeyGroupGen}
	\vspace{-2mm}
	\begin{multicols}{2}
		\algsetup{linenosize=\tiny}
		\scriptsize
		\begin{algorithmic}[1]
			\renewcommand{\algorithmicrequire}{\textbf{Input:}}
			\renewcommand{\algorithmicensure}{\textbf{Output:}}
			\REQUIRE $n, mk$
			\ENSURE a $(2n-1)$-array $\mathsf{KT}$
			\STATE let $\mathsf{KT}$ be an array, $|\mathsf{KT}|=2n-1$ 
			\STATE $\mathsf{KT}[0] = \mathcal{H}(mk)$
			\IF{$n \equiv 1$}
			\RETURN $\mathsf{KT}$
			\ENDIF
			\IF{$n > 1$}
			\FOR{$i$ in $[0, n-2]$}
			\STATE $\mathsf{KT}[2i+1] = \mathcal{H}(\mathsf{KT}[i] || 0)$
			\STATE $\mathsf{KT}[2 i+2] = \mathcal{H}(\mathsf{KT}[i] || 1)$
			\ENDFOR
			\ENDIF
			
			\RETURN $\mathsf{KT}$ 
		\end{algorithmic}
	\end{multicols}
	\vspace{-4mm}
\end{algorithm}

At the end of this phase, $\mathcal{P}$ owns a master key $mk$, the key tree $\mathsf{KT}$, and the Merkle tree $\mathsf{MT}$ while $\mathcal{D}$ receives the encrypted content chunks and is ready to deliver.

\smallskip
\noindent
{\bf Phase II for Deliver.} The consumer $\mathcal{C}$, the provider $\mathcal{P}$, and the deliverer $\mathcal{D}$ interact with the contract $\mathcal{G}_d$ in this phase as:

\begin{itemize}
	\item The consumer $\mathcal{C}$ would execute as follows:
	\begin{itemize}
		
		\item Asserts $\Sigma \equiv \mathsf{ready}$, runs $(pk_{\mathcal{C}},sk_{\mathcal{C}}) \leftarrow \mathsf{SIG.KGen}(1^{\lambda})$ and $(vpk_{\mathcal{C}},vsk_{\mathcal{C}}) \leftarrow \mathsf{VPKE.KGen}(1^{\lambda})$, and sends $(\mathsf{consume}, pk_{\mathcal{C}}, vpk_{\mathcal{C}})$ to $\mathcal{G}_{d}$;
		
		\item Upon receiving the message $(\mathsf{mtree}, \mathsf{MT}, \sigma^{\mathsf{MT}}_{\mathcal{P}})$ from $\mathcal{P}$ where
		$\mathsf{Verify}(\mathsf{MT}, \sigma^{\mathsf{MT}}_{\mathcal{P}},pk_{\mathcal{P}})\equiv1$ and $\mathsf{root}(\mathsf{MT})\equiv\mathsf{root}_m$, stores the Merkle tree $\mathsf{MT}$ and then activates the receiver $\mathcal{R}$ in the $\mathsf{VFD}$ subroutine by invoking $\mathcal{R}.\mathsf{recv}()$ and instantiating the external validation function $\Psi(i, c_i, \sigma_{c_i})$ as $\mathsf{Verify}(i||c_i, \sigma_{c_i}, pk_\mathcal{P})$, and then waits for the execution of $\mathsf{VFD}$ to return the delivered chunks $((c_1, \sigma_{c_1}), (c_2, \sigma_{c_2}),\cdots)$ and stores them; upon receiving the whole $n$-size sequence after executing the $\mathsf{VFD}$ module, sends $(\mathsf{delivered})$ to $\mathcal{G}_{d}$;
		\item Waits for $(\mathsf{revealing}, \mathsf{ctr})$ from $\mathcal{G}_{d}$ to enter the next phase.
	\end{itemize}
	
	\item The provider $\mathcal{P}$ executes as follows during this phase: upon receiving $(\mathsf{initiated}, pk_{\mathcal{C}}, vpk_{\mathcal{C}})$ from $\mathcal{G}_{d}$, asserts $\Sigma \equiv \mathsf{initiated}$, and sends $(\mathsf{mtree}, \mathsf{MT}, \sigma^{\mathsf{MT}}_{\mathcal{P}})$ to $\mathcal{C}$, and then enters the next phase.

	\item The deliverer $\mathcal{D}$ executes as follows during this phase:
	\begin{itemize}
		\item Upon receiving $(\mathsf{initiated}, pk_{\mathcal{C}}, vpk_{\mathcal{C}})$ from $\mathcal{G}_{d}$: asserts $\Sigma \equiv \mathsf{initiated}$, and then activates the sender $\mathcal{S}$ in the $\mathsf{VFD}$ module by invoking $\mathcal{S}.\mathsf{send}()$ and instantiating the external validation function $\Psi(i, c_i, \sigma_{c_i})$ as $\mathsf{Verify}(i||c_i, \sigma_{c_i}, pk_\mathcal{P})$, and feeds $\mathsf{VFD}$ module with input $((c_1,\sigma_{c_1}),\dots,(c_n,\sigma_{c_n}))$;
		\item Upon receiving $(\mathsf{\mathsf{getVFDProof}})$ from $\mathcal{G}_d$, sends the latest receipt, namely $(\mathsf{receipt}, i, \sigma^{i}_{\mathcal{C}})$ to $\mathcal{G}_d$;
		\item Waits for  $(\mathsf{revealing},\mathsf{ctr})$ from $\mathcal{G}_{d}$ to halt.
	\end{itemize}
	
\end{itemize}

At the end of this phase, $\mathcal{C}$ receives the sequence of encrypted chunks $(c_1, c_2,\dots)$, and $\mathcal{D}$ receives the payment for the bandwidth contribution of delivering chunks, and the contract records the number of delivered chunks $\mathsf{ctr}$.

\begin{algorithm}[!t]
	\caption{$\mathsf{RevealKeys}$ algorithm}
	\label{alg:RevealKey}
	\vspace{-2mm}
	\begin{multicols}{2}
		\algsetup{linenosize=\tiny}
		\scriptsize
		\begin{algorithmic}[1]
			\renewcommand{\algorithmicrequire}{\textbf{Input:}}
			\renewcommand{\algorithmicensure}{\textbf{Output:}}
			\REQUIRE $n, \mathsf{ctr},$ and $mk$ 
			\ENSURE $rk$, an array containing the minimum number of elements in $\mathsf{KT}$ that suffices to recover the $\mathsf{ctr}$ keys from $\mathsf{KT}[n-1]$  to $\mathsf{KT}[n+\mathsf{ctr}-2]$
			\STATE let $rk$ and $ind$ be empty arrays
			\STATE $\mathsf{KT} \leftarrow \mathsf{GenSubKeys}(n, mk)$
			\IF{$\mathsf{ctr}\equiv 1$}
			\STATE $rk$ appends $(n-1,\mathsf{KT[n-1]})$
			\RETURN $rk$
			\ENDIF
			\FOR{$i$ in $[0, \mathsf{ctr}-1]$}
			\STATE $ind[i]$ = $n-1+i$ 
			\ENDFOR
			\WHILE{$true$}
			\STATE let $t$ be an empty array
			\FOR{$j$ in $[0, \lfloor|ind|/2\rfloor-1]$}
			\STATE $p_l=(ind[2j]-1)/2$
			\STATE $p_r=(ind[2j+1]-2)/2$
			\\  $\triangleright$ merge elements with the same parent node in $\mathsf{KT}$
			\IF{$p_l \equiv p_r $}
			\STATE $t$ appends $p_l$ 
			\ELSE 
			\STATE  $t$ appends $ind[2j]$
			\STATE  $t$ appends $ind[2j+1]$
			\ENDIF
			\ENDFOR
			\IF{$|ind|$ is odd}
			\STATE $t$ appends $ind[|ind|-1]$
			\ENDIF 
			\IF{$|ind| \equiv |t|$}
			\STATE break
			\ENDIF
			\STATE $ind = t$
			\ENDWHILE
			\FOR{$x$ in $[0,|ind|-1]$}
			\STATE $rk$ appends $(ind[x], \mathsf{KT}[ind[x]])$
			\ENDFOR
			\RETURN $rk$
		\end{algorithmic}
	\end{multicols}
	\vspace{-3mm}
\end{algorithm}

\begin{algorithm}[!t]
	\caption{$\mathsf{ValidateRKeys}$ algorithm}
	\label{alg:ValidateRKeys}
	\vspace{-2mm}
	\begin{multicols}{1}
		\algsetup{linenosize=\tiny}
		\scriptsize
		\begin{algorithmic}[1]
			\renewcommand{\algorithmicrequire}{\textbf{Input:}}
			\renewcommand{\algorithmicensure}{\textbf{Output:}}
			\REQUIRE $n$, $\mathsf{ctr}$ and $erk$
			\ENSURE $true$ or $false$ indicating that whether the correct number (i.e., $\mathsf{ctr}$) of decryption keys can be recovered
			\IF{$n \equiv \mathsf{ctr}$ and $|erk| \equiv 1$ and the position of $erk[0] \equiv 0$}
			\RETURN $true$ \COMMENT{root of $\mathsf{KT}$}
			\ENDIF
			\STATE Initialize $chunks\_index$ as a set of numbers $\{n-1,\dots,n+\mathsf{ctr}-2\}$
			\FOR{each $(i, \_)$ in $erk$}
			\STATE $d_i = {\log(n) - \lfloor \log(i+1) \rfloor}$
			\STATE $l_i = i$, $r_i = i$
			\IF{$d_i \equiv 0$}
			\STATE $chunks\_index$ removes $i$
			\ELSE
			\WHILE{$(d_i\text{-}\text{-}) > 0$}
			\STATE $l_i = 2l_i+1$
			\STATE $r_i = 2r_i+2$
			\ENDWHILE
			\ENDIF 
			
			\STATE $chunks\_index$ removes the elements from $l_i$ to $r_i$
			\ENDFOR
			\IF{$chunks\_index\equiv\emptyset$}
			\RETURN $true$
			\ENDIF
			\RETURN $false$
		\end{algorithmic}
	\end{multicols}
	\vspace{-3mm}
\end{algorithm}

\begin{algorithm}[!t]
	\caption{$\mathsf{RecoverKeys}$ algorithm}
	\label{alg:ExtractKey}
	\vspace{-2mm}
	\begin{multicols}{1}
		\algsetup{linenosize=\tiny}
		\scriptsize
		\begin{algorithmic}[1]
			\renewcommand{\algorithmicrequire}{\textbf{Input:}}
			\renewcommand{\algorithmicensure}{\textbf{Output:}}
			\REQUIRE $n, \mathsf{ctr},$ and $rk$ 
			\ENSURE a $\mathsf{ctr}$-sized array $ks$ 
			\STATE let $ks$ be an empty array
			\FOR{each $(i, \mathsf{KT}[i])$ in $rk$}
			\STATE $n_i = 2^{(\log n - \lfloor \log(i+1) \rfloor)}$
			\STATE $v_i=\mathsf{GenSubKeys}$($n_i$, $\mathsf{KT}[i]$)
			\STATE $ks$ appends $v_i[n_i-1: 2n_i-2]$
			\ENDFOR
			\RETURN $ks$
		\end{algorithmic}
	\end{multicols}
	\vspace{-3mm}
\end{algorithm}

\smallskip
\noindent
{\bf Phase III for Reveal.} This phase is completed by   $\mathcal{P}$ and $\mathcal{C}$ in the assistance of the arbiter contract $\mathcal{G}_{d}$, which proceeds as: 

\begin{itemize}
	\item The provider $\mathcal{P}$ proceeds as follows during this phase:
	\begin{itemize}
		\item Asserts $\Sigma \equiv \mathsf{revealing}$, executes Alg.~\ref{alg:RevealKey}, namely $rk \leftarrow \mathsf{RevealKeys}(n, \mathsf{ctr}, mk)$ to generate the revealed elements $rk$, and encrypt $rk$ by running $erk \leftarrow \mathsf{VEnc}_{vpk_{\mathcal{C}}}(rk)$, as exemplified by Fig.~\ref{fig:key_derivation}b, and then sends $(\mathsf{revealKeys}, erk)$ to $\mathcal{G}_{d}$; waits for $(\mathsf{sold})$ from $\mathcal{G}_{d}$ to halt.
	\end{itemize}
	
	\item The {\em consumer} $\mathcal{C}$ in this phase would first assert $\Sigma \equiv \mathsf{revealing}$ and wait for $(\mathsf{revealed}, erk)$ from $\mathcal{G}_d$ to execute the following:
	\begin{itemize}
		
		\item Runs Alg.~\ref{alg:ValidateRKeys}, namely $\mathsf{ValidateRKeys}(n, \mathsf{ctr}, erk)$ to preliminarily check whether the revealed elements $erk$ can recover the correct number (i.e, $\mathsf{ctr}$) of keys. If $false$ is returned, sends $(\mathsf{wrongRK})$ to $\mathcal{G}_{d}$ and halts;
		\item If $\mathsf{ValidateRKeys}(n, \mathsf{ctr}, erk) \equiv true$, decrypts $erk$ to obtain $rk \leftarrow \mathsf{VDec}_{vsk_{\mathcal{C}}}(erk)$, and then runs Alg.~\ref{alg:ExtractKey}, i.e., $ks = (k_1,\cdots,k_{\mathsf{ctr}}) \leftarrow \mathsf{RecoverKeys}(n, \mathsf{ctr}, rk)$, as exemplified by Fig.~\ref{fig:key_derivation}c, to recover the chunk keys. Then $\mathcal{C}$ uses these keys to decrypt $(c_1,\cdots,c_{\mathsf{ctr}})$ to obtain $(m_1',\cdots,m_{\mathsf{ctr}}')$, where $m_i' = \mathsf{SDec}_{k_i}(c_i),i\in[\mathsf{ctr}]$, and checks whether for every $m_i' \in (m_1',\cdots,m_{\mathsf{ctr}}')$, $\mathcal{H}(m_i')$ is the $i$-th leaf node in Merkle tree $\mathsf{MT}$ received from $\mathcal{P}$ in the {\em Deliver} phase. If all are consistent, meaning that $\mathcal{C}$ receives all the desired chunks and there is no dispute, $\mathcal{C}$ outputs $(m_1',\cdots,m_{\mathsf{ctr}}')$, and then waits for $(\mathsf{sold})$ from $\mathcal{G}_{d}$ to halt. Otherwise, $\mathcal{C}$ can raise complaint by: choosing one inconsistent position (e.g., the $i$-th chunk), and computes $(rk,\pi_{\mathsf{VD}})\leftarrow\mathsf{ProvePKE}_{vsk_{\mathcal{C}}}(erk)$ and $\pi^{i}_{\mathsf{M}}\leftarrow \mathsf{GenMTP}(\mathsf{MT},\mathcal{H}(m_i))$, and then  sends $(\mathsf{PoM}, i, j, c_i, \sigma_{c_i}, \mathcal{H}(m_i), \pi^{i}_{\mathsf{MT}},rk,erk,\pi_{\mathsf{VD}})$ to the contract $\mathcal{G}_{d}$, where $i$ is the index of the incorrect chunk to be proved; $j$ is the index of the element in $erk$ that can induce the key $k_i$ for the position $i$;  $c_i$ and $\sigma_{c_i}$ are the $i$-th encrypted chunk and its signature received in the {\em Deliver} phase; $\mathcal{H}(m_i)$ is the value of the $i$-th leaf node in $\mathsf{MT}$; $\pi^{i}_{\mathsf{MT}}$ is the Merkle proof for $\mathcal{H}(m_i)$; $rk$ is decryption result from $erk$; $erk$ is the encrypted revealed key; $\pi_{\mathsf{VD}}$ is the verifiable decryption proof attesting to the correctness of decrypting $erk$ to $rk$. 
	\end{itemize}
	
\end{itemize}

\smallskip
\noindent
\textbf{Dispute resolution}. For the sake of completeness, the details of $\PoM$ subroutine is presented in Alg.~\ref{alg:ValidatePoM}, which allows the consumer to prove that it decrypts a  chunk inconsistent to the   digest $\mathsf{root}_m$. The time complexity is $O(\log n)$, which is critical to achieve the efficiency requirement. Additionally, we consider a natural case where an honest consumer $\mathcal{C}$ would not complain to the contract if receiving valid content.

\smallskip
\noindent
{\bf Design highlights}. We would like to highlight some design details in $\Pi_{\mathsf{FD}}$: (i) the $rk$ is an array containing several revealed elements, which are in the form of $(position, value)$. The $erk$ shares the similar structure where the $position$ is same and $value$ is encrypted from the corresponding $rk.value$. The $position$ is the index in $\mathsf{KT}$; (ii) to reduce the on-chain cost, the contract only stores the 256-bit hash of the $erk.value$ while emits the actual $erk$ as event logs~\cite{Lu-et-al-2020-ArXiv}. During the dispute resolution, $\mathcal{C}$ submits the $j$-th $erk$ element, and the contract would check the consistency of the submitted $erk$ with its on-chain hash; (iii) Alg.~\ref{alg:ValidateRKeys} allows the judge contract to perform preliminary check on whether the revealed elements can recover the desired number (i.e., $\mathsf{ctr}$) of decryption keys, without directly executing the relatively complex contract part of $\mathsf{ValidatePoM}$ (i.e., Alg.~\ref{alg:ValidatePoM}).

\smallskip
\noindent
{\bf Repeatable delivery}. The protocol $\Pi_\mathsf{FD}$ supports repeatable delivery, meaning that once a delivery session completes, the provider $\mathcal{P}$ can invoke the contract (by sending $(\mathsf{reset}$) to $\mathcal{G}_d$) to reset to ready state, so that new consumers can join in and start a new protocol instance. Such a $\theta$-time repeatable delivery mechanism can amortize the costs of contracts deployment and preparation (i.e., delegating encrypted chunks to a deliverer). Once $\theta$ decreases to 0, the provider $\mathcal{P}$ can either deploy a new contract (thus residing at a new contract address) or utilize the same contract address while re-run the {\em Prepare} phase. $\mathcal{P}$ may not need to delegate the encrypted chunks if a previously participating deliverer joins in. 

\smallskip
\noindent
{\bf Dynamic deposits adjustment}. An interesting extension is with respect to the financial deposits that $\mathcal{P}$ provides. Specifically, $\Pi_\mathsf{FD}$ needs $\mathcal{P}$ to deposit the payment $(\theta\cdot n\cdot\bitcoinA_{\mathcal{P}})$ to incentivize successful delivery and $(\theta\cdot\bitcoinA_{\mathsf{pf}})$ to discourage $\mathcal{P}$'s potential misbehavior. Such a deposit is locked-up in contract and unaccessible, which poses potential loss for $\mathcal{P}$, for example, the provider may face an opportunity cost in the form of forgone returns that they could have accrued in alternative investments. Hence, integrating mechanisms~\cite{Harz-et-al-2019-CCS} of dynamic adjustment of cryptocurrency deposits can be an future extension, which also applies to the streaming setting in Section~\ref{sec:ProtocolDesign_Streaming}.

\begin{algorithm}[!t]
	\caption{$\mathsf{ValidatePoM}$ algorithm}
	\label{alg:ValidatePoM}
	\algsetup{linenosize=\tiny}
	\scriptsize
	\begin{algorithmic}[1]
		\renewcommand{\algorithmicrequire}{\textbf{Input:}}
		\renewcommand{\algorithmicensure}{\textbf{Output:}}
		\REQUIRE $(i, j, c_i, \sigma_{c_i}, \mathcal{H}(m_i),\pi^{i}_{\mathsf{MT}},rk,erk,\pi_{\mathsf{VD}})$\\ $(\mathsf{root}_m,n,erk_\mathsf{hash},pk_{\mathcal{P}},vpk_{\mathcal{C}})$ are stored in the contract and hence accessible
		\ENSURE $true$ or $false$
		\STATE assert $j \in [0, |erk|-1] $
		
		\STATE assert $\mathcal{H}(erk) \equiv erk_\mathsf{hash}$
		
		\STATE assert $\mathsf{VerifyPKE}_{vpk_{\mathcal{C}}}(erk, rk, \pi_{\mathsf{VD}})\equiv 1$
		
		\STATE assert $\mathsf{Verify}(i||c_i, \sigma_{c_i}, pk_\mathcal{P}) \equiv 1$
		
		\STATE assert $\mathsf{VerifyMTP}(\mathsf{root}_m, i, \pi^{i}_{\mathsf{MT}}, \mathcal{H}(m_i)) \equiv 1$

		\STATE $k_i$ = $\mathsf{RecoverChunkKey}(i, j, n, rk)$
		\STATE assert $k_i \neq \bot$ 
		\STATE $ m_i' = \mathsf{SDec}(c_i, k_i)$
		\STATE assert $\mathcal{H}(m_i') \neq \mathcal{H}(m_i)$
		\RETURN $false$ in case of any assertion error or $true$ otherwise
	\end{algorithmic}
\end{algorithm}

\begin{algorithm}[!t]
	\caption{$\mathsf{RecoverChunkKey}$ algorithm}
	\label{alg:ExtractChkKey}
	\vspace{-2mm}
	\begin{multicols}{1}
		\algsetup{linenosize=\tiny}
		\scriptsize
		\begin{algorithmic}[1]
			\renewcommand{\algorithmicrequire}{\textbf{Input:}}
			\renewcommand{\algorithmicensure}{\textbf{Output:}}
			\REQUIRE $(i, j, n, rk)$
			\ENSURE $k_i$ or $\perp$
			\STATE $(x,y) \leftarrow rk[j]$\\ \COMMENT{parse the $j$-th element in $rk$ to get the key $x$ and the value $y$}
			\STATE let $k\_path$ be an empty stack
			\STATE $ind = n+i-2$ \COMMENT{index in $\mathsf{KT}$}
			\IF{$ind < x$}
			\RETURN $\perp$
			\ENDIF
			\IF{$ind \equiv x$}
			\RETURN $y$ \COMMENT{$k_i = y$}
			\ENDIF
			\WHILE{$ind > x$}
			\STATE $k\_path$ pushes $0$  if $ind$ is odd
			\STATE $k\_path$ pushes $1$  if $ind$ is even
			\STATE $ind = \lfloor (ind - 1) / 2 \rfloor$
			\ENDWHILE
			\STATE let $b = |k\_path|$
			\WHILE{$(b\text{-\text{-}}) > 0$}
			\STATE pop $k\_path$ to get the value $t$
			\STATE $k_i = \mathcal{H}(y||t)$
			\ENDWHILE
			\RETURN $k_i$
		\end{algorithmic}
	\end{multicols}
	\vspace{-3mm}
\end{algorithm}

\subsection{Analyzing $\mathsf{FairDownload}$ Protocol}

Now we provide the detailed proofs that the protocol $\Pi_{\mathsf{FD}}$ satisfies the design goals.

\noindent
\begin{lemma}
	\label{lemma:downloading_completeness}
	Conditioned that all parties $\mathcal{P}$, $\mathcal{D}$ and $\mathcal{C}$ are honest, $\Pi_{\mathsf{FD}}$ satisfies the completeness property in the synchronous authenticated network and stand-alone model.
\end{lemma}

\noindent{\em Proof.} 
The completeness of $\Pi_{\mathsf{FD}}$ is immediate to see: when all three participating parties honestly follow the protocol, the provider $\mathcal{P}$ gets a net income  of $n\cdot(\bitcoinA_{\mathcal{C}}-\bitcoinA_{\mathcal{P}})$;
the deliverer $\mathcal{D}$ obtains the well-deserved payment of $n\cdot\bitcoinA_{\mathcal{P}}$;
the consumer $\mathcal{C}$ receives the valid content $m$, i.e., $\phi(m)\equiv1$.

\noindent
\begin{lemma}
	\label{lemma:downloading_fairness}
	In the synchronous authenticated model and stand-alone setting, conditioned that the underlying cryptographic primitives are secure, $\Pi_{\mathsf{FD}}$ satisfies the fairness requirement even when at most two parties of $\mathcal{P}$, $\mathcal{D}$ and $\mathcal{C}$ are corrupted by non-adaptive P.P.T. adversary $\adv$.
\end{lemma}

\noindent{\em Proof.} The fairness for each party in $\Pi_{\mathsf{FD}}$ can be reduced to the underlying cryptographic building blocks, which can be analyzed as follows:
\begin{itemize}
	\item \underline{\smash{\em Consumer Fairness.}} Consumer fairness means that the honest $\mathcal{C}$ only needs to pay proportional to what it {\em de facto} obtains even though malicious $\mathcal{P}^*$ and $\mathcal{D}^*$ may collude with each other. This case can be modeled as an adversary $\adv$ corrupts both $\mathcal{P}$ and $\mathcal{D}$ to provide and deliver the content to the honest $\mathcal{C}$. During the {\em Deliver} phase, the $\mathsf{VFD}$ subroutine ensures that $\mathcal{C}$ receives the sequence $(c_1,\sigma_{c_1}),\dots,(c_\ell,\sigma_{c_\ell})$, $\ell\in [n]$ though $\mathcal{A}$ may maliciously abort. Later $\mathcal{A}$ can only claim payment from the contract of $\ell\cdot\bitcoinA_{\mathcal{P}}$, which is paid by the $\mathcal{A}$ itself due to the collusion. During the {\em Reveal} phase, if $\mathcal{A}$ reveals correct elements in $\mathsf{KT}$ to recover the $\ell$ decryption keys, then $\mathcal{C}$ can decrypt to obtain the valid $\ell$ chunks. Otherwise, $\mathcal{C}$ can raise complaint by sending the $(\mathsf{wrongRK})$ and further $(\mathsf{PoM})$ to the contract and gets refund. It is obvious to see that $\mathcal{C}$ either pays for the $\ell$ valid chunks or pays nothing. The fairness for the consumer is guaranteed unless $\mathcal{A}$ can: (i) break $\mathsf{VFD}$ to forge $\mathcal{C}$'s signature; (ii) find Merkle tree collision, namely find another chunk $m_i'\neq m_i$ in position $i$ of $m$ to bind to the same $\mathsf{root}_m$ so that $\mathcal{A}$ can fool the contract to reject $\mathcal{C}$'s complaint (by returning $false$ of $\mathsf{ValidatePoM}$) while indeed sent wrong chunks; (iii) manipulate the execution of smart contract in blockchain. However, according to the security guarantee of the underlying signature scheme, the second-preimage resistance of hash function in Merkle tree, and that the smart contract is modeled as an ideal functionality, the probability to break $\mathcal{C}$'s fairness is negligible. Therefore, the consumer fairness being secure against the collusion of malicious $\mathcal{P}^*$ and $\mathcal{D}^*$ is guaranteed. 
	
	\item \underline{\smash{\em Deliverer Fairness.}} Deliverer fairness states that the honest $\mathcal{D}$ receives the payment proportional to the expended bandwidth even though the malicious $\mathcal{P}^*$ and $\mathcal{C}^*$ may collude with each other. This amounts to the case that $\mathcal{A}$ corrupts both $\mathcal{P}$ and $\mathcal{C}$ and try to reap $\mathcal{D}$'s bandwidth contribution without paying. In the $\mathsf{VFD}$ subroutine, considering $\mathcal{D}$ delivers $\ell$ chunks, then it can correspondingly obtain either $\ell$ $(\ell\in [n])$ or $\ell-1$ (i.e., $\mathcal{A}$ stops sending the $\ell$-th receipt) receipts acknowledging the bandwidth contribution. Later $\mathcal{D}$ can use the latest receipt containing $\mathcal{C}$'s signature to claim payment $\ell\cdot\bitcoinA_{\mathcal{P}}$ or $(\ell-1)\cdot\bitcoinA_{\mathcal{P}}$ from the contract. At most $\mathcal{D}$ may waste bandwidth for delivering one chunk-validation pair of $O(\eta)$ bits. To break the security, $\mathcal{A}$ has to violate the contract functionality (i.e., control the execution of smart contract in blockchain), which is of negligible probability. Therefore, the deliverer fairness being secure against the collusion of malicious $\mathcal{P}^*$ and $\mathcal{C}^*$ is ensured.
	
	\item \underline{\smash{\em Provider Fairness.}} Provider fairness indicates that the honest $\mathcal{P}$ receives the payment proportional to the number of valid content chunks that $\mathcal{C}$ receives. The malicious $\mathcal{D}^*$ can collude with the malicious $\mathcal{C}^*$ or simply create multiple fake $\mathcal{C}^*$ (i.e., Sybil attack), and then cheat $\mathcal{P}$ without real delivery. These cases can be modeled as an adversary $\mathcal{A}$ corrupts both $\mathcal{D}$ and $\mathcal{C}$. $\mathcal{A}$ can break the fairness of the honest $\mathcal{P}$ from two aspects by: (i) letting $\mathcal{P}$ pay for the delivery without truly delivering any content; (ii) obtaining the content without paying for $\mathcal{P}$. For case (i), $\mathcal{A}$ can claim that $\ell$ ($\ell\in [n]$) chunks have been delivered and would receive the payment $\ell\cdot\bitcoinA_{\mathcal{P}}$ from the contract. Yet this procedure would also update $\mathsf{ctr} := \ell$ in the contract, which later allows $\mathcal{P}$ to receive the payment $\ell\cdot\bitcoinA_{\mathcal{C}}$ after $\mathcal{T}_{\mathsf{dispute}}$ expires unless $\mathcal{A}$ can manipulate the execution of smart contract, which is of negligible probability. Hence, $\mathcal{P}$ can still obtain the well-deserved payment $\ell\cdot(\bitcoinA_\mathcal{C} - \bitcoinA_{\mathcal{P}})$. For case (ii), $\mathcal{A}$ can either try to decrypt the delivered chunks by itself without utilizing the revealing keys from $\mathcal{P}$, or try to fool the contract to accept the $\mathsf{PoM}$ and therefore repudiate the payment for $\mathcal{P}$ though $\mathcal{P}$ honestly reveals chunk keys. The former situation can be reduced to the need of violating the semantic security of the underlying encryption scheme and the pre-image resistance of cryptographic hash functions, and the latter requires $\mathcal{A}$ to forge $\mathcal{P}$'s signature, or break the soundness of the verifiable decryption scheme, or control the execution of the smart contract. Obviously, the occurrence of aforementioned situations are in negligible probability. Overall, the provider fairness being secure against the collusion of malicious $\mathcal{D}^*$ and $\mathcal{C}^*$ is assured.
\end{itemize}

In sum, $\Pi_{\mathsf{FD}}$ strictly guarantees the fairness for $\mathcal{P}$ and $\mathcal{C}$, and the unpaid delivery for $\mathcal{D}$ is bounded to $O(\eta)$ bits. The fairness requirement of $\Pi_{\mathsf{FD}}$ is satisfied.

\noindent
\begin{lemma}
	\label{lemma:downloading_confidentiality}
	In the synchronous authenticated network and stand-alone model, 
	$\Pi_{\mathsf{FD}}$ satisfies the confidentiality property against malicious deliverer corrupted by non-adaptive P.P.T. adversary $\adv$.
\end{lemma}

\noindent{\em Proof.} This property states that on input all protocol scripts and the corrupted deliverer's private input and all internal states, 
it is still computationally infeasible for the adversary to output the provider's input content. In $\Pi_{\mathsf{FD}}$, each chunk delegated to $\mathcal{D}$ is encrypted using symmetric encryption scheme before delivery by encryption key derived from Alg. \ref{alg:KeyGroupGen}. The distribution of encryption keys and uniform distribution cannot be distinguished by the P.P.T. adversary. Furthermore, the revealed on-chain elements $erk$ for recovering some chunks' encryption keys are also encrypted utilizing the consumer $\mathcal{C}$'s pubic key, which can not be distinguished from uniform distribution by the adversary. Additionally, $\mathcal{C}$ receives the Merkle tree $\mathsf{MT}$ of the content $m$ before the verifiable fair delivery ($\mathsf{VFD}$) procedure starts. 
Thus to break the confidentiality property, the adversary $\adv$ has to violate any of the following conditions: (i) the pre-image resistance of Merkle tree, which can be further reduce to the pre-image resistance of cryptographic hash function; and (ii) the security of the public key encryption scheme, essentially requiring at least to solve decisional Diffie-Hellman  problem. The probability of violating the aforementioned security properties is negligible, and therefore, $\Pi_{\mathsf{FD}}$ satisfies the confidentiality property against malicious deliverer corrupted by $\adv$.

\noindent
\begin{lemma}
	\label{lemma:downloading_timeliness}
	If at least one of the three parties $\mathcal{P}$, $\mathcal{D}$, $\mathcal{C}$ is honest and others are corrupted by non-adaptive P.P.T. adversary $\mathcal{A}$, $\Pi_{\mathsf{FD}}$ satisfies the timeliness property in the synchronous authenticated network and stand-alone model.
\end{lemma}

\noindent{\em Proof.} 
Timeliness states that the honest parties in the protocol $\Pi_{\mathsf{FD}}$ terminates in $O(n)$ synchronous rounds, where $n$ is the number of content chunks, and when the protocol completes or aborts, the fairness and confidentiality are always preserved. As the guarantee of confidentiality can be straightforwardly derived due to the lemma~\ref{lemma:downloading_confidentiality} even if malicious parties abort, we only focus on the assurance of  fairness. Now we elaborate the following termination cases for the protocol $\Pi_{\mathsf{FD}}$ with the arbiter contract $\mathcal{G}_d$ and at least one honest party:

\noindent
\underline{\smash{\em No abort.}} If all parties of $\mathcal{P}$, $\mathcal{D}$ and $\mathcal{C}$ are honest, the protocol $\Pi_{\mathsf{FD}}$ terminates in the {\em Reveal} phase, after $\mathcal{T}_{\mathsf{dispute}}$ expires. The {\em Prepare} phase and the {\em Reveal} phase need $O(1)$ synchronous rounds, and the {\em Deliver} phase requires $O(n)$ rounds where $n$ is the number of content chunks, yielding totally $O(n)$ rounds for the protocol $\Pi_{\mathsf{FD}}$ to terminate and the fairness is guaranteed at completion since each party obtains the well-deserved items.

\noindent
\underline{\smash{\em Aborts in the Prepare phase.}} This phase involves the interaction between the provider $\mathcal{P}$, the deliverer $\mathcal{D}$, and the arbiter contract $\mathcal{G}_d$. It is obvious this phase can terminate in $O(1)$ rounds if any party maliciously aborts or the honest party does not receive response after $\mathcal{T}_{\mathsf{round}}$ expires. Besides, after each step in this phase, the fairness for both $\mathcal{P}$ and $\mathcal{D}$ is preserved no matter which one of them aborts, meaning that $\mathcal{P}$ does not lose any coins in the $\mathsf{ledger}$ or leak any content chunks, while $\mathcal{D}$ does not waste any bandwidth resource.

\noindent
\underline{\smash{\em Aborts in the Deliver phase.}} This phase involves the provider $\mathcal{P}$, the deliverer $\mathcal{D}$, the consumer $\mathcal{C}$, and the arbiter contract $\mathcal{G}_d$. It can terminate in $O(n)$ rounds. After $\mathcal{C}$ sends $(\mathsf{consume})$ message to the contract, and then other parties aborts, $\mathcal{C}$ would get its deposit back once $\mathcal{T}_{\mathsf{round}}$ times out. The $\mathsf{VFD}$ procedure in this phase only involves $\mathcal{D}$ and $\mathcal{C}$, and the fairness is guaranteed whenever one of the two parties aborts, as analyzed in lemma~\ref{lemma:VFD_completeness_fairness}. The timer $\mathcal{T}_{\mathsf{deliver}}$ in contract indicates that the whole $n$-chunk delivery should be completed within such a time period, or else $\mathcal{G}_d$ would continue with the protocol by informing $\mathcal{D}$ to claim payment and update $\mathsf{ctr}$ after $\mathcal{T}_{\mathsf{deliver}}$ times out. $\mathcal{D}$ is motivated not to maliciously abort until receiving the payment from the contract. At the end of this phase, $\mathcal{D}$ completes its task in the delivery session, while for $\mathcal{P}$ and $\mathcal{C}$, they are motivated to enter the next phase and the fairness for them at this point is guaranteed, i.e., $\mathcal{P}$ decreases coins by $\mathsf{ctr}\cdot\bitcoinA_\mathcal{P}$ in $\mathsf{ledger}$, but the contract has also updated $\mathsf{ctr}$, which allows $\mathcal{P}$ to receive $\mathsf{ctr}\cdot\bitcoinA_{\mathcal{C}}$ from the $\mathsf{ledger}$ if keys are revealed honestly, and $\mathcal{C}$ obtains the encrypted chunks while does not lose any coins in $\mathsf{ledger}$.

\noindent
\underline{\smash{\em Aborts in the Reveal phase.}} This phase involves the provider $\mathcal{P}$, the consumer $\mathcal{C}$, and the arbiter contract $\mathcal{G}_d$. It can terminate in $O(1)$ rounds after the contract sets the state as $\mathsf{sold}$ or $\mathsf{not\_sold}$. If $\mathcal{C}$ aborts after $\mathcal{P}$ reveals the chunk keys on-chain, $\mathcal{P}$ can wait until $\mathcal{T}_{\mathsf{dispute}}$ times out and attain the deserved payment $\mathsf{ctr}\cdot\bitcoinA_{\mathcal{C}}$. If $\mathcal{P}$ reveals incorrect keys and then aborts, $\mathcal{C}$ can raise complaint within $\mathcal{T}_{\mathsf{dispute}}$ by sending message $(\mathsf{wrongRK})$ and further $(\mathsf{PoM})$ to get refund. Hence, the fairness for either $\mathcal{P}$ and $\mathcal{C}$ is guaranteed no matter when and which one aborts maliciously in this phase.


\noindent
\begin{lemma}
	\label{lemma:downloading_efficiency}
	In the synchronous authenticated network and stand-alone model, for any non-adaptive P.P.T. adversary $\adv$, $\Pi_{\mathsf{FD}}$ meets the efficiency requirement that: the communication complexity is bounded to $O(n)$; the on-chain cost is bounded to $\widetilde{O}(1)$; the messages sent by the provider $\mathcal{P}$ after preparation are bounded to $n\cdot \lambda$ bits, where $n$ is the number of chunks and $\lambda$ is a small cryptographic parameter, and $n\cdot \lambda$ is much less than the content size $|m|$.
\end{lemma}

\noindent{\em Proof.} 
The analysis regarding the non-trivial efficiency property can be conducted in the following three aspects:
\begin{itemize}
	\item \underline{\smash{\em Communication Complexity.}} In the {\em Prepare} phase, $\mathcal{P}$ delegates the signed encrypted chunks to $\mathcal{D}$, where the communication complexity is $O(n)$. Typically this phase only needs to be executed once for the same content. In the {\em Deliver} phase, $\mathcal{P}$ sends the content Merkle tree $\mathsf{MT}$ to $\mathcal{C}$, and $\mathcal{D}$ activates the $\mathsf{VFD}$ subroutine to deliver the content chunks to $\mathcal{C}$. The communication complexity in this phase is also $O(n)$. In the {\em Reveal} phase, the revealed elements for recovering $\mathsf{ctr}$ keys is {\em at most} $O(\log{n})$. Finally, if dispute happens, the communication complexity of sending $\mathsf{PoM}$ (mostly due to the merkle proof $\pi^{i}_{\mathsf{MT}}$) to the contract is $O(\log{n})$. Therefore, the communication complexity of the protocol $\Pi_{\mathsf{FD}}$ is $O(n)$.
	
	\item \underline{\smash{\em On-chain Cost.}} In the \textit{optimistic} case where no dispute occurs, the on-chain costs of $\Pi_{\mathsf{FD}}$ include: (i) the functions (i.e., $\mathsf{start}$, $\mathsf{join}$ and $\mathsf{prepared}$) in the {\em Prepare} phase are all $O(1)$; (ii) in the {\em Deliver} phase, the $\mathsf{consume}$ and $\mathsf{delivered}$ functions are $O(1)$. Note that in the $\mathsf{delivered}$ function, the cost of signature verification is $O(1)$ since $\mathcal{D}$ only needs to submit the latest $\mathsf{receipt}$ containing one signature of $\mathcal{C}$; (iii) the storage cost for revealed elements (i.e., $erk$) is {\em at most} $O(\log{n})$, where $n$ is the number of chunks. Hence, the overall on-chain cost is {\em at most} $O(\log n)$, namely $\widetilde{O}(1)$. In the \textit{pessimistic} case where dispute happens, the on-chain cost is only related to the delivered chunk size $\eta$ no matter how large the content size $|m|$ is (the relationship between the chunk size and costs in different modes is depicted in Section~\ref{sec:ImplementationandEvaluation}).
	
	\item \underline{\smash{\em Message Volume for $\mathcal{P}$.}} Considering that the contract is deployed and the deliverer is ready to deliver. Every time when a new consumer joins in, a new delivery session starts. The provider $\mathcal{P}$ shows up twice for: (i) sending the Markle tree $\mathsf{MT}$, which is in complexity of $O(\log n)$, to $\mathcal{C}$ in the {\em Deliver} phase, and (ii) revealing $erk$, which is in complexity of {\em at most} $O(\log n)$, to $\mathcal{C}$ in the {\em Reveal} phase. The total resulting message volume $O(\log n)$ can be represented as $n\cdot\lambda$ bits, where $\lambda$ is a small cryptographic parameter, and $n\cdot\lambda$ is obviously much less than the content size of $|m|$.
\end{itemize}

\begin{theorem}
	\label{thm:downloading}
	Conditioned on that the underlying cryptographic primitives are secure, the protocol $\mathsf{FairDownload}$ satisfies the completeness, fairness, confidentiality against deliverer, timeliness and non-trivial efficiency properties in the synchronous authenticated network, $\mathcal{G}^{\mathsf{ledger}}_{d}$-hybrid and stand-alone model.
\end{theorem}

\noindent{\em Proof.} 
Lemmas~\ref{lemma:downloading_completeness},~\ref{lemma:downloading_fairness},~\ref{lemma:downloading_confidentiality},~\ref{lemma:downloading_timeliness}, and~\ref{lemma:downloading_efficiency} complete the proof.



\section{\systems{}: Fair p2p Streaming}
\label{sec:ProtocolDesign_Streaming}
In this section, we present the p2p fair delivery protocol $\Pi_{\mathsf{FS}}$, allowing {\em view-while-delivery} in the streaming setting.

\subsection{\systems{} Overview}

\begin{figure}[!t]
	\centering
	\includegraphics[width=.49\textwidth]{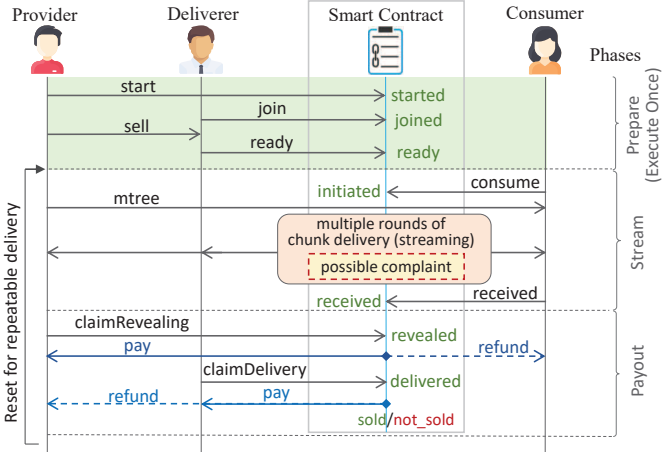}
	\vspace{-6mm}
	\caption{The overview of $\mathsf{FairStream}$ protocol $\Pi_{\mathsf{FS}}$. The dispute may arise in a certain round in the {\em Stream} phase, and the messages $(\mathsf{claimDelivery})$ and $(\mathsf{claimRevealing})$ may be sent to the contract in a different order.}
	\label{fig:streaming_protocol_overview}
	\vspace{-2mm}
\end{figure}

As depicted in Fig.~\ref{fig:streaming_protocol_overview}, our protocol $\Pi_{\mathsf{FS}}$ works as three phases, i.e., {\em Prepare}, {\em Stream}, and {\em Payout}, at a high level. The core ideas for $\Pi_{\mathsf{FS}}$ are: 

\begin{itemize}
	\item Same as the {\em Prepare} phase in $\Pi_{\mathsf{FD}}$, initially the content provider $\mathcal{P}$ would deploy the smart contract, encrypt content chunks, sign the encrypted chunks and delegate to the deliverer $\mathcal{D}$.
	\item The streaming process consists of $O(n)$ communication rounds, where $n$ is the number of chunks. In each round, the consumer $\mathcal{C}$ would receive an encrypted chunk from $\mathcal{D}$ and a decryption key from $\mathcal{P}$; any party may abort in a certain round due to, e.g., untimely response or invalid message; especially, in case erroneous chunk is detected during streaming, $\mathcal{C}$ can complain and get compensated with a valid and short (i.e., $O(\eta + \lambda)$ bits) proof; 
	\item Eventually all parties enter the {\em Payout} phase, where $\mathcal{D}$ and $\mathcal{P}$ can claim the deserved payment by submitting the latest receipt signed by the consumer before a timer maintained in contract expires; the contract determines the final internal state $\mathsf{ctr}$, namely the number of delivered chunks or revealed keys, as the {\em larger} one of the indexes in $\mathcal{P}$ and $\mathcal{D}$'s receipts. If no receipt is received from $\mathcal{P}$ or $\mathcal{D}$ before the timer expires, the contract would treat the submitted index for that party as 0. Such a design is critical to ensure fairness as analyzed in Section~\ref{sec:FS_analysis}.
\end{itemize}

\begin{figure}[!t]
	\centering
	\includegraphics[width=.45\textwidth]{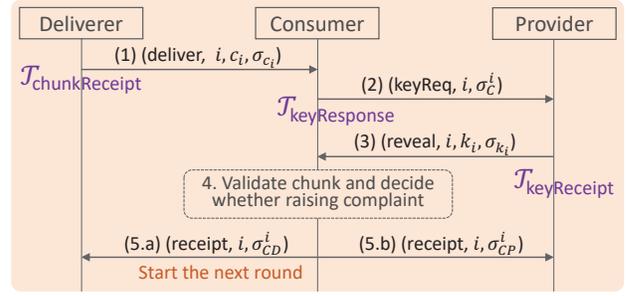}
	\caption{The concrete message flow of one round chunk delivery in the {\em Stream} phase of $\Pi_{\mathsf{FS}}$. All these messages are sent off-chain.}
	\label{fig:streaming_round}
	\vspace{-2mm}
\end{figure}

Fig.~\ref{fig:streaming_round} illustrates the concrete message flow of one round chunk delivery during the {\em Stream} phase. We highlight that a black-box call of the $\mathsf{VFD}$ module is not applicable to the streaming setting since $\mathsf{VFD}$ only allows the consumer $\mathcal{C}$ to obtain the encrypted chunks, which brings the advantage that the provider $\mathcal{P}$ merely needs to show up once to reveal a minimum number of elements and get all chunk keys to be recovered. However, the streaming procedure demands much less latency of retrieving each content chunk, leading to the intuitive design to let $\mathcal{C}$ receive both an encrypted chunk and a corresponding chunk decryption key in one same round. $\mathcal{P}$ is therefore expected to keep online and reveal each chunk key to $\mathcal{C}$. Overall, the protocol design in $\Pi_{\mathsf{FS}}$ requires relatively more involvement of the provider $\mathcal{P}$ compared with the downloading setting, but the advantage is that instead of downloading all chunks in $O(n)$ rounds before viewing, the consumer $\mathcal{C}$ now can retrieve each chunk with $O(1)$ latency. All other properties including each party's fairness, the on-chain computational cost, and the deliverer's communication complexity remain the same as those in the downloading setting. 

\subsection{Arbiter Contract $\mathcal{G}_{s}^{\mathsf{ledger}}$ for Streaming}

The arbiter contract $\mathcal{G}_{s}^{\mathsf{ledger}}$ (abbr. $\mathcal{G}_{s}$) illustrated in Fig.~\ref{fig:streaming_contract_ideal_functionality} is a stateful ideal functionality that can access to $\mathsf{ledger}$ functionality to facilitate the fair content delivery via streaming. The timer $\mathcal{T}_{\mathsf{receive}}$ ensures that when any party maliciously aborts or the consumer $\mathcal{C}$ receives invalid chunk during the streaming process, the protocol $\Pi_{\mathsf{FS}}$ can smoothly continue and enter the next phase. The dispute resolution in contract is relatively simpler than the downloading setting since no verifiable decryption is needed. The timer $\mathcal{T}_{\mathsf{finish}}$ indicates that both $\mathcal{D}$ and $\mathcal{P}$ are supposed to send the request of claiming their payment before $\mathcal{T}_{\mathsf{finish}}$ times out, and therefore it is natural to set $\mathcal{T}_{\mathsf{finish}} > \mathcal{T}_{\mathsf{receive}}$. Once $\mathcal{T}_{\mathsf{finish}}$ expires, the contract determines the final $\mathsf{ctr}$ by choosing the maximum index in $\mathcal{P}$ and $\mathcal{D}$'s receipts, namely $\mathsf{ctr}_{\mathcal{P}}$ and $\mathsf{ctr}_{\mathcal{D}}$, respectively, and then distributes the well-deserved payment for each party. Once the delivery session completes, the provider $\mathcal{P}$ can invoke the contract by sending $(\mathsf{reset})$ to $\mathcal{G}_s$ to reset to the ready state and continue to receive new requests from consumers.


\begin{figure*}[!t]
	\centering
	\footnotesize
	\fbox{%
		\parbox{.96\linewidth}{%
			\vspace{-2mm}
			\begin{center}
				{\bf The  Arbiter Contract Functionality $\mathcal{G}_{s}^{\mathsf{ledger}}$ for p2p Streaming}
				\vspace{1mm}
				
				The contract  $\mathcal{G}_{s}$ can access to $\mathsf{ledger}$, and it interacts with $\mathcal{P}$, $\mathcal{D}$, $\mathcal{C}$ and the adversary $\adv$. It locally stores $\theta$, $n$, $\mathsf{root}_m$, $\bitcoinA_{\mathcal{P}}$, $\bitcoinA_{\mathcal{C}}$, ${\bitcoinA}_{\mathsf{pf}}$, $\mathsf{ctr}_{\mathcal{D}}$, $\mathsf{ctr}_\mathcal{P}$, $\mathsf{ctr}$ (all $\mathsf{ctr}_{\mathcal{D}}$, $\mathsf{ctr}_\mathcal{P}$, $\mathsf{ctr}$ are initialized as 0), $pk_{\mathcal{P}}, pk_{\mathcal{D}}, pk_{\mathcal{C}}$, the penalty flag $\mathsf{plt}$ (initialized by $false$),  the state $\Sigma$ and three timers \\ $\mathcal{T}_{\mathsf{round}}$ (implicitly), $\mathcal{T}_{\mathsf{receive}}$, $\mathcal{T}_{\mathsf{finish}}$.
			\end{center}
			\vspace{-4mm}
			
			\begin{multicols}{2}
				\begin{flushleft}
					
					\noindent
					\xrfill[0.5ex]{0.5pt} {} {\bf Phase 1: Prepare} \xrfill[0.5ex]{0.5pt}
					
					\begin{itemize}
						\item[$\bullet$] This phase is the same as the {\em Prepare} phase in $\mathcal{G}_{d}$.
					\end{itemize}

					\noindent
					\xrfill[0.5ex]{0.5pt} {} {\bf Phase 2: Stream}  \xrfill[0.5ex]{0.5pt}
					
					\begin{itemize}
						\item[$\bullet$] {\color{blue}On receive} $(\mathsf{consume}, pk_{\mathcal{C}})$ from $\mathcal{C}$:
						\begin{itemize}[-]
							\item assert $\theta > 0$
							\item[-] assert $\mathsf{ledger}[\mathcal{C}]\ge n\cdot\bitcoinA_{\mathcal{C}}$ $\wedge$ $\Sigma\equiv\mathsf{ready}$
							\item[-] store $pk_{\mathcal{C}}$ and let $\mathsf{ledger}[\mathcal{C}]:=\mathsf{ledger}[\mathcal{C}]-n\cdot\bitcoinA_{\mathcal{C}}$
							\item[-] start two timers $\mathcal{T}_{\mathsf{receive}}$, and $\mathcal{T}_{\mathsf{finish}}$ 
							\item let $\Sigma := \mathsf{initiated}$ and send $(\mathsf{initiated}, pk_{\mathcal{C}})$ to all entities
						\end{itemize}
						
						\item[$\bullet$] {\color{blue}On receive} $(\mathsf{received})$ from $\mathcal{C}$ before $\mathcal{T}_{\mathsf{receive}}$ times out:
						\begin{itemize}[-]
							\item[-] assert current time $\mathcal{T} < \mathcal{T}_{\mathsf{receive}}$ and $\Sigma\equiv\mathsf{initiated}$
							\item[-] let $\Sigma := \mathsf{received}$ and send $(\mathsf{received})$ to all entities
						\end{itemize}
						
						\item[$\bullet$] {\color{blue}Upon} $\mathcal{T}_{\mathsf{receive}}$ times out:
						\begin{itemize}[-]
							\item[-] assert current time $\mathcal{T} \geq \mathcal{T}_{\mathsf{receive}}$ and $\Sigma\equiv\mathsf{initiated}$
							\item[-] let $\Sigma := \mathsf{received}$ and send $(\mathsf{received})$ to all entities
						\end{itemize}
						
						{\ } \\ 
						
						\item[] {\color{purple}$\triangleright$ Below is to resolve dispute during streaming in $\Pi_{\mathsf{FS}}$}
						\vspace{0.5mm}
						
						\item[$\bullet$] {\color{blue}On receive} $(\mathsf{PoM}, i, c_i, \sigma_{c_i}, k_i, \sigma_{k_i}, \mathcal{H}(m_i), \pi^{i}_{\mathsf{MT}})$ from $\mathcal{C}$ before $\mathcal{T}_{\mathsf{receive}}$ expires:
						\begin{itemize}[-]
							\item assert current time $\mathcal{T} < \mathcal{T}_{\mathsf{receive}}$ and $\Sigma\equiv\mathsf{initiated}$
							\item assert $\mathsf{Verify}(i||c_i, \sigma_{c_i}, pk_{\mathcal{P}}) \equiv 1$
							\item assert $\mathsf{Verify}(i||k_i, \sigma_{k_i}, pk_{\mathcal{P}}) \equiv 1$
							\item assert $\mathsf{VerifyMTP}(\mathsf{root}_m, i, \pi^{i}_{\mathsf{MT}}, \mathcal{H}(m_i)) \equiv 1$
							\item $m_i' = \mathsf{SDec}(c_i, k_i)$
							\item assert $\mathcal{H}(m_i') \neq \mathcal{H}(m_i)$
							\item let $\mathsf{plt} := true$
							\item let $\Sigma := \mathsf{received}$ and send $(\mathsf{received})$ to all entities
						\end{itemize}
						
					\end{itemize}

					\noindent
					\xrfill[0.5ex]{0.5pt} {} {\bf Phase 3: Payout}  \xrfill[0.5ex]{0.5pt}
					
					\begin{itemize}
						
						\item[$\bullet$] {\color{blue}On receive} $(\mathsf{claimDelivery}, i, \sigma^{i}_{\mathcal{C}\mathcal{D}})$ from $\mathcal{D}$:
						\begin{itemize}
					
							\item[-] assert current time $\mathcal{T} < \mathcal{T}_{\mathsf{finish}}$
							
							\item[-] assert $i\equiv n$ or $\Sigma\equiv\mathsf{received}$ or $\Sigma\equiv\mathsf{payingRevealing}$

							\item[-] assert $\mathsf{ctr} \equiv 0$ and $0 < i \leq n$
							
							\item[-] assert $\mathsf{Verify}(\mathsf{receipt}||i||pk_{\mathcal{C}}||pk_{\mathcal{D}}, \sigma^{i}_{\mathcal{C}\mathcal{D}}, pk_{\mathcal{C}}) \equiv 1$

							\item[-] let $\mathsf{ctr}_{\mathcal{D}} := i$, $\Sigma := \mathsf{payingDelivery}$, and then send $(\mathsf{payingDelivery})$ to all entities
						\end{itemize}
						
						\item[$\bullet$] {\color{blue}On receive} $(\mathsf{claimRevealing}, i, \sigma^{i}_{\mathcal{C}\mathcal{P}})$ from $\mathcal{P}$:
						\begin{itemize}
			
							\item[-] assert current time $\mathcal{T} < \mathcal{T}_{\mathsf{finish}}$
							
							\item[-] assert $i \equiv n$ or $\Sigma\equiv\mathsf{received}$ or $\Sigma\equiv\mathsf{payingDelivery}$ 
							
							\item[-] assert $\mathsf{ctr} \equiv 0$ and $0 < i \leq n$
							
							\item[-] assert $\mathsf{Verify}(\mathsf{receipt}||i||pk_{\mathcal{C}}||pk_{\mathcal{P}}, \sigma^{i}_{\mathcal{C}\mathcal{P}}, pk_{\mathcal{C}}) \equiv 1$
							
							\item[-] let $\mathsf{ctr}_{\mathcal{P}} := i$, $\Sigma := \mathsf{payingRevealing}$, and then send $(\mathsf{payingRevealing})$ to all entities
			
						\end{itemize}
						
						\item[$\bullet$] {\color{blue}Upon} $\mathcal{T}_{\mathsf{finish}}$ times out:
						\begin{itemize}
							\item[-] assert current time $\mathcal{T} \geq \mathcal{T}_{\mathsf{finish}}$
							
							\item[-] let $\mathsf{ctr} := \max\{\mathsf{ctr}_{\mathcal{D}}, \mathsf{ctr}_\mathcal{P}\}$
							
							\item[-] let $\mathsf{ledger}[\mathcal{D}]:=\mathsf{ledger}[\mathcal{D}]+\mathsf{ctr}\cdot\bitcoinA_{\mathcal{P}}$
							
							\item[-] if $\mathsf{plt}$: \\
							{\ } let $\mathsf{ledger}[\mathcal{P}]:= \mathsf{ledger}[\mathcal{P}]+ (n-\mathsf{ctr})\cdot\bitcoinA_{\mathcal{P}} + \mathsf{ctr}\cdot\bitcoinA_{\mathcal{C}}$ \\
							
							{\ } let $\mathsf{ledger}[\mathcal{C}]:=\mathsf{ledger}[\mathcal{C}]+ (n-\mathsf{ctr})\cdot\bitcoinA_{\mathcal{C}} + \bitcoinA_{\mathsf{pf}}$
							
							\item[-] else: \\
							
							{\ } let $\mathsf{ledger}[\mathcal{P}]:=\mathsf{ledger}[\mathcal{P}]+ (n-\mathsf{ctr})\cdot\bitcoinA_{\mathcal{P}} + \mathsf{ctr}\cdot\bitcoinA_{\mathcal{C}} + \bitcoinA_{\mathsf{pf}}$
							
							{\ } let $\mathsf{ledger}[\mathcal{C}]:=\mathsf{ledger}[\mathcal{C}]+ (n-\mathsf{ctr})\cdot\bitcoinA_{\mathcal{C}}$
							
							\item[-] if $\mathsf{ctr} > 0$: \\ {\ } let $\Sigma := \mathsf{sold}$ and send $(\mathsf{sold})$ to all entities
							\item[-] else let $\Sigma := \mathsf{not\_sold}$ and send $(\mathsf{not\_sold})$ to all entities
							
						\end{itemize}
						
						{\ } \\
						{\color{violet}$\triangleright$ Reset to the ready state for repeatable delivery}
						\item[$\bullet$] {\color{blue}On receive} $(\mathsf{reset})$ from $\mathcal{P}$:
						\begin{itemize}[-]
							\item assert $\Sigma\equiv \mathsf{sold}$ or $\Sigma\equiv \mathsf{not\_sold}$
							\item set $\mathsf{ctr}$, $\mathsf{ctr}_{\mathcal{D}}$, $\mathsf{ctr}_{\mathcal{P}}$, $\mathcal{T}_{\mathsf{receive}}$, $\mathcal{T}_{\mathsf{finish}}$ as 0
							\item nullify $pk_{\mathcal{C}}$
							\item let $\theta := \theta - 1$ and $\Sigma := \mathsf{ready}$ 
							\item send $(\mathsf{ready})$ to all entities
						\end{itemize}
						
					\end{itemize}
					
				\end{flushleft}
			\end{multicols}
			\vspace{-4mm}
		}
	}
    
	\caption{The streaming-setting arbiter functionality $\mathcal{G}_{s}^{\mathsf{ledger}}$. ``Sending to all entities" captures that the smart contract is transparent to the public.}\label{fig:streaming_contract_ideal_functionality}
    \vspace{-4mm}
\end{figure*}

\subsection{$\Pi_{\mathsf{FS}}$: \systems{} Protocol}
\label{sec:ft_streaming_protocol_details}

\smallskip
\noindent
{\bf Phase I for Prepare.} This phase executes the same as the {\em Prepare} phase in the $\Pi_{\mathsf{FD}}$ protocol.

\smallskip
\noindent
{\bf Phase II for Stream.} The consumer $\mathcal{C}$, the deliverer $\mathcal{D}$ and the provider $\mathcal{P}$ interact with the contract $\mathcal{G}_s$ in this phase as:

\begin{itemize}
	\item The consumer $\mathcal{C}$ interested in the content with digest $\mathsf{root}_m$ would initialize a variable $x := 1$ and then:
	\begin{itemize}
		\item Asserts $\Sigma \equiv \mathsf{ready}$, runs $(pk_{\mathcal{C}},sk_{\mathcal{C}}) \leftarrow \mathsf{SIG.KGen}(1^{\lambda})$, and sends $(\mathsf{consume}, pk_{\mathcal{C}})$ to $\mathcal{G}_{s}$;
		\item Upon receiving the message $(\mathsf{mtree}, \mathsf{MT}, \sigma^{\mathsf{MT}}_{\mathcal{P}})$ from $\mathcal{P}$, asserts
		$\mathsf{Verify}(\mathsf{MT}, \sigma^{\mathsf{MT}}_{\mathcal{P}},pk_{\mathcal{P}})\equiv1 \wedge \mathsf{root}(\mathsf{MT})\equiv\mathsf{root}_m$, and stores the Merkle tree $\mathsf{MT}$, or else halts; 
		
		\item Upon receiving the message $(\mathsf{deliver}, i, c_i, \sigma_{c_i})$ from $\mathcal{D}$, checks whether $i \equiv x \wedge \mathsf{Verify}(i||c_i, \sigma_{c_i}, pk_{\mathcal{P}}) \equiv 1$, if hold, starts (for $i\equiv 1$) a timer $\mathcal{T}_{\mathsf{keyResponse}}$ or resets (for $1 <i\leq n$) it, sends $(\mathsf{keyReq}, i, \sigma^{i}_{\mathcal{C}})$ where $\sigma^{i}_{\mathcal{C}} \leftarrow \mathsf{Sign}(i||pk_{\mathcal{C}}, sk_{\mathcal{C}})$ to $\mathcal{P}$ (i.e., the step (2) in Fig.~\ref{fig:streaming_round}). If failing to check or $\mathcal{T}_{\mathsf{keyResponse}}$ times out, halts;
		
		\item Upon receiving the message $(\mathsf{reveal}, i, k_i, \sigma_{k_i})$ from $\mathcal{P}$ before $\mathcal{T}_{\mathsf{keyResponse}}$ times out, checks whether $i \equiv x \wedge \mathsf{Verify}(i||k_i, \sigma_{k_i}, pk_{\mathcal{P}}) \equiv 1$, if failed, halts. Otherwise, starts to validate the content chunk based on received $c_i$ and $k_i$: decrypts $c_i$ to obtain $m_i'$, where $m_i'=\mathsf{SDec}_{k_i}(c_i)$, and then checks whether $\mathcal{H}(m_i')$ is consistent with the $i$-th leaf node in the Merkle tree $\mathsf{MT}$, if inconsistent, sends $(\mathsf{PoM}, i, c_i, \sigma_{c_i}, k_i, \sigma_{k_i}, \mathcal{H}(m_i), \pi^{i}_{\mathsf{MT}})$ to $\mathcal{G}_{s}$. If it is consistent, sends the receipts $(\mathsf{receipt}, i, \sigma^{i}_{\mathcal{C}\mathcal{D}})$ to $\mathcal{D}$ and $(\mathsf{receipt}, i, \sigma^{i}_{\mathcal{C}\mathcal{P}})$ to $\mathcal{P}$, where $\sigma^{i}_{\mathcal{C}\mathcal{D}} \leftarrow \mathsf{Sign}(\mathsf{receipt}||i||pk_{\mathcal{C}}||pk_{\mathcal{D}}, sk_{\mathcal{C}})$ and $\sigma^{i}_{\mathcal{C}\mathcal{P}} \leftarrow \mathsf{Sign}(\mathsf{receipt}||i||pk_{\mathcal{C}}||pk_{\mathcal{P}}, sk_{\mathcal{C}})$, and sets $x := x + 1$, and then waits for the next $(\mathsf{deliver})$ message from $\mathcal{D}$. Upon $x$ is set to be $n+1$, sends $(\mathsf{received})$ to $\mathcal{G}_s$;  
		\item Waits for the messages $(\mathsf{received})$ from $\mathcal{G}_s$ to halt.
	\end{itemize}
	
	\item The deliverer $\mathcal{D}$ initializes a variable $y := 1$ and executes as follows in this phase:
	\begin{itemize}
		\item Upon receiving $(\mathsf{initiated}, pk_{\mathcal{C}})$ from $\mathcal{G}_s$, sends the message $(\mathsf{deliver}, i, c_i, \sigma_{c_i}), i = 1$ to $\mathcal{C}$ and starts a timer $\mathcal{T}_{\mathsf{chunkReceipt}}$;
		\item Upon receiving the message $(\mathsf{receipt}, i, \sigma^{i}_{\mathcal{C}\mathcal{D}})$ from $\mathcal{C}$ before $\mathcal{T}_{\mathsf{chunkReceipt}}$ times out, checks whether $\mathsf{Verify}(\mathsf{receipt}||i||pk_{\mathcal{C}}||pk_\mathcal{D}, \sigma^{i}_{\mathcal{C}\mathcal{D}}, pk_{\mathcal{C}}) \equiv 1 \wedge i \equiv y$ or not, if succeed, continues with the next iteration: sets $y := y+1$, sends $(\mathsf{deliver}, i, c_i, \sigma_{c_i}), i = y$ to $\mathcal{C}$, and resets  $\mathcal{T}_{\mathsf{chunkReceipt}}$ (i.e., the step (1) in Fig.~\ref{fig:streaming_round}); otherwise $\mathcal{T}_{\mathsf{chunkReceipt}}$ times out, enters the next phase.
		
	\end{itemize}
	
	\item The provider $\mathcal{P}$ initializes a variable $z:=1$ and executes as follows in this phase:
	\begin{itemize}
		\item Upon receiving $(\mathsf{initiated}, pk_{\mathcal{C}})$ from $\mathcal{G}_{s}$: asserts $\Sigma \equiv \mathsf{initiated}$,  and sends  $(\mathsf{mtree}, \mathsf{MT}, \sigma^{\mathsf{MT}}_{\mathcal{P}})$ to $\mathcal{C}$;
		
		\item Upon receiving $(\mathsf{keyReq}, i, \sigma^{i}_{\mathcal{C}})$ from $\mathcal{C}$, checks whether $i \equiv z \wedge \mathsf{Verify}(i||pk_{\mathcal{C}}, \sigma^{i}_{\mathcal{C}}, pk_{\mathcal{C}}) \equiv 1$, if succeed, sends $(\mathsf{reveal}, i, k_i, \sigma_{k_i})$, where $\sigma_{k_i} \leftarrow \mathsf{Sign}(i||k_i, sk_{\mathcal{P}})$, to $\mathcal{C}$ and starts (for $i \equiv 1$) a timer $\mathcal{T}_{\mathsf{keyReceipt}}$ or resets (for $1 < i \leq n$) it (i.e., the step (3) in Fig.~\ref{fig:streaming_round}), otherwise enters the next phase;
		
		\item On input  $(\mathsf{receipt}, i, \sigma^{i}_{\mathcal{C}\mathcal{P}})$ from $\mathcal{C}$ before $\mathcal{T}_{\mathsf{keyReceipt}}$ expires, checks $\mathsf{Verify}(\mathsf{receipt}||i||pk_{\mathcal{C}}||pk_{\mathcal{P}}, \sigma^{i}_{\mathcal{C}\mathcal{P}}, pk_{\mathcal{C}}) \equiv 1$  $\wedge$   $i\equiv z$ or not, if succeed, sets $z = z+1$. Otherwise $\mathcal{T}_{\mathsf{keyReceipt}}$ times out, enters the next phase.
	\end{itemize}
\end{itemize}

\smallskip
\noindent
{\bf Phase III for Payout.} The provider $\mathcal{P}$ and the deliverer $\mathcal{D}$ interact with the contract $\mathcal{G}_s$ in this phase as:

\begin{itemize}
	\item The provider $\mathcal{P}$ executes as follows in this phase:
	
	\begin{itemize}
		\item Upon receiving $(\mathsf{received})$ or $(\mathsf{delivered})$ from $\mathcal{G}_s$, or receiving the $n$-th $\mathsf{receipt}$ from $\mathcal{C}$ (i.e., $z$ is set to be $n+1$), sends $(\mathsf{claimRevealing}, i, \sigma^{i}_{\mathcal{C}\mathcal{P}})$ to $\mathcal{G}_s$;
		\item Waits for $(\mathsf{revealed})$ from $\mathcal{G}_s$ to halt.
	\end{itemize}
	
	\item The deliverer $\mathcal{D}$ executes as follows during this phase:
	\begin{itemize}
		\item Upon receiving $(\mathsf{received})$ or $(\mathsf{revealed})$ from $\mathcal{G}_s$, or receiving the $n$-th $\mathsf{receipt}$ from $\mathcal{C}$ (i.e., $y$ is set to be $n+1$), sends $(\mathsf{claimDelivery}, i, \sigma^{i}_{\mathcal{C}\mathcal{D}})$ to $\mathcal{G}_s$;
		
		\item Waits for  $(\mathsf{delivered})$ from $\mathcal{G}_{s}$ to halt.
	\end{itemize}
	
\end{itemize}

\subsection{Analyzing $\mathsf{FairStream}$ Protocol} 
\label{sec:FS_analysis}

\noindent
\begin{lemma}
	\label{lemma:streaming_completeness}
	Conditioned that all parties $\mathcal{P}$, $\mathcal{D}$ and $\mathcal{C}$ are honest, $\Pi_{\mathsf{FS}}$ satisfies the completeness property in the synchronous authenticated network model and stand-alone setting.
\end{lemma}

\noindent{\em Proof.}
If all parties $\mathcal{P}$, $\mathcal{D}$ and $\mathcal{C}$ are honest to follow the protocol, the completeness is obvious to see: the provider $\mathcal{P}$ receives a net income of $n\cdot(\bitcoinA_{\mathcal{C}}-\bitcoinA_{\mathcal{P}})$; the deliverer $\mathcal{D}$ obtains the payment of $n\cdot\bitcoinA_\mathcal{P}$; the consumer $\mathcal{C}$ pays for $n\cdot\bitcoinA_{\mathcal{C}}$ and attains the valid content $m$ with $\phi(m)\equiv 1$.

\noindent
\begin{lemma}
	\label{lemma:streaming_fairness}
	In the synchronous authenticated network model and stand-alone setting, conditioned that the underlying cryptographic primitives are secure, $\Pi_{\mathsf{FS}}$ meets the fairness requirement even when at most two parties of $\mathcal{P}$, $\mathcal{D}$ and $\mathcal{C}$ are corrupted by non-adaptive P.P.T. adversary $\adv$.
\end{lemma}

\noindent{\em Proof.}
The fairness for each party can be reduced to the underlying cryptographic building blocks. Specifically,
\begin{itemize}
	\item \underline{\smash{\em Consumer Fairness.}} The consumer fairness means that the honest $\mathcal{C}$ needs to pay proportional to what it {\em de facto} receives even though malicious $\mathcal{P}^*$ and $\mathcal{D}^*$ may collude with each other. This case can be modeled as a non-adaptive P.P.T. adversary $\adv$ corrupts $\mathcal{P}$ and $\mathcal{D}$ to provide and deliver the content to $\mathcal{C}$. During the {\em Stream} phase, $\mathcal{C}$ can stop sending back the receipts any time when an invalid chunk is received and then raise complaint to the contract to get compensation. Considering that $\mathcal{C}$ receives a sequence of $(c_1, \sigma_{c_1}), \cdots, (c_\ell, \sigma_{c_\ell}), \ell \in [n]$ though $\mathcal{A}$ may abort maliciously. Then it is ensured that $\adv$ can {\em at most} get $\ell$ receipts and claim payment of $\ell\cdot \bitcoinA_{\mathcal{P}}$ and $\ell\cdot\bitcoinA_{\mathcal{C}}$, where the former is paid by $\mathcal{A}$ itself due to collusion. Overall, $\mathcal{C}$ either pays $\ell\cdot\bitcoinA_{\mathcal{C}}$ and obtains $\ell$ valid chunks or pays nothing. To violate the fairness for $\mathcal{C}$, $\mathcal{A}$ has to break the security of signature scheme, i.e., forge $\mathcal{C}$'s signature. The probability is negligible due to the EU-CMA property of the underlying signature scheme. Therefore, the consumer fairness being against the collusion of malicious $\mathcal{P}^*$ and $\mathcal{D}^*$ is ensured. Note that breaking the security of the Merkle tree (i.e., finding another chunk $m_i'\neq m_i$ in position $i$ of $m$ to bind to the same $\mathsf{root}_m$ so as to fool the contract to reject $\mathcal{C}$'s $\mathsf{PoM}$) or controlling the execution of smart contract in blockchain, which are of negligible probability due to the second-preimage resistance of hash function in Merkle tree and the fact that contract is modeled as an ideal functionality, can only repudiate the penalty fee $\bitcoinA_{\mathsf{pf}}$ and would not impact $\mathcal{C}$'s fairness in the streaming setting.
	
	\item \underline{\smash{\em Deliverer Fairness.}} The deliverer fairness states that the honest $\mathcal{D}$ receives the payment proportional to the contributed bandwidth even though the malicious $\mathcal{P}^*$ and $\mathcal{C}^*$ may collude with each other. This case can be modeled as the non-adaptive P.P.T. adversary $\mathcal{A}$ corrupts both $\mathcal{P}$ and $\mathcal{C}$ to reap $\mathcal{D}$'s bandwidth resource without paying. In the {\em Stream} phase, if the honest $\mathcal{D}$ delivers $\ell$ chunks, then it is guaranteed to obtain $\ell$ or $\ell-1$ (i.e., $\mathcal{A}$ does not respond with the $\ell$-th receipt) receipts. In the {\em Payout} phase, $\mathcal{A}$ cannot lower the payment for the honest $\mathcal{D}$ since $\mathcal{D}$ can send the $\ell$-th or $(\ell-1)$-th receipt to the contract, which would update the internal state $\mathsf{ctr}_{\mathcal{D}}$ as $\ell$ or $\ell - 1$. Once $\mathcal{T}_{\mathsf{finish}}$ times out, $\mathcal{D}$ can receive the well-deserved payment of $\ell\cdot\bitcoinA_{\mathcal{P}}$ or $(\ell-1)\cdot\bitcoinA_{\mathcal{P}}$ from the contract, and {\em at most} waste bandwidth for delivering one chunk of size $\eta$.  To violate the fairness for $\mathcal{D}$, $\mathcal{A}$ has to control the execution of smart contract to refuse $\mathcal{D}$'s request of claiming payment though the request is valid. The probability to control the contract functionality in blockchain is negligible, and therefore the deliverer fairness being secure against the collusion of malicious $\mathcal{P}^*$ and $\mathcal{C}^*$ is assured.   
	
	\item \underline{\smash{\em Provider Fairness.}} The provider fairness indicates that the honest $\mathcal{P}$ receives the payment proportional to the number of valid chunks that $\mathcal{C}$ receives. The malicious $\mathcal{D}^*$ and $\mathcal{C}^*$ may collude with each other or $\mathcal{D}^*$ can costlessly create multiple fake $\mathcal{C}^*$ (i.e., Sybil attack), and then cheat $\mathcal{P}$ without truly delivering the content. These cases can be modeled as a non-adaptive P.P.T. adversary $\adv$ corrupts both $\mathcal{D}$ and $\mathcal{C}$. There are two situations $\mathcal{P}$'s fairness would be violated: (i) $\mathcal{A}$ claims payment (paid by $\mathcal{P}$) without real delivery; (ii) $\mathcal{A}$ obtains content chunks without paying for $\mathcal{P}$. For case (i), $\mathcal{A}$ would try to maximize the payment paid by $\mathcal{P}$ by increasing the $\mathsf{ctr}_{\mathcal{D}}$ via the $(\mathsf{claimDelivery})$ message sent to the contract. However, the $\mathcal{G}_s$ would update the counter $\mathsf{ctr}$ as $\max\{\mathsf{ctr}_{\mathcal{D}},\mathsf{ctr}_{\mathcal{P}}\}$ in contract after $\mathcal{T}_{\mathsf{finish}}$ times out, and the intention that $\mathcal{A}$ tries to maximize $\mathsf{ctr}_{\mathcal{D}}$ would correspondingly maximize $\mathsf{ctr}$. Considering that $\mathcal{A}$ wants to claim the payment of $\ell\cdot\bitcoinA_{\mathcal{P}}, \ell\in [n]$ by letting the $(\mathsf{claimDelivery})$ message contain the index of $\ell$ while no content is actually delivered, essentially the honest $\mathcal{P}$ can correspondingly receive the payment of $\ell\cdot\bitcoinA_{\mathcal{C}}$, and therefore a well-deserved net income of $\ell\cdot(\bitcoinA_{\mathcal{C}}-\bitcoinA_{\mathcal{P}})$, unless $\mathcal{A}$ can manipulate the execution of smart contract. For case~(ii), on one hand, each content chunk is encrypted before receiving the corresponding chunk key from $\mathcal{P}$. Hence, $\mathcal{A}$ has to violate the semantic security of the underlying symmetric encryption scheme to break the provider fairness, which is of negligible probability. On the other hand, during the streaming procedure,  $\mathcal{P}$ can always stop revealing the chunk key to $\mathcal{A}$ if no valid receipt for the previous chunk key is responded in time. {\em At most} $\mathcal{P}$ would lose one content chunk of size $\eta$ and receive well-deserved payment using the latest receipt. To violate the fairness, $\mathcal{A}$ again has to control the execution of smart contract, which is of negligible probability, to deny the payment for $\mathcal{P}$ though the submitted receipt is valid. Therefore, the provider fairness against the collusion of malicious $\mathcal{D}^*$ and $\mathcal{C}^*$ is guaranteed. 
\end{itemize}

In sum, the fairness for $\mathcal{C}$ is strictly ensured in $\Pi_{\mathsf{FS}}$, while for $\mathcal{P}$ and $\mathcal{D}$, the unpaid revealed content for $\mathcal{P}$ and the unpaid bandwidth resource of delivery are bounded to $O(\eta)$ bits. i.e., $\Pi_{\mathsf{FS}}$ satisfies the defined fairness property. 

\noindent
\begin{lemma}
	\label{lemma:streaming_confidentiality}
	In the synchronous authenticated network and stand-alone model, 
	$\Pi_{\mathsf{FS}}$ satisfies the confidentiality property against malicious deliverer corrupted by non-adaptive P.P.T. adversary $\adv$.
\end{lemma}

\noindent{\em Proof.} The confidentiality indicates that the deliverer $\mathcal{D}$ cannot learn any useful information about the content $m$ besides a-priori known knowledge within a delivery session. It can be modeled as a non-adaptive P.P.T. adversary corrupts $\mathcal{D}$. In $\Pi_{\mathsf{FS}}$, the possible scripts of leaking information of $m$ include: (i) the encrypted content chunks delegated to $\mathcal{D}$; and (ii) the Merkle tree $\mathsf{MT}$ of the content $m$. To break the confidentiality property, $\mathcal{A}$ has to violate the pre-image resistance of cryptographic hash functions (for the encryption scheme and $\mathsf{MT}$), which is of negligible probability. Hence, the confidentiality property against the malicious deliverer can be ensured.  

\noindent
\begin{lemma}
	\label{lemma:streaming_timeliness}
	If at least one of the three parties $\mathcal{P}$, $\mathcal{D}$ and $\mathcal{C}$ is honest and others are corrupted by non-adaptive P.P.T. adversary $\adv$, $\Pi_{\mathsf{FS}}$ meets the timeliness property in the synchronous authenticated network and stand-alone model.
\end{lemma}

\noindent{\em Proof.}
The timeliness means that the honest parties in $\Pi_{\mathsf{FS}}$ can terminate in $O(n)$ synchronous rounds, where $n$ is the number of content chunks, and when the protocol completes or aborts, the fairness and confidentiality are always preserved. Similarly, we focus on the analysis of fairness since the guarantee of confidentiality can be straightforwardly derived in light of the lemma~\ref{lemma:streaming_confidentiality} even if malicious parties abort. We distinguish the following termination cases for $\Pi_{\mathsf{FS}}$ with the arbiter contract $\mathcal{G}_s$ and at least one honest party:

\noindent
\underline{\smash{\em No abort.}} If all of $\mathcal{P}$, $\mathcal{D}$ and $\mathcal{C}$ are honest, the protocol $\Pi_{\mathsf{FS}}$ terminates in the {\em Payout} phase, after $\mathcal{T}_{\mathsf{finish}}$ times out. Both the {\em Prepare} and {\em Payout} phases can be completed in $O(1)$ rounds, while the {\em Stream} phase needs $O(n)$ rounds, where $n$ is the number of content chunks, resulting in $O(n)$ rounds for the protocol $\Pi_{\mathsf{FS}}$ to terminate and the fairness for all parties at completion are ensured as they obtain the well-deserved items.

\noindent
\underline{\smash{\em Aborts in the Prepare phase.}} The analysis for this phase is the same as the $\Pi_{\mathsf{FD}}$ protocol in lemma~\ref{lemma:downloading_timeliness}.

\noindent
\underline{\smash{\em Aborts in the Stream phase.}} This phase involves the provider $\mathcal{P}$, the deliverer $\mathcal{D}$, the consumer $\mathcal{C}$ and the arbiter contract $\mathcal{G}_s$, and it would terminate in $O(n)$ rounds due to the following cases: (i) $\mathcal{C}$ receives all the chunks and sends the $(\mathsf{received})$ message to contract; (ii) any party aborts during the streaming, and then the timer $\mathcal{T}_{\mathsf{receive}}$ times out in contract; (iii) $\mathcal{C}$ successfully raises complaint of $\mathcal{P}$'s misbehavior. During streaming, if $\mathcal{D}$ aborts, for example, after receiving the $\ell$-th receipt for chunk delivery, then $\mathcal{C}$ is guaranteed to have received $\ell$ encrypted chunks at that time point. If $\mathcal{P}$ aborts, for example, after receiving the $\ell$-th receipt for key revealing, then $\mathcal{C}$ is assured to have received $\ell$ keys for decryption at that time point. If $\mathcal{C}$ aborts, in the worst case, after receiving the $\ell$-th encrypted chunk from $\mathcal{D}$ and the $\ell$-th key from $\mathcal{P}$, at that time point, $\mathcal{D}$ is ensured to have obtained $\ell - 1$ receipts for the bandwidth contribution, while $\mathcal{P}$ is guaranteed to have received $\ell - 1$ receipts for key revealing, which means the fairness for $\mathcal{D}$ and $\mathcal{P}$ is still preserved according to the fairness definition, i.e., the unpaid delivery resource for $\mathcal{D}$ and the unpaid content for $\mathcal{P}$ are bounded to one chunk of $O(\eta)$ bits.

\noindent
\underline{\smash{\em Aborts in the Payout phase.}}
This phase involves the provider $\mathcal{P}$, the deliverer $\mathcal{D}$ and the arbiter contract $\mathcal{G}_s$, and it can terminate in $O(1)$ rounds. The fairness for the honest one is not impacted no matter when the other party aborts since $\mathcal{P}$ and $\mathcal{D}$ are independently claim the payment from contract. After $\mathcal{T}_\mathsf{finish}$ times out,  the contract would automatically distribute the payment to all parties according to the internal state $\mathsf{ctr}$.

\noindent
\begin{lemma}
	\label{lemma:streaming_efficiency}
	In the synchronous authenticated network model and stand-alone setting, for any non-adaptive P.P.T. adversary $\adv$, $\Pi_{\mathsf{FS}}$ satisfies the efficiency requirement: the communication complexity is bounded to $O(n)$; the on-chain cost is bounded to $\widetilde{O}(1)$; the messages transferred by the provider $\mathcal{P}$ after the setup phase are bounded to $n\cdot\lambda$ bits, where $n$ is the number of chunks and $\lambda$ is a cryptographic parameter, and $n\cdot \lambda$ is much less than the content size $|m|$.
\end{lemma}

\noindent{\em Proof.}
The analysis of efficiency guarantee in $\Pi_{\mathsf{FS}}$ can be conducted in the following three perspectives:
\begin{itemize}
	\item \underline{\smash{\em Communication Complexity.}} The {\em Prepare} phase is the same as the downloading setting, and therefore the time complexity is $O(n)$. In the {\em Stream} phase, $\mathcal{P}$ sends the Merkle tree $\mathsf{MT}$ of $m$ and meanwhile $\mathcal{D}$ starts to deliver the delegated $n$ chunks to $\mathcal{C}$. If dispute happens during streaming, the complexity of sending $\mathsf{PoM}$ is $O(\log n)$. Overall the communication complexity of this phase is $O(n)$. In the {\em Payout} phase, the $(\mathsf{claimDelivery})$ and $(\mathsf{claimRevealing})$ messages sent by $\mathcal{P}$ and $\mathcal{D}$ to contract is in $O(1)$. Hence, the total communication complexity of $\Pi_{\mathsf{FS}}$ is $O(n)$.
	
	\item \underline{\smash{\em On-chain Costs.}} The {\em Prepare} phase yields on-chain costs of $O(1)$, which is same as the downloading setting. In the {\em Stream} phase, the on-chain cost of the $\mathsf{consume}$ function is $O(1)$ and the multiple rounds of content delivery (i.e., the streaming process) are executed off-chain. When dispute occurs during streaming, the on-chain cost is $O(\log n)$ (for verifying the Merkle proof), leading to a total on-chain costs of $O(\log n)$. In the {\em Payout} phase, the on-chain costs is $O(1)$ since $\mathcal{P}$ and $\mathcal{D}$ only need to submit the latest receipt consisting of one signature. Overall, the on-chain cost of $\Pi_{\mathsf{FS}}$ is $O(\log n)$, namely $\widetilde{O}(1)$.
	
	\item \underline{\smash{\em Message Volume for $\mathcal{P}$.}} Considering that the contract is deployed and the deliverer is ready to deliver. Every time when a new consumer joins in, a new delivery session starts. The messages that $\mathcal{P}$ needs to send include: (i) the Merkle tree $\mathsf{MT}$ of $m$ in the {\em Stream} phase is $O(\log n)$; (ii) the $n$ chunk keys revealed to $\mathcal{C}$ is $O(n)$. Note that the message volume decrease from $n$ chunks to $n$ keys (e.g., 32 KB for a chunk v.s. 256 bits for a chunk key); (iii) the $(\mathsf{claimRevealing})$ message for claiming payment, which is $O(1)$ since only the latest receipt containing one signature needs to be submitted to $\mathcal{G}_s$. Overall, the resulting message volume can be represented as $n\cdot \lambda$, where $\lambda$ is a small cryptographic parameter, which is much smaller than the content size $|m|$. 
\end{itemize}

\begin{theorem}
	\label{thm:stream}
	Conditioned that the underlying cryptographic primitives are secure, the protocol $\mathsf{FairStream}$ satisfies the completeness, fairness, confidentiality against deliverer, timeliness, and non-trivial efficiency properties in the synchronous authenticated network, $\mathcal{G}^{\mathsf{ledger}}_s$-hybrid and stand-alone model.
\end{theorem}

\noindent{\em Proof.} Lemmas~\ref{lemma:streaming_completeness},~\ref{lemma:streaming_fairness},~\ref{lemma:streaming_confidentiality},~\ref{lemma:streaming_timeliness}, and~\ref{lemma:streaming_efficiency} complete the proof.

\vspace{1mm}

Besides, we have the following corollary to characterize the latency relationship between $\mathsf{FairDownload}$ and $\mathsf{FairStream}$.

\noindent
\begin{corollary}
	\label{lemma:streaming_less_latency}
	In the synchronous authenticated setting without corruptions, the honest consumer $\mathcal{C}$ in $\Pi_{\mathsf{FS}}$ can: (i) retrieve the first chunk in $O(1)$ communication rounds once activating the Stream phase; (ii) retrieve every $(i+1)$-th content chunk in $O(1)$ communication rounds once the $i$-th content chunk has delivered. This yields less retrieval latency compared to that all chunks retrieved by the consumer in $\Pi_\mathsf{FD}$ delivers in $O(n)$ rounds after the Deliver phase is activated.
\end{corollary}

\noindent{\em Proof.} In $\Pi_\mathsf{FD}$, the honest consumer $\mathcal{C}$ is able to obtain the keys only after the completion of the verifiable fair delivery module to decrypt the received chunks, meaning that the latency of retrieving the raw content chunks is in $O(n)$ communication rounds. While for $\Pi_{\mathsf{FS}}$, in  each round of streaming, the honest $\mathcal{C}$ can obtain one encrypted chunk from the deliverer $\mathcal{D}$ as well as one decryption key from the provider $\mathcal{P}$, and consequently the retrieval latency, though entailing relatively more involvement of $\mathcal{P}$, is only in $O(1)$ communication rounds.

\vspace{1mm}

It is worth pointing out in $\mathsf{FairStream}$, $\mathcal{P}$ and $\mathcal{D}$ are only allowed to claim the payment {\em after} the $\mathsf{received}$ state in contract, which indicates that either $\mathcal{C}$ has received all the valid chunks or any party has aborted during the streaming procedure. Typically the number of delivered chunks $\mathsf{ctr}$ and therefore the payment amount to $\mathcal{D}$ (i.e., $\mathsf{ctr}\cdot\bitcoinA_{\mathcal{P}}$) and $\mathcal{P}$ (i.e., $\mathsf{ctr}\cdot\bitcoinA_{\mathcal{C}}$) would {\em not} be very small. If considering another strategy that allows $\mathcal{P}$ and $\mathcal{D}$ to claim the payment any time during the streaming, the payment amount may be small, e.g., in pennies. In that case, it is feasible to introduce the payment channels~\cite{Malavolta-et-al-2017-CCS,Dziembowski-et-al-2019-SP} to handle micropayments~\cite{Decker-et-al-2015-Springger} and improve efficiency. Such a strategy can be an interesting future extension.

{\bf Extension for delivering  from any specific chunk}.
The protocol $\mathsf{FairStream}$ (as well as $\mathsf{FairDownload}$) can be easily tuned to transfer the content from the middle instead of the beginning.
Specifically, for the downloading setting, one can simply let the content provider reveal the elements that are able to recover a {\em sub-tree} of the key derivation tree $\mathsf{KT}$ for decrypting the transferred chunks. The complaint of incorrect decryption key follows the same procedure in~\S\ref{sec:ProtocolDesign_Downloading}. For the streaming setting, it is more straightforward as each chunk ciphertext and its decryption key are uniquely identified by the index and can be obtained in $O(1)$ rounds by the consumer, who can immediately complain to contract in the presence of an incorrect decryption result.


\section{Implementation and Evaluations}
\label{sec:ImplementationandEvaluation}
To shed some light on the feasibility of \systemd{} and \systems{}, we implement, deploy and evaluate them in the {\em Ethereum Ropsten} network. The arbiter contract is implemented in Solidity and split into \textit{Optimistic} and \textit{Pessimistic} modules, where the former is executed when no dispute occurs while the later is additionally called if dispute happens. Note that the contracts are only deployed once and may be used for multiple times to facilitate many deliveries, which amortizes the cost of deployment.

\smallskip
\noindent
{\bf Cryptographic instantiations.} The hash function is  \textit{keccak256}  and the digital signature  is via ECDSA over secp256k1 curve. The  encryption of each  chunk $m_i$ with key $k_i$ is instantiated as: parse $m_i$ into $t$ 32-byte blocks $(m_{i,1}, \dots, m_{i,t})$ and output   $c_i = (m_{i,1}  \oplus \mathcal{H}(k_i||1), \dots, m_{i,t}  \oplus \mathcal{H}(k_i||t))$.  {The decryption is same to the encryption.} 
We construct public key encryption scheme based on ElGamal: Let $\mathcal{G}=\langle{g}\rangle$ to be $G_1$ group over {\em alt-bn128} curve \cite{EIP-196} of prime order $q$, where $g$ is group generator; The private key $k \stackrel{R}{\longleftarrow} \mathbb{Z}_{q}$, the public key $h = g^k$, the encryption $\mathsf{VEnc}_{h}(m)=(c_1,c_2)=(g^r,m\cdot g^{kr})$ where $r\stackrel{R}{\longleftarrow} \mathbb{Z}_{q}$ and $m$ is encoded into $\mathcal{G}$  with Koblitz's method~\cite{Koblitz-1987-Mathematics}, and the decryption $\mathsf{VDec}_{k}((c_1,c_2))= c_2/c_1^{k}$. To augment ElGamal for verifiable decryption, we adopt Schnorr protocol \cite{Schnorr-1989-CTAC} for Diffie-Hellman tuples with using Fiat-Shamir transform \cite{Fiat-Shamir-1986-Crypto} in the random oracle model. Specifically, $\mathsf{ProvePKE}_{k}((c_1,c_2))$ is as: run $\mathsf{VDec}_{k}((c_1,c_2))$ to obtain $m$. Let $x \stackrel{R}{\longleftarrow} \mathbb{Z}_{q}$, and compute $A=g^x$, $B=c_1^x$, $C=\mathcal{H}(g||A||B||h||c_1||c_2||m)$, $Z=x+kC$, $\pi=(A,B,Z)$, and output $(m,\pi)$; $\mathsf{VerifyPKE}_{h}((c_1,c_2), m,\pi)$ is as: parse $\pi$ to obtain $(A,B,Z)$, compute $C'=\mathcal{H}(g||A||B||h||c_1||c_2||m)$, and verify $(g^Z \equiv A\cdot h^{C'}) \wedge (m^{C'}\cdot c_1^{Z} \equiv B\cdot c_2^{C'})$, and output $1/0$ indicating the verification succeeds or fails.

\smallskip
\noindent
\subsection{Evaluating \systemd{}}

Table~\ref{tab:gas_costs} presents the on-chain costs for all functions in $\Pi_{\mathsf{FD}}$. For the recent  violent fluctuation of Ether price, we adopt a gas price at 10 Gwei to ensure over half of the mining power in Ethereum would   mine this transaction\footnote{https://ethgasstation.info/.}, and an exchange rate of 259.4 USD per Ether, which is the average market price of Ether between Jan./1st/2020 and Nov./3rd/2020 from coindesk\footnote{https://www.coindesk.com/price/ethereum/.}. We stress that utilizing other cryptocurrencies such as Ethereum classic\footnote{https://ethereumclassic.org/.} can much further decrease the price for execution. The price also applies to the streaming setting.

\begin{figure}[!htpb]
\vspace{-4mm}
	\centering
	\subfloat[Costs for various chunk size \label{fig:chunk_size_cost}]{\includegraphics[width=.48\linewidth]{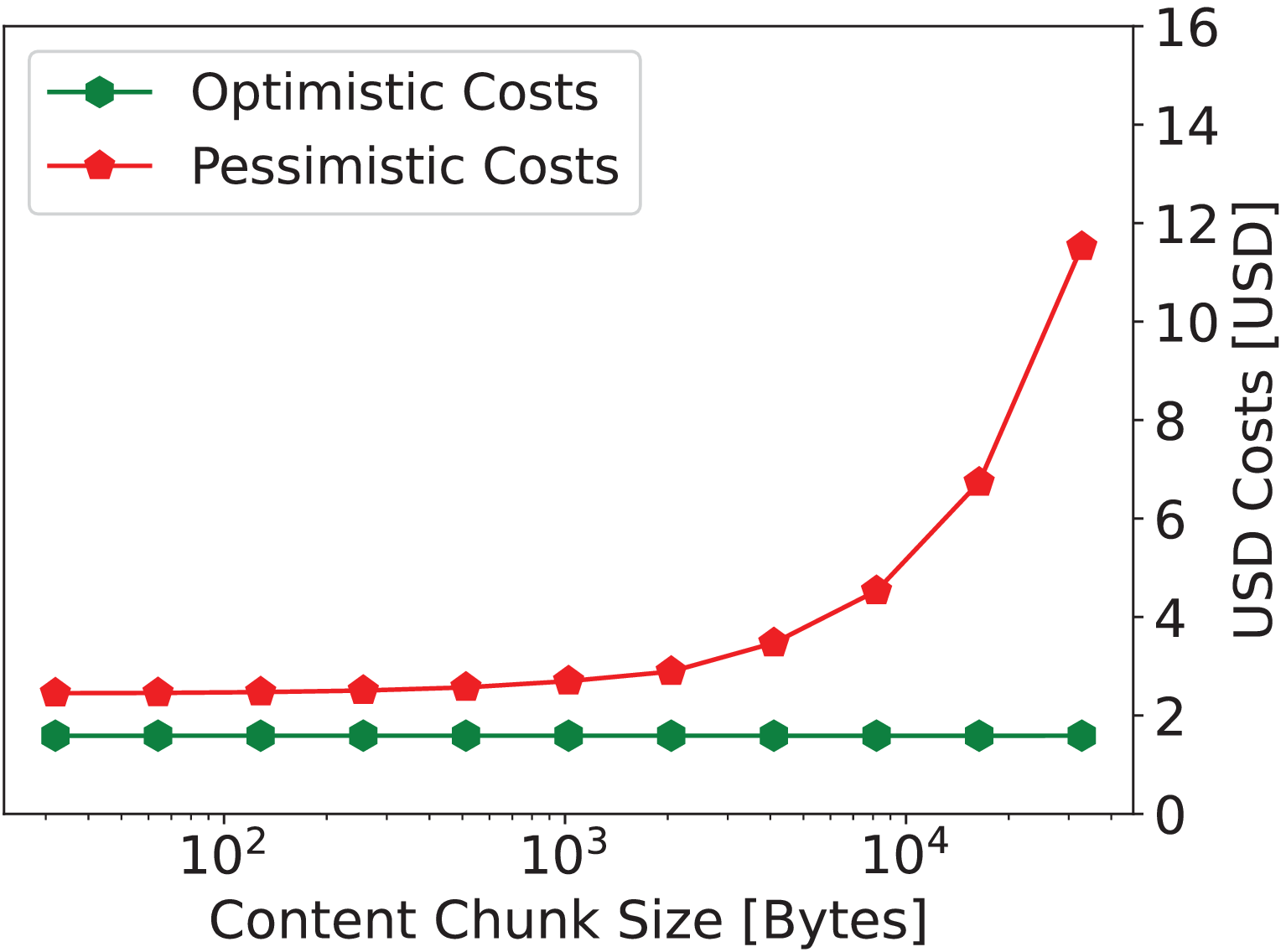}}
	\hspace{1mm}
	\subfloat[Costs for $erk$ revealing cost \label{fig:reveal_cost}]{\includegraphics[width=.48\linewidth]{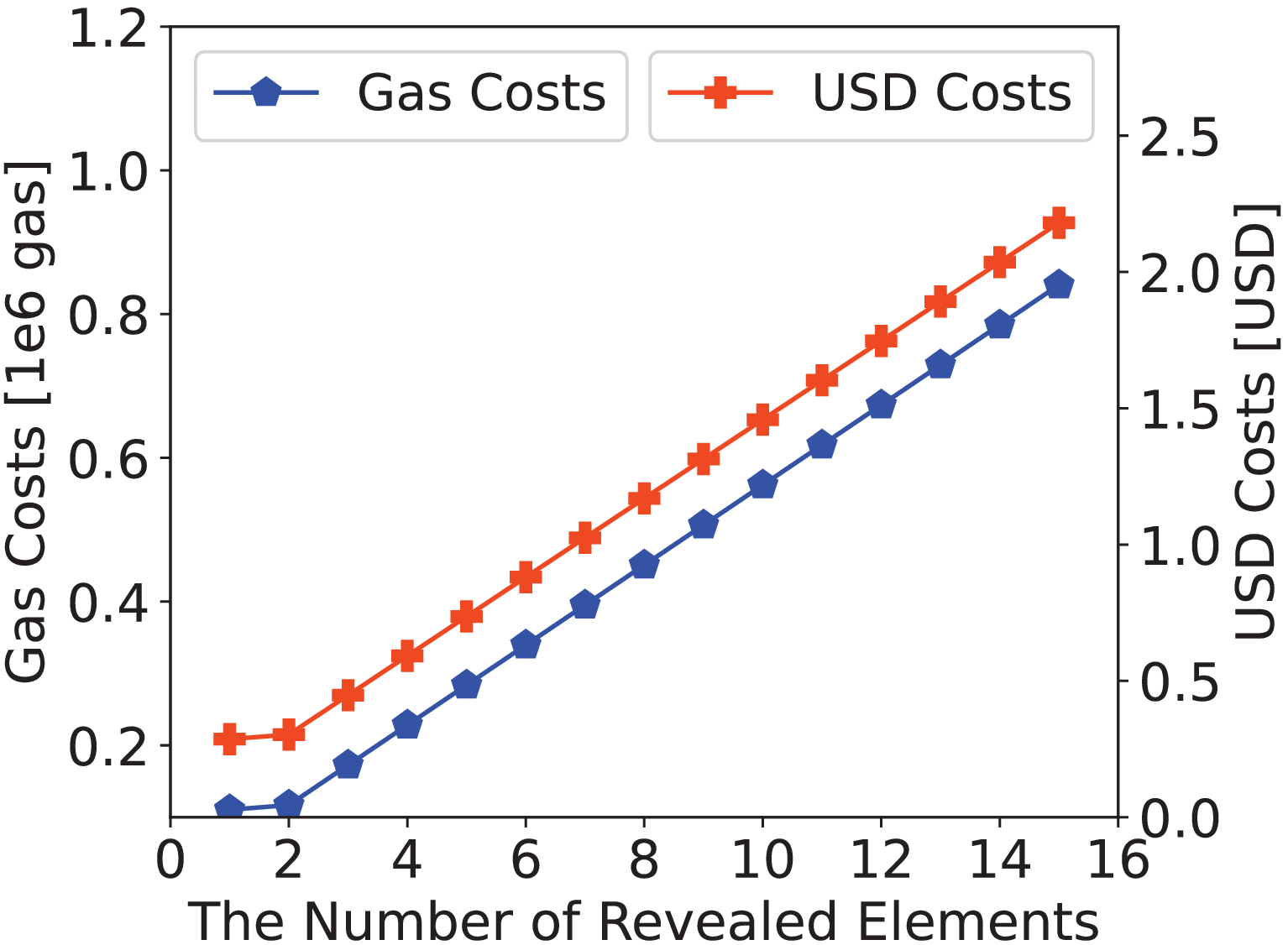}}
	\vspace{-1mm}
	\caption{Experiment results for the $\mathsf{FairDownload}$ protocol (averaged over 5 independent runs).}
	\vspace{-2mm}
\end{figure}

\smallskip
\noindent
{\bf Cost of optimistic case.} Without complaint the protocol $\Pi_{\mathsf{FD}}$ only executes the functions in {\em Deliver} and {\em Reveal} phases when a new consumer joins in, yielding the total cost of 1.032 USD for all involved parties except the one-time cost for deployment and the {\em Prepare} phase. Typically, such an on-chain cost is {\em constant} no matter how large the content size or the chunk size are, as illustrated in Figure~\ref{fig:chunk_size_cost}. In a worse case, up to $\log n$ elements in Merkle tree need to be revealed. In that case, Figure~\ref{fig:reveal_cost} depicts the relationship between the number of revealed elements and the corresponding costs.

\begin{table}[!t]
	\centering
	\begin{scriptsize}
		\caption{The on-chain costs of all functions in $\mathsf{FairDownload}$}
		\vspace{-2mm}
		\centering
		\label{tab:gas_costs}
		\setlength{\tabcolsep}{0.5em}
		{\renewcommand{\arraystretch}{1.2}
		\begin{tabular}{ c | c | c | c | c }
			\hline
			\hline
			{\em Phase} & {\em Function}& {\em Caller} & {\em Gas Costs} & {USD Costs} \\
			\hline
			Deploy & (Optimistic) &  $\mathcal{P}$  & 2 936 458 & 7.617 \\
			\hline
			Deploy & (Pessimistic) &  $\mathcal{P}$  & 2 910 652 & 7.550 \\
			\hline
			\hline
			\multirow{3}{*}{Prepare}
			& \multicolumn{1}{c|}{$\mathsf{start}$} 
			& \multicolumn{1}{c|}{$\mathcal{P}$} 
			& \multicolumn{1}{c|}{ 110 751} 
			& \multicolumn{1}{c}{0.287} \\\cline{2-5}
			& 
			\multicolumn{1}{c|}{$\mathsf{join}$} 
			& \multicolumn{1}{c|}{$\mathcal{D}$} 
			& \multicolumn{1}{c|}{ 69 031} 
			& \multicolumn{1}{c}{0.179}  \\\cline{2-5}
			& 
			\multicolumn{1}{c|}{$\mathsf{prepared}$} & \multicolumn{1}{c|}{$\mathcal{D}$} & \multicolumn{1}{c|}{ 34 867} & \multicolumn{1}{c}{0.090} \\\cline{2-5}
			\hline
			\multirow{3}{*}{Deliver}
			&
			\multicolumn{1}{c|}{$\mathsf{consume}$} & \multicolumn{1}{c|}{$\mathcal{C}$} & \multicolumn{1}{c|}{117 357} & \multicolumn{1}{c}{0.304} \\\cline{2-5}
			&
			\multicolumn{1}{c|}{$\mathsf{delivered}$} & \multicolumn{1}{c|}{$\mathcal{C}$} & \multicolumn{1}{c|}{57 935} & \multicolumn{1}{c}{0.150}  \\\cline{2-5}
			& 
			\multicolumn{1}{c|}{$\mathsf{verifyVFDProof}$} & \multicolumn{1}{c|}{$\mathcal{D}$} & \multicolumn{1}{c|}{56 225} & \multicolumn{1}{c}{0.146}  \\
			\hline
			\multirow{2}{*}{Reveal}
			& \multicolumn{1}{c|}{$\mathsf{revealKeys}$} & \multicolumn{1}{c|}{$\mathcal{P}$} & \multicolumn{1}{c|}{113 041} & \multicolumn{1}{c}{0.293} \\\cline{2-5}
			& 
			\multicolumn{1}{c|}{$\mathsf{payout}$} & \multicolumn{1}{c|}{$\mathcal{G}_{d}$} & \multicolumn{1}{c|}{53 822} & \multicolumn{1}{c}{0.139}  \\\cline{2-5}
			\hline
			\hline
			\multirow{2}{*}{\makecell[c]{Dispute Resolution}}
			& \multicolumn{1}{c|}{$\mathsf{wrongRK}$} & \multicolumn{1}{c|}{$\mathcal{C}$} & \multicolumn{1}{c|}{23 441} & \multicolumn{1}{c}{0.061} \\\cline{2-5}
			& 
			\multicolumn{1}{c|}{$\mathsf{PoM}$} & \multicolumn{1}{c|}{$\mathcal{C}$} & \multicolumn{1}{c|}{389 050} & \multicolumn{1}{c}{1.009}  \\
			\hline
			\hline
		\end{tabular}
		}
	\end{scriptsize}
	\vspace{-4mm}
\end{table}

\smallskip
\noindent
{\bf Cost of pessimistic case.} When complaint arises, the arbiter contract involves to resolve dispute. The cost of executing $\mathsf{wrongRK}$ function relates to the concrete values of $n$, $\mathsf{ctr}$ and $|erk|$, and in Table~\ref{tab:gas_costs}, the cost is evaluated on $n\equiv \mathsf{ctr}\equiv 512$, and $|erk|\equiv 1$. The cost of $\mathsf{PoM}$ function validating misbehavior varies by the content chunk size $\eta$, as depicted in Figure~\ref{fig:chunk_size_cost} pessimistic costs. The results demonstrate that the on-chain costs increase linearly in the chunk size (mostly due to chunk decryption in contract).

\subsection{Evaluating \systems{}}

\begin{table}[!t]
	\centering
	\begin{scriptsize}
		\caption{The on-chain costs of all functions in $\mathsf{FairStream}$}
		\vspace{-2mm}
		\centering
		\label{tab:gas_costs_streaming}
		\setlength{\tabcolsep}{1.0em}
		{\renewcommand{\arraystretch}{1.2}
		\begin{tabular}{ c | c | c | c | c }
			\hline
			\hline
			{\em Phase} & {\em Function}& {\em Caller} & {\em Gas Costs} & {USD Costs} \\
			\hline
			Deploy & (Optimistic) &  $\mathcal{P}$  & 1 808 281 & 4.691 \\
			\hline
			Deploy & (Pessimistic) &  $\mathcal{P}$  & 1 023 414 & 2.655 \\
			\hline
			\hline
			\multirow{3}{*}{Prepare}
			& \multicolumn{1}{c|}{$\mathsf{start}$} 
			& \multicolumn{1}{c|}{$\mathcal{P}$} 
			& \multicolumn{1}{c|}{ 131 061} 
			& \multicolumn{1}{c}{0.340} \\\cline{2-5}
			& 
			\multicolumn{1}{c|}{$\mathsf{join}$} 
			& \multicolumn{1}{c|}{$\mathcal{D}$} 
			& \multicolumn{1}{c|}{ 54 131} 
			& \multicolumn{1}{c}{0.140}  \\\cline{2-5}
			& 
			\multicolumn{1}{c|}{$\mathsf{prepared}$} & \multicolumn{1}{c|}{$\mathcal{D}$} & \multicolumn{1}{c|}{ 34 935} & \multicolumn{1}{c}{0.091} \\\cline{2-5}
			\hline
			\multirow{4}{*}{Stream}
			&
			\multicolumn{1}{c|}{$\mathsf{consume}$} & \multicolumn{1}{c|}{$\mathcal{C}$} & \multicolumn{1}{c|}{95 779} & \multicolumn{1}{c}{0.248} \\\cline{2-5}
			&
			\multicolumn{1}{c|}{$\mathsf{received}$} & \multicolumn{1}{c|}{$\mathcal{C}$} & \multicolumn{1}{c|}{39 857} & \multicolumn{1}{c}{0.103}  \\\cline{2-5}
			& 
			\multicolumn{1}{c|}{$\mathsf{receiveTimeout}$} & \multicolumn{1}{c|}{$\mathcal{G}_s$} & \multicolumn{1}{c|}{39 839} & \multicolumn{1}{c}{0.103}  \\\cline{2-5}
			& 
			\multicolumn{1}{c|}{$\mathsf{PoM}$} & \multicolumn{1}{c|}{$\mathcal{C}$} & \multicolumn{1}{c|}{90 018} & \multicolumn{1}{c}{0.234}  \\
			\hline
			\multirow{3}{*}{Payout}
			&
			\multicolumn{1}{c|}{$\mathsf{claimDelivery}$} & \multicolumn{1}{c|}{$\mathcal{D}$} & \multicolumn{1}{c|}{67 910} & \multicolumn{1}{c}{0.176} \\\cline{2-5}
			&
			\multicolumn{1}{c|}{$\mathsf{claimRevealing}$} & \multicolumn{1}{c|}{$\mathcal{P}$} & \multicolumn{1}{c|}{67 909} & \multicolumn{1}{c}{0.176}  \\\cline{2-5}
			&
			\multicolumn{1}{c|}{$\mathsf{finishTimeout}$} & \multicolumn{1}{c|}{$\mathcal{G}_s$} & \multicolumn{1}{c|}{88 599} & \multicolumn{1}{c}{0.230}  \\
			\hline
			\hline
		\end{tabular}
		}
	\end{scriptsize}
	\vspace{-4mm}
\end{table}

\begin{figure*}[!htpb]
	\centering
	\subfloat[Bandwidths among entities in the testing experiment \label{fig:streaming_bandwidth}]{\includegraphics[width=.18\linewidth]{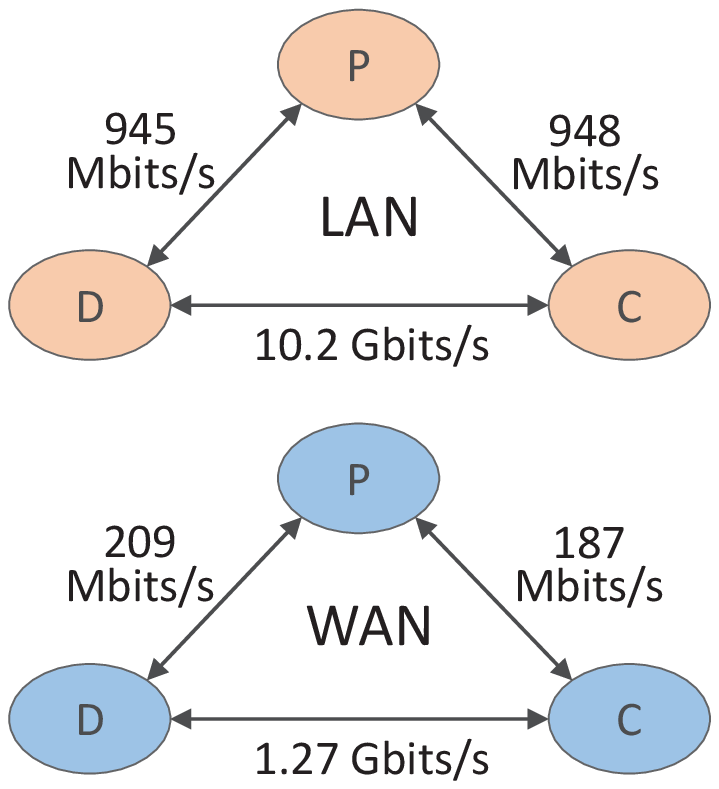}}
	\hspace{0.8mm}
	\subfloat[Time costs of streaming 512 content chunks in LAN \label{fig:streaming_LAN}]{\includegraphics[width=.26\linewidth]{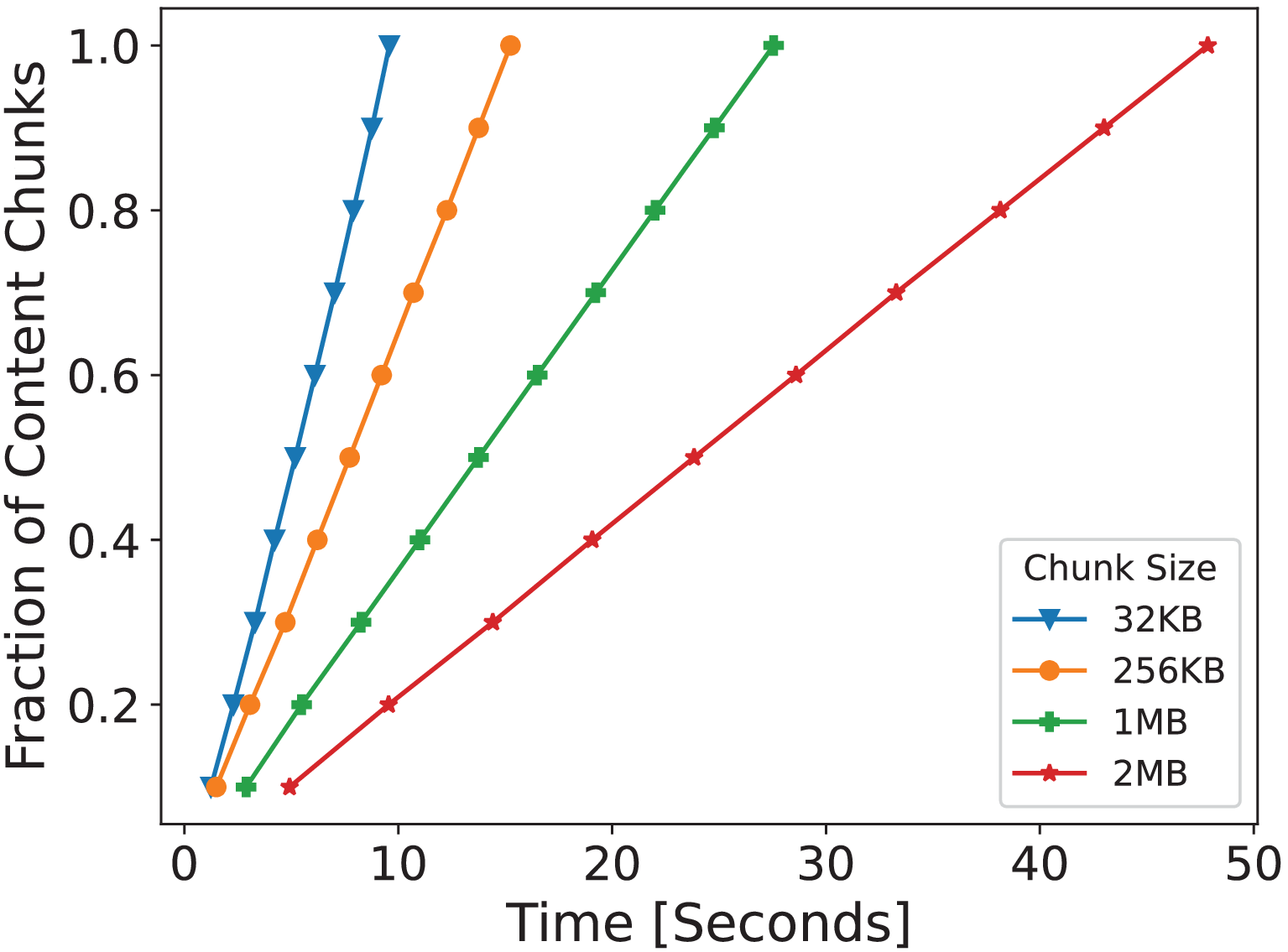}}
	\hspace{0.8mm}
	\subfloat[Time costs of streaming 512 content chunks in WAN \label{fig:streaming_WAN}]{\includegraphics[width=.26\linewidth]{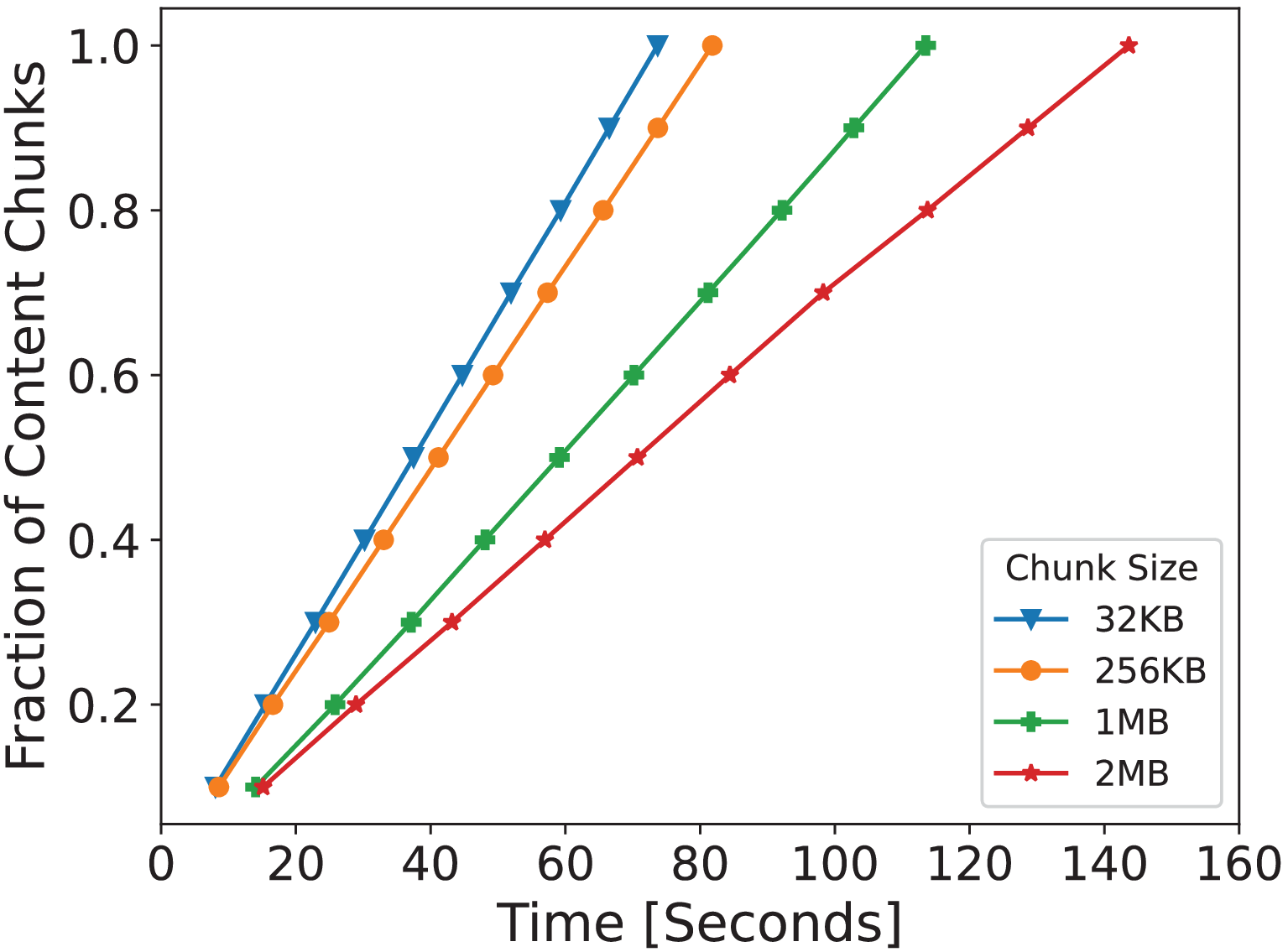}}
	\hspace{0.8mm}
	\subfloat[Average time costs and the corresponding bitrate for various chunk size \label{fig:streaming_result}]{\includegraphics[width=.26\linewidth]{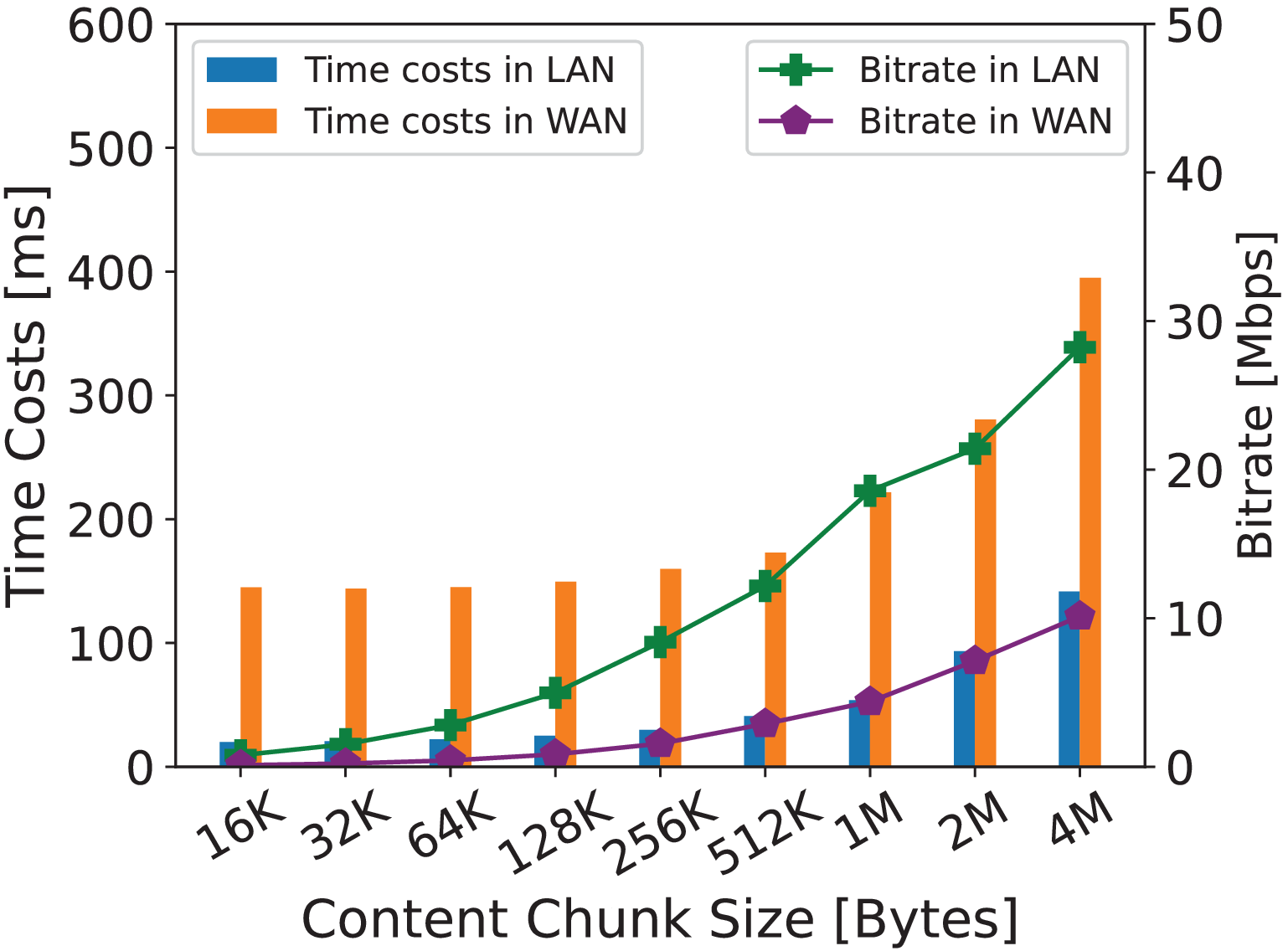}}
    \vspace{-2mm}
	\caption{The performance of $\mathsf{FairStream}$ protocol in the LAN and WAN testing environments (averaged over 5 independent runs).}
	\vspace{-2mm}
\end{figure*}

Table~\ref{tab:gas_costs_streaming} illustrates the on-chain costs of all functions in ${\mathsf{FairStream}}$. As the deployment of contract and the {\em Prepare} phase can be executed only {\em once}, we discuss the costs in both optimistic and pessimistic modes after a new consumer participates in, i.e., starting from the {\em Stream} phase. Specifically,

\smallskip
\noindent
{\bf Costs of optimistic case.} When no dispute occurs, the $\Pi_{\mathsf{FS}}$ protocol executes the functions in {\em Stream} and {\em Payout} phases except the $\mathsf{PoM}$ function for verifying proof of misbehavior, yielding a total cost of 0.933 USD for all involved parties. Note that only one of the $(\mathsf{received})$ and $(\mathsf{receiveTimeout})$ functions would be invoked. Meanwhile, the $(\mathsf{claimDelivery})$ and $(\mathsf{claimRevealing})$ functions may be called in different orders. The costs in the optimistic mode is {\em constant} regardless of the content size and chunk size.

\smallskip
\noindent
{\bf Costs of pessimistic case.} When complaint arises, the total on-chain cost is 1.167 USD for all involved parties during a delivery session. The cost of the $\mathsf{PoM}$ function: (i) increases slightly in number of chunks $n$, since it computes $O(\log n)$ hashes to verify the Merkle tree proof; (ii) increase linearly in the the content chunk size $\eta$ due to the chunk decryption in contract, which follows a similar trend to Fig.~\ref{fig:chunk_size_cost} pessimistic costs but with lower costs since no verification of verifiable decryption proof is needed.

\smallskip
\noindent
{\bf Streaming efficiency.} To demonstrate feasibility of using $\Pi_{\mathsf{FS}}$ for p2p streaming, we evaluate the efficiency for streaming 512 content chunks with various chunk size. Fig.~\ref{fig:streaming_bandwidth} shows the experimental bandwidth among parties in LAN (i.e., three VM instances on three servers residing on the same rack connected with different switches, where servers are all Dell PowerEdge R740 and each is equipped with 2 Intel(R) Xeon(R) CPU Silver 4114 processors, 256 GB (16 slots$\times$16 GB/slot) 2400MHz DDR4 RDIMM memory and 8 TB (8 slots$\times$1TB/slot) 2.5 inch SATA hard drive. Each VM on the servers has the same configuration of 8 vCPUs, 24 GB memory and 800 GB hard drive) and WAN (i.e., three {\em Google cloud} VM instances are initialized in {\em us-east4-c}, {\em us-east1-b} and {\em europe-north1-a}, respectively. Each VM is configured with 2 vCPUs, 4 GB memory and 10 GB hard drive). Considering that $\mathcal{P}$ owns information to choose the proper deliverer $\mathcal{D}$ to ensure better delivery quality (e.g., less delay from $\mathcal{D}$ to $\mathcal{C}$), the link between $\mathcal{D}$ and $\mathcal{C}$ is therefore evaluated in a higher bandwidth environment. Figure~\ref{fig:streaming_LAN} and~\ref{fig:streaming_WAN} illustrate the experiment results of consecutively streaming 512 content chunks in both LAN and WAN and the corresponding time costs. We can derive the following observations: (i) obviously the time costs increase due to the growth of chunk size; (ii) the delivery process remains stable with only slight fluctuation, as reflected by the slope for each chunk size in Figure~\ref{fig:streaming_LAN} and~\ref{fig:streaming_WAN}. Furthermore, Fig.~\ref{fig:streaming_result} depicts the average time costs for each chunk (over the 512 chunks) and the corresponding bitrate. The results show that the bitrate can reach 10 Mpbs even in the public network, which is potentially sufficient to support high-quality content streaming. E.g., the video bitrate for HD 720 and HD 1080 are {\em at most} 4 Mbps and 8 Mbps, respectively~\cite{Bitrate_Bandwidth_IBM-2020-WebPage}.

\ignore{
	\begin{table}[!htpb]
		\centering
		\begin{scriptsize}
			\caption{The time costs of streaming 512 chunks in LAN (unit: seconds)}
			\centering
			\label{tab:streaming_time_cost_LAN}
			\begin{tabular}{ c | c | c | c | c }
				\hline
				Fraction of delivered chunks& 32 KB & 256 KB & 1 MB & 2 MB\\
				\hline\hline
				10\% & 1.24  & 1.49 & 2.89 & 4.92 \\
				\hline
				20\% & 2.29  & 3.07 & 5.48 & 9.55 \\
				\hline
				30\% & 3.30  & 4.72 & 8.26 & 14.42 \\
				\hline
				40\% & 4.22  & 6.22 & 11.02 & 19.07 \\
				\hline
				50\% & 5.18  & 7.73 & 13.75 & 23.83 \\
				\hline
				60\% & 6.11  & 9.23 & 16.50 & 28.59 \\
				\hline
				70\% & 7.03  & 10.71 & 19.23 & 33.28 \\
				\hline
				80\% & 7.91  & 12.28 & 22.01 & 38.15 \\
				\hline
				90\% & 8.77  & 13.76 & 24.78 & 38.15 \\
				\hline
				100\% & 9.59  & 15.25 & 27.55 & 47.84 \\
				\hline\hline
			\end{tabular}
		\end{scriptsize}
	\end{table}
}

\ignore{
	\begin{table}[!htpb]
		\centering
		\begin{scriptsize}
			\caption{The time costs of streaming 512 chunks in WAN (unit: seconds)}
			\centering
			\label{tab:streaming_time_cost_WAN}
			\begin{tabular}{ c | c | c | c | c }
				\hline
				Fraction of delivered chunks& 32 KB & 256 KB & 1 MB & 2 MB\\
				\hline\hline
				10\% & 8.06  & 8.58 & 14.03 & 15.15 \\
				\hline
				20\% & 15.46  & 16.61 & 25.82 & 28.93 \\
				\hline
				30\% & 22.92  & 24.92 & 37.13 & 43.17 \\
				\hline
				40\% & 30.21  & 33.03 & 48.08 & 56.97 \\
				\hline
				50\% & 37.48  & 41.17 & 59.11 & 70.69 \\
				\hline
				60\% & 44.72  & 49.27 & 70.17 & 84.42 \\
				\hline
				70\% & 51.94  & 57.36 & 81.17 & 98.27 \\
				\hline
				80\% & 59.31  & 65.62 & 92.16 & 113.75 \\
				\hline
				90\% & 66.51  & 73.72 & 102.82 & 128.65 \\
				\hline
				100\% & 73.69  & 81.81 & 116.74 & 143.64 \\
				\hline\hline
			\end{tabular}
		\end{scriptsize}
	\end{table}
}


\section{Related Work}
\label{sec:RelatedWork}

Here we review the pertinent technologies and discuss their insufficiencies in the specific context of p2p content delivery. Table~\ref{tab:p2p_cdn_comparison} summarizes the advantages provided by our protocol when compared with other representative related works.

\smallskip
\noindent{\bf P2P information exchange schemes.} Many works~\cite{Piatek-et-al-2007-NSDI,Cohen-2003-P2P,Sherman-et-al-2012-TON,Kamvar-et-al-2003-WWW,Sirivianos-2007-Usenix,Shin-et-al-2017-TON,Levin-et-al-2008-SIGCOMM} focused on the basic challenge to incentivize users in the p2p network to voluntarily exchange information. However, these schemes have not been notably successful in combating free-riding problem and strictly ensuring the fairness. Specifically, the schemes in BitTorrent \cite{Cohen-2003-P2P}, BitTyrant~\cite{Piatek-et-al-2007-NSDI},
FairTorrent \cite{Sherman-et-al-2012-TON}, PropShare \cite{Levin-et-al-2008-SIGCOMM} support direct reciprocity (i.e., the willingness for participants to continue exchange basically depends on their past direct interactions, e.g., the {\em Tit-for-Tat} mechanism in BitTorrent) for participants, which cannot accommodate the {\em asymmetric} interests (i.e., participants have distinct types of resources such as bandwidth and cryptocurrencies to trade between each others) in the p2p content delivery setting. For indirect reciprocity (e.g., reputation, currency, credit-based) mechanisms including Eigentrust~\cite{Kamvar-et-al-2003-WWW}, Dandelion~\cite{Sirivianos-2007-Usenix}, they suffer from Sybil attacks, e.g., a malicious peer could trivially generate a sybil peer and ``deliver to himself" and then rip off the credits. We refer readers to~\cite{Shin-et-al-2017-TON} for more discussions about potential attacks to existing p2p information exchange schemes. For T-chain~\cite{Shin-et-al-2017-TON}, it still considers rational attackers and cannot strictly ensure the delivery fairness as an adversary can waste a lot of bandwidth of deliverers though the received content is encrypted. 

More importantly, all existing schemes, to our knowledge, are presented in the non-cooperative game-theoretic setting, which means that they only consider independent attackers free ride spontaneously without communication of their strategies, and the attackers are rational with the intentions to maximize their own benefits. However, such rational assumptions are elusive to guarantee the fairness for parties in the ad-hoc systems accessible by all malicious entities. 
Our protocol, on the contrary, assures the delivery fairness in the cryptographic sense. Overall, our protocol can rigorously guarantee fairness for all participating parties, i.e., deliverers with delivery fairness, providers and consumers with exchange fairness. Also, the fairness in the p2p information exchange setting is typically measured due to the discrepancy between the number of pieces uploaded and received over a long period~\cite{Joe-et-al-2016-ICDCS} for a participant. If we examine each concrete delivery session, there is no guarantee of fairness. This further justifies that the p2p information exchange schemes are not directly suitable for the specific p2p content delivery setting.

\begin{table*}[!t]
	\centering
	\begin{scriptsize}
		\caption{Comparison of different related representative approaches}
		\vspace{-2mm}
		\centering
		\label{tab:p2p_cdn_comparison}
		\setlength{\tabcolsep}{1.0em}
		{\renewcommand{\arraystretch}{1.2}
		\begin{tabular}{ c | c | c | c | c | c | c }
			\hline
			\hline
			\multicolumn{2}{c|}{\backslashbox{Schemes}{Features}} &
			\multicolumn{1}{c|}{\makecell[c]{What to exchange? \\ (Incentive type)}} & 
			\multicolumn{1}{c|}{\makecell[c]{Delivery Fairness\\ c.f., Sec.4}} & 
			\multicolumn{1}{c|}{\makecell[c]{Confidentiality\\ c.f., Sec.4}}  & 
			\multicolumn{1}{c|}{\makecell[c]{Exchange Fairness\\c.f., Sec.4}} &
			\multicolumn{1}{c}{\makecell[c]{On-chain Costs,\\$n$ is the \# of content chunks}} \\
			\hline
			\hline
			\multicolumn{1}{c|}{\multirow{3}{*}{\makecell[c]{\\ \\P2P Information\\ Exchange}}}
			& \multicolumn{1}{c|}{BitTorrent~\cite{Cohen-2003-P2P}}
			& \multicolumn{1}{c|}{\makecell*[{}{c}]{Files $\leftrightarrow$ Files\\(Tit-for-Tat)}}
			& \multicolumn{1}{c|}{$\times$}
			& \multicolumn{1}{c|}{$\times$}
			& \multicolumn{1}{c|}{Not fully}
			& \multicolumn{1}{c}{n/a} \\[4pt]\cline{2-7}
			& \multicolumn{1}{c|}{Dandelion~\cite{Sirivianos-2007-Usenix}}
			& \multicolumn{1}{c|}{\makecell*[{}{c}]{Files $\leftrightarrow$ Credits\\ (Reputation)}}
			& \multicolumn{1}{c|}{$\times$}
			& \multicolumn{1}{c|}{$\times$}
			& \multicolumn{1}{c|}{Not fully}
			& \multicolumn{1}{c}{n/a} \\[4pt]\cline{2-7}
			& \multicolumn{1}{c|}{T-Chain~\cite{Shin-et-al-2017-TON}}
			& \multicolumn{1}{c|}{\makecell*[{}{c}]{Files $\leftrightarrow$ Files \\ (Tit-for-Tat)}}
			& \multicolumn{1}{c|}{$\times$}
			& \multicolumn{1}{c|}{$\surd$}
			& \multicolumn{1}{c|}{Not fully}
			& \multicolumn{1}{c}{n/a} \\[4pt]\cline{2-7}
			
			
			\hline
			\multirow{3}{*}{\makecell[c]{\\ Decentralized\\Content Delivery}}
			& \multicolumn{1}{c|}{Gringotts~\cite{Goyal-et-al-2019-Usenix}}
			& \multicolumn{1}{c|}{\makecell{Bandwidth $\leftrightarrow$ Coins\\(Monetary)}}
			& \multicolumn{1}{c|}{\makecell*[{}{c}]{{\em multiple} chunks' deliveries\\ not paid in worst cases}}
			& \multicolumn{1}{c|}{$\times$}
			& \multicolumn{1}{c|}{$\times$}
			& \multicolumn{1}{c}{$O(n)$} \\[4pt]\cline{2-7}
			& \multicolumn{1}{c|}{CacheCash~\cite{Aalmashaqbeh-2019-Columbia}}
			& \multicolumn{1}{c|}{\makecell*[{}{c}]{Bandwidth $\leftrightarrow$ Coins\\(Monetary)}}
			& \multicolumn{1}{c|}{\makecell*[{}{c}]{all chunks' deliveries\\ not paid in worst cases}}
			& \multicolumn{1}{c|}{$\times$}
			& \multicolumn{1}{c|}{$\times$}
			& \multicolumn{1}{c}{$[o(1),O(n)]$} \\[4pt]\cline{2-7}
			& \multicolumn{1}{c|}{\textbf{Our Protocols}}
			& \multicolumn{1}{c|}{\makecell*[{}{c}]{Bandwidth/Files $\leftrightarrow$ Coins\\(Monetary)}}
			& \multicolumn{1}{c|}{\makecell*[{}{c}]{{\em one} chunk's delivery\\ not paid in worst cases}}
			& \multicolumn{1}{c|}{$\surd$}
			& \multicolumn{1}{c|}{$\surd$}
			& \multicolumn{1}{c}{$\Tilde{O}(1)$} \\
			\hline
			\hline
			
		\end{tabular}
		}
	\end{scriptsize}
	\vspace{-4mm}
\end{table*}

\smallskip
\noindent{\bf Fair exchange and fair MPC.} There are also intensive works on fair exchange protocols in cryptography.  It is well-known that a fair exchange protocol cannot be designed to provide complete exchange fairness without a trusted third party (TTP)~\cite{Pagnia-1999-TUD-BS}, which is a specific impossible result of the general impossibility of fair multi-party computation (MPC) without honest majority \cite{Cleve-1983-STOC}. Some traditional ways hinge on a TTP~\cite{Micali-2003-PODC,Asokan-et-al-2000-JSAC, Belenkiy-et-al-2007-WPES,Kupccu--et-al-2010-RSAC} to solve this problem, which has been reckon hard to find such a TTP in practice. 
To avoid the available TTP requirement, some other studies \cite{Blum-1983-STOC,Damgaard-1995-Cryptology,Pinkas-2003-EUROCRYPT,Garay-et-al-2006-TCC} rely on the ``gradual release'' approach, in which the parties act in turns to release their private values bit by bit,
such that even if one malicious party aborts, 
the honest party can recover the desired output by investing computational resources (in form of CPU time) comparable to that of the adversary uses.
Recently, the blockchain offers an attractive way to instantiate a non-private TTP, and a few results \cite{Maxwell-2016-BitcoinCore,Dziembowski-et-al-2018-CCS,Eckey-et-al-2019-Arxiv,Kiayias-et-al-2016-Crypto, Choudhuri-et-al-2017-CCS,Bentov-and-Kumaresan-2014-Crypto} leverage this innovative decentralized infrastructure to facilitate fair exchange and fair MPC despite the absence of honest majority. Unfortunately, all above fair exchange and fair MPC protocols fail to guarantee {\em delivery fairness} in the specific p2p content delivery setting, as they cannot capture the fairness property for the special exchanged item (i.e., bandwidth), as earlier discussed in Section~\ref{sec:Introduction}.

\smallskip
\noindent{\bf State channels.} A state channel establishes a private p2p medium, managed by pre-set rules, allowing the involved parties to update state unanimously by exchanging authenticated state transitions off-chain~\cite{Gudgeon-et-al-2020-FC}. Though our protocols can be reckon as the application of payment channel networks (PCNs) (or more general state channels~\cite{Miller-et-al-2019-FC}) yet there are two key differences: i) fairness in state channels indicates that an honest party (with valid state transition proof) can always withdraw the agreed balance from the channel~\cite{Gudgeon-et-al-2020-FC}, while our protocols, dwelling on the delivery fairness in the specific context of p2p content delivery, ensure the bandwidth contribution can be quantified and verified to generate such a valid state transition proof; ii) state channels essentially allow any two parties to interact, while our protocols target the interaction among any three parties with a totally different payment paradigm~\cite{Aalmashaqbeh-2019-Columbia} for p2p content delivery.

\smallskip
\noindent{\bf Decentralized content delivery.} There exist some systems that have utilized the idea of exchanging bandwidth for rewards to incentivize users' availability or honesty such as Dandelion~\cite{Sirivianos-2007-Usenix}, Floodgate~\cite{Nair-et-al-2008-ICCCN}. However, different drawbacks impede their practical adoption, as discussed in~\cite{Aalmashaqbeh-2019-Columbia}. Here we elaborate the comparison with two protocols, i.e., Gringotts~\cite{Goyal-et-al-2019-Usenix}, CacheCash~\cite{Aalmashaqbeh-2019-Columbia}, that target the similar p2p content delivery scenario. 

{\em Application Scenario.} Typically, the p2p content delivery setting involves asymmetric exchange interests of participants, i.e., the consumers expect to receive a specific content identified by a certain digest in time, while the providers and the deliverers would share their content (valid due to the digest) and bandwidth in exchange of well-deserved payments/credits, respectively. 
Unfortunately, Gringotts and CacheCash fail to capture this usual scenario, and cannot support the content providers to sell content over the p2p network, due to the lack of content confidentiality and exchange fairness. In greater detail, both Gringotts and CacheCash delegate a copy of raw content to the deliverers, which results in a straightforward violation of exchange fairness, i.e., a malicious consumer can pretend to be or collude with a deliverer to obtain the plaintext content without paying for the provider.

{\em Delivery Fairness.} Gringotts typically requires the deliverer to receive a receipt (for acknowledging the resource contribution) only after multiple chunks are delivered, which poses the risk of losing bandwidth for delivering multiple chunks. For CacheCash, a set of deliverers are selected to distribute the chunks in parallel, which may cause the loss of bandwidth for all chunks in the worst case. Our protocols ensures that the unfairness of delivery is bounded to one chunk of size $\eta$.

{\em On-chain Costs.} Gringotts stores all chunk delivery records on the blockchain, and therefore the on-chain costs is in $O(n)$. While for CacheCash, the deliverers can obtain {\em lottery tickets} (i.e., similar to ``receipts") from the consumer after each ``valid" chunk delivery. The on-chain costs is highly pertinent to the winning probability $p$ of tickets. E.g., $p = \frac{1}{n}$ means that on average the deliverer owns a winning ticket after $n$ chunks are delivered, or $p = 1$ indicates that the deliverer receives a winning ticket after each chunk delivery, leading to at most $O(n)$ on-chain costs of handling redeem transactions. For our protocols, the on-chain costs is bounded to $\Tilde{O}(1)$.

Additionally, Gringotts allows streaming of content chunks, which functions similarly to our $\mathsf{FairStream}$ protocol, and CacheCash demands to download all the chunks, which applies to the similar scenario as our $\mathsf{FairDownload}$ protocol.

\vspace{1mm}
\section{Conclusion and Future Works}
\label{sec:Conclusion}
We present the first two fair p2p content delivery protocols atop blockchains to support {\em fair p2p downloading} and {\em fair p2p streaming}, respectively. They enjoy strong fairness guarantees to protect any of the content provider, the content consumer, and the content deliverer from being ripped off by other colluding parties. Detailed complexity analysis and extensive experiments of prototype implementations are performed and demonstrate that the proposed protocols are highly efficient. 

Yet still, the area is largely unexplored and has a few immediate follow-up, for example: (i) in order to realize maximized delivery performance, it is desirable to design a mechanism of adaptively choosing deliverers during each delivery task; (ii) it is also enticing to leverage the off-chain payment channels to handle possible micropayments and further reduce the on-chain cost; (iii) to better preserve the digital rights of sold contents against pirating consumers, some digital rights management (DRM) schemes can be introduced.

\bibliographystyle{./IEEEtran}
{\footnotesize
\bibliography{./references}
}

\end{document}